\documentclass[11pt,a4paper]{article}
\usepackage[T1]{fontenc}
\usepackage{lmodern}
\usepackage[a4paper,left=14mm,right=14mm,top=20mm,bottom=21mm,footskip=10mm]{geometry}
\usepackage{amsmath,amssymb,amsfonts,amsthm}
\usepackage{graphicx,booktabs,multirow,array,makecell}
\usepackage{adjustbox}
\usepackage{algorithm,algpseudocode}
\usepackage[dvipsnames,svgnames]{xcolor}
\usepackage{cite,chapterbib}
\usepackage{url}
\usepackage{etoolbox}
\usepackage{microtype}
\usepackage[font=small,labelfont=bf,labelsep=period]{caption}
\usepackage[hidelinks,bookmarksnumbered]{hyperref}
\hypersetup{pdftitle={Exact Mixed-Locality Encodings for Resource-Constrained QAOA},
 pdfauthor={Jingcong Liu, Yiding Su, Ruojun Zhang, Hafiz Muhammad Waseem, Bo Chen, Carsten Maple}}
\graphicspath{{figures/}}
\newtheorem{theorem}{Theorem}
\newtheorem{proposition}{Proposition}
\newtheorem{corollary}{Corollary}

\newtheorem{remark}{Remark}

\newcommand{\B}{\mathbb{B}}
\newcommand{\R}{\mathbb{R}}
\newcommand{\PR}{P_{\mathrm{R}}}

\newcommand{\Jres}{J_{\mathrm{res}}}
\newcommand{\Efib}{E_{\mathrm{fib}}}
\newcommand{\norminf}[1]{\lVert #1 \rVert_{\infty}}
\DeclareMathOperator*{\argmin}{arg\,min}
\DeclareMathOperator{\msupp}{msupp}

\def\BibTeX{{\rm B\kern-.05em{\sc i\kern-.025em b}\kern-.08em
    T\kern-.1667em\lower.7ex\hbox{E}\kern-.125emX}}

\renewcommand{\thesection}{\Roman{section}}

\makeatletter
\newcommand{\authorrefmark}[1]{\textsuperscript{#1}}
\newcommand{\address}[2][]{\expandafter\gdef\csname affil#1\endcsname{#2}}
\newcommand{\tfootnote}[1]{\gdef\fundingnote{#1}}
\newcommand{\corresp}[1]{\gdef\correspondence{#1}}
\newcommand{\MainTitle}[1]{%
  \twocolumn[{\@twocolumnfalse
  \pdfbookmark[0]{Exact Mixed-Locality Encodings for Resource-Constrained QAOA}{manuscript}
  \begin{center}
    \parbox{.9\textwidth}{\centering\LARGE\bfseries\@title\par}\par\vspace{1em}
    {\large\let\uppercase\@firstofone\@author\par}\vspace{.8em}
    \begingroup\footnotesize
    \@for\n:=1,2,3,4\do{\textsuperscript{\n}\csname affil\n\endcsname\par}
    \vspace{.45em}\correspondence\par
    \endgroup
  \end{center}#1}]\small
  \begin{NoHyper}\begingroup\renewcommand{\thefootnote}{}\footnotetext{\fundingnote}\endgroup\end{NoHyper}}
\newenvironment{keywords}{\begin{quotation}\small\noindent\textbf{Keywords.}\ }{\end{quotation}}
\newcommand{\PARstart}[2]{#1#2}
\AddToHook{env/thebibliography/begin}{\if@twocolumn\footnotesize\raggedright\else\small\fi}
\apptocmd{\thebibliography}{\if@twocolumn\setlength{\itemsep}{2pt plus .3pt}\fi}{}{}
\NewDocumentCommand{\Figure}{O{tbp} d() O{} m +m}{%
  \in@{\textwidth}{#3}\ifin@\def\figenv{figure*}\else\def\figenv{figure}\fi
  \begin{\figenv}[!tbp]\centering
  \includegraphics[#3,height=.68\textheight,keepaspectratio]{#4}
  \caption{#5}\end{\figenv}}
\NewDocumentEnvironment{IEEEbiography}{O{} m +b}
  {\par\addvspace{.85\baselineskip}\noindent
   \begin{minipage}[c]{1in}#1\end{minipage}\hfill
   \begin{minipage}[c]{\dimexpr\linewidth-1in-1.3em\relax}\small
     \textbf{#2} #3\end{minipage}\par}{}

\newcommand{\MainContentsSetup}{\let\originaladdtocontents\addtocontents
  \renewcommand{\addtocontents}[2]{\ifstrequal{##1}{toc}{}{\originaladdtocontents{##1}{##2}}}}
\renewcommand*\l@subsection{\@dottedtocline{2}{1.5em}{3.6em}}
\newcommand{\SupplementSetup}{%
  \onecolumn
  \newgeometry{left=22mm,right=22mm,top=23mm,bottom=24mm,footskip=12mm}%
  \normalsize
  \let\addtocontents\originaladdtocontents
  \renewcommand{\thesection}{S\arabic{section}}
  \@for\cnt:=section,subsection,subsubsection,figure,table,equation,algorithm,theorem,proposition,corollary,lemma\do{%
     \setcounter{\cnt}{0}}
  \@for\cnt:=figure,table,equation,algorithm,theorem,proposition,corollary,lemma\do{%
     \expandafter\edef\csname the\cnt\endcsname{S\noexpand\arabic{\cnt}}}
  \@for\cnt:=section,figure,table,equation,algorithm,theorem,proposition,corollary,lemma\do{%
     \expandafter\edef\csname theH\cnt\endcsname{supp.\noexpand\arabic{\cnt}}}
  \pdfbookmark[0]{Supplementary Information}{supplement}
  \let\originaltableofcontents\tableofcontents
  \renewcommand{\tableofcontents}{\begingroup
    \edef\restoresuppcontext{\noexpand\gdef\noexpand\the@ipfilectr{\the@ipfilectr}}%
    \small\originaltableofcontents\restoresuppcontext\endgroup}
  \renewcommand{\maketitle}{\begin{center}
    {\LARGE\bfseries\@title\par}\vspace{1em}
    {\normalsize\@author\par}\end{center}}}
\makeatother

\begin{document}
\MainContentsSetup
\title{Exact Mixed-Locality Encodings for Resource-Constrained QAOA}

\author{\uppercase{Jingcong Liu}\authorrefmark{1},
\uppercase{Yiding Su}\authorrefmark{1},
\uppercase{Ruojun Zhang}\authorrefmark{2},
\uppercase{Hafiz Muhammad Waseem}\authorrefmark{3},\\
\uppercase{Bo Chen}\authorrefmark{4},
and \uppercase{Carsten Maple}\authorrefmark{3}}

\address[1]{Department of Mathematics, University of Warwick, Coventry CV4 7AL,
United Kingdom}
\address[2]{School of Electronics and Computer Science, University of Southampton,
Southampton SO17 1BJ, United Kingdom}
\address[3]{Warwick Manufacturing Group, University of Warwick, Coventry CV4 7AL,
United Kingdom}
\address[4]{Warwick Business School, University of Warwick, Coventry CV4 7AL,
United Kingdom}

\tfootnote{This work was supported in part by the National Hub for Edge AI, UK Research and Innovation (UKRI), through Research Grant EP/Y028813/1, in part by the UKRI Cross Council under Grant UKRI3734, and in part by the Academic Centre of Excellence in Cyber Security Research, University of Warwick, under Grant EP/R007195/1.}

\corresp{Corresponding author: Hafiz Muhammad Waseem (e-mail: HM.Waseem@warwick.ac.uk).}
\MainTitle{
\begin{abstract}
Higher-order objectives possess many exact Hamiltonian encodings, and these encodings differ in the qubits, penalties, gates, and depth that they require. Representation selection is therefore an algorithmic decision for the quantum approximate optimisation algorithm on resource-constrained hardware. This work develops a pointwise-exact family of representations in which selected higher-order interactions are replaced by shared Rosenberg auxiliaries while the remaining interactions stay native. A sharp single-step penalty condition and a closed form for cubic residuals certify exactness at every stage of such a selective reduction. Within the direct-pair class, a static feasibility theory identifies infinite families of objectives for which a mixed representation meets joint budgets that exclude both the native and the fully reduced endpoint. A separate construction extends this separation to gate and depth limits under a fixed synthesis. Compilation on 12-qubit line, ring, and grid targets finds certified mixed-only regions for 23 of 30 weighted maximum three-satisfiability instances, all 30 constructed instances, and 2 of 30 cubic spin-glass instances under the stated class and penalty rule. A separate routing experiment retains 52.9\% of mixed-only budget points across five seeds within fixed candidate sets. In noiseless depth-one evaluation, the native encoding attains the best mean objective under the tested protocols, and the joint resource-and-quality test finds no mixed-only success at the primary budget. Resource feasibility is thus a sound reason to select a mixed encoding, whereas solution quality requires a separate assessment.
\end{abstract}

}

\section{Introduction}
\label{sec:introduction}

\PARstart{Q}{uantum} processors are approaching the scale at which combinatorial optimisation becomes a credible application. However, contemporary NISQ-era devices impose strict limits on the number of qubits, the connectivity between them, and the circuit depth that can be executed before noise degrades fidelity~\cite{Weidenfeller2022,Harrigan2021,Abbas2024}. Within these limits, the quantum approximate optimisation algorithm (QAOA) is the principal variational method, and its cost Hamiltonian is derived directly from the objective to be minimised \cite{Farhi2014,Zhou2020}. Many objectives of practical interest, including clause penalties in satisfiability, hypergraph objectives, and many-body spin models, contain interactions among three or more binary variables \cite{BorosHammer2002,Hadfield2021}. Such an objective can be represented by many algebraically equivalent Hamiltonians, each of which is an encoding, or representation, of the objective. These representations differ in the number of qubits they occupy, the interaction graphs they induce, the penalty coefficients they require, and the compiled circuits they produce. The choice among them is therefore an algorithmic decision rather than a matter of notation. It governs whether a given instance can be executed on a given device and how well it performs once executed.

Two established strategies form the endpoints of this design space. Native encoding retains the higher-order objective and synthesises its cost evolution directly, so it needs no auxiliary qubits and no artificial penalties. It has outperformed quadratic encodings on several tested problem families~\cite{Glos2022,CampbellDahl2022,Stein2023,Bell2025,Koch2026}. A standard parity gadget for a Pauli term of weight $k$ uses $2(k-1)$ CNOT gates before routing~\cite{CampbellDahl2022,Verchere2023}, and the compiled cost further depends on connectivity, routing, and circuit optimisation. Full quadratization replaces every higher-order term by a quadratic function on an enlarged register, and minimisation over the auxiliaries recovers the original objective pointwise~\cite{Rosenberg1975,BorosGruber2014,Dattani2019}. Individual phase gadgets become simpler, but the reduction adds qubits, couplings, and penalties. Recent work on this trade-off includes circuit optimisation~\cite{Verchere2023}, automatic quadratization~\cite{Schmidbauer2024,Schmidbauer2026}, hardware-aware auxiliary selection~\cite{Rovara2026}, and approximate reductions that modify the objective~\cite{Dragoi2026}. These methods address different representation classes and resource objectives, and some methods aimed at quadratic output also permit partial substitution.

Native and fully reduced forms belong to a broader family of exact representations. A mixed representation replaces selected higher-order interactions with shared auxiliaries and retains the others in their original form. Partial substitution itself is not new. The question addressed here is which exact representation fits the available resources, and in particular whether a mixed design can meet budgets that exclude both endpoints. This question is studied within a stated direct-pair class, and three issues require care. First, a shared penalty must control the combined residual of every selected term rather than the coefficient of any single term. Second, later substitutions may act on earlier auxiliaries, so exactness must survive composition. Third, selection must account for qubits, gates, depth, and penalty scale together, since a design that saves one resource may exhaust another. Pointwise exactness preserves the classical optimum but does not uniquely determine the energies of inconsistent auxiliary assignments, so it guarantees neither equal QAOA dynamics nor equal solution quality. Warm-start initialisation on the enlarged register is therefore assessed as a separate question.

The construction applies to objectives of arbitrary degree, whereas the resource results and the primary experiments concern cubic objectives and direct pair substitutions. This paper makes four contributions.
\begin{enumerate}

\item \textbf{Exact selective family.} A sharp penalty condition for a shared Rosenberg substitution is proved to be necessary as well as sufficient, and it is shown to remain valid when substitutions are applied in sequence. For cubic residuals, the threshold has a closed form in the signed coefficient sums, so any partial reduction can be certified exact without evaluating every assignment.

\item \textbf{Feasibility boundaries under joint budgets.} A static capacity characterises direct-pair feasibility under budgets on auxiliary count, penalty, auxiliary degree, and retained terms. Its closed form for unit-weight pair-star forests yields infinite families for which only a mixed representation is feasible, and a separate construction extends the separation to gate and depth limits under the reference synthesis.

\item \textbf{Compiled resource selection.} Exact candidates are compiled under a fixed protocol, checked against joint budgets 
on physical width, gates, depth, and penalty, and ranked by a declared resource score.
Fixed candidate sets support paired comparisons on line, ring, and grid targets, together with separate checks of endpoint exclusion and routing sensitivity.

\item \textbf{Separate assessment of solution quality.} 
Noiseless depth-one QAOA tests compare the original objective across representations and decompose the improvement of a warm-started circuit into a preparation gain and an evolution gain. A joint test then checks whether a resource-feasible mixed design can also meet a stated quality target when neither endpoint does.

\end{enumerate}

The main contribution is representation selection in an exact design space under joint budgets, with the static boundaries, the fixed-synthesis construction, and the finite compiler experiments each holding within its own stated scope. None of them establishes a universally preferred encoding. The native representation performs best on average in the tested QAOA protocols, although the best representation for an individual instance can change, so resource feasibility and solution quality are interpreted as distinct criteria throughout.

The remainder of the paper is organised as follows. Section~\ref{sec:related} reviews related work and positions the framework relative to the closest strategies. Section~\ref{sec:reduction} develops the exact selective family and its penalty thresholds. Section~\ref{sec:selection} defines the compilation model, the resource score and budgets, the feasibility boundaries, the search procedures, and the warm starts. Section~\ref{sec:experiments} reports the four experiments, Section~\ref{sec:discussion} interprets the results and states their limitations, and Section~\ref{sec:conclusion} concludes the paper.

\section{Related Work}
\label{sec:related}

The representation problem addressed in this paper is connected to prior research on classical quadratization, on higher-order encodings and their circuit synthesis for QAOA, on hardware-aware reformulation and compilation, and on the initialisation of variational circuits from classical relaxations. Recent reviews of variational quantum algorithms and of quantum optimisation identify problem encoding and circuit resources as central open issues \cite{Cerezo2021,Blekos2024,Abbas2024}. This section reviews the four areas and positions the present framework relative to the closest existing strategies, which Table~\ref{tab:closest_work} summarises.

\subsection{Quadratization of Pseudo-Boolean Functions}
\label{sec:related:quadratization}

Pseudo-Boolean optimisation and the multilinear representation of Boolean functions are established subjects \cite{BorosHammer2002}. Rosenberg proved that any pseudo-Boolean function can be reduced to a quadratic function over an enlarged variable set by replacing a product of two variables with a new variable and adding a penalty that enforces consistency at every minimiser \cite{Rosenberg1975}. Boros and Gruber surveyed this and later constructions, described the aggregative treatment of terms that share a common product, and derived the penalty condition $M>\max|A|$ for a substitution applied to all occurrences of the product \cite{BorosGruber2014}. Dattani collected quadratization methods with emphasis on their use in quantum annealing \cite{Dattani2019}. In computer vision, Kolmogorov and Zabih characterised the energy functions that graph cuts can minimise~\cite{KolmogorovZabih2004}. Graph-cut constructions also reduce higher-order binary energies \cite{FreedmanDrineas2005,Ishikawa2011}.

Subsequent work bounded the number of auxiliary variables required for general and structured function classes \cite{AnthonyEtAl2017,Anthony2016,BorosCramaRodriguezHeck2020}. Optimised substitution choices have been studied for  fourth-degree objectives \cite{VermaLewis2020} and cubic objectives arising from Max-3SAT \cite{VermaLewisKochenberger2021}, while recursive reformulations relate monomial decomposition, consistency penalties, and convexification \cite{CramaEtAl2025}.
These methods use different representation classes and resource objectives, and some also permit partial substitution. Section~\ref{sec:reduction} uses the established Rosenberg mechanism to define an exact family for resource-constrained selection, without claiming partial substitution as a new operation.

\subsection{Higher-Order Encodings and QAOA Circuit Synthesis}
\label{sec:related:higherorder}

QAOA encodes an objective as a diagonal cost Hamiltonian and alternates cost evolution with mixing operations \cite{Farhi2014,Zhou2020}. Hadfield characterised the representation of Boolean and real functions as Hamiltonians and showed that the Pauli weight of a diagonal encoding equals the degree of the multilinear polynomial \cite{Hadfield2021}. Several works compared native higher-order encodings with quadratic encodings and reported advantages for the higher-order form on particular problem families \cite{Glos2022,CampbellDahl2022,Stein2023,Bell2025}. Koch et al. recently reported substantial resource savings from higher-order encodings relative to QUBO formulations \cite{Koch2026}. Higher-order phases decompose into one-qubit and two-qubit gates through parity gadgets \cite{CampbellDahl2022}, and circuit-level optimisation reduces the resulting overhead \cite{Verchere2023}. Phase-gadget synthesis and the CNOT complexity of CNOT-phase circuits provide the compilation background for the reference synthesis used here \cite{CowtanPhase2020,CowtanUCC2020,Amy2018}, and hardware-native decompositions of multi-qubit Pauli rotations offer an alternative \cite{Sriluckshmy2023}. Analytical studies characterise QAOA performance on hypergraph and spin-glass models \cite{Basso2022,BassoHighDepth2022}. Hardware experiments demonstrate higher-order Ising cost layers for hardware-compatible instances on the tested devices \cite{PelofskeShortDepth2024,Pelofske2024}. Approximate quadratization modifies the implemented Hamiltonian to reduce circuit cost and accepts a controlled loss of objective fidelity \cite{Dragoi2026}.

These studies examine different representation choices and circuit costs, with their stated scopes compared in Table~\ref{tab:closest_work}. This paper instead studies joint resource feasibility within an exact selective family, including the conditions under which both endpoints are excluded.

\begin{table*}[!t]
\caption{Scope of the cited methods. ``Not studied'' refers to the cited formulation. The last column concerns native, full and mixed feasibility under joint budgets. Exactness means recovery after auxiliary minimisation; it does not imply equal QAOA dynamics.}
\label{tab:closest_work}
\centering
\scriptsize
\setlength{\tabcolsep}{2.3pt}
{
\begin{tabular}{@{}p{0.2\textwidth}p{0.08\textwidth}p{0.14\textwidth}p{0.19\textwidth}p{0.20\textwidth}p{0.14\textwidth}@{}}
\toprule
Method & Pointwise exact & Mixed output & Joint reduced-set / pair choice & Resource objective & Endpoint phase theorem \\
\midrule
Rosenberg substitution ~\cite{Rosenberg1975,BorosGruber2014} & Yes & Partial use possible & Local construction & Not a compiler objective & Not studied \\
Compact quadratization \cite{BorosCramaRodriguezHeck2020} & Yes & Quadratic output & Construction dependent & Auxiliary bounds & Not studied \\
Automatic full reduction \cite{Schmidbauer2024,Schmidbauer2026} & Yes & Full-Q target & Substitution choice & Transformation and circuit costs & Not studied \\
Hardware-aware selection \cite{Rovara2026} & Yes & Full-Q target & Auxiliary selection & Graph structure and mapped depth & Not studied \\
Approximate reduction \cite{Dragoi2026} & Not in general & Ansatz dependent & Different ansatz space & Qubit and circuit cost & Not studied \\
\textbf{This work} & Yes & Explicit output & Both $D$ and $\rho$ & Fixed-compiler joint budgets & Within direct-pair class \\
\bottomrule
\end{tabular}}
\end{table*}

\subsection{Hardware-Aware Reformulation and Compilation}
\label{sec:related:hardware}

A compact logical register does not guarantee a compact physical circuit. Weidenfeller et al. showed that routing can be the largest cost in QAOA implementations on sparse superconducting devices~\cite{Weidenfeller2022}. Hardware experiments on non-planar problems confirmed the cost of mapping to planar connectivity~\cite{Harrigan2021}, and scaling analyses quantified how gate counts and depth grow with problem size on near-term devices~\cite{Lotshaw2022}. Qubit mapping and routing have been studied extensively as compiler problems \cite{Li2019,Zulehner2019,Murali2019}, compilation flows specific to QAOA exploit the commutation of cost terms to reduce depth \cite{Alam2020}, and connectivity has a measurable effect on algorithm performance \cite{Holmes2020}. Depth lower bounds and depth optimisation for QAOA circuits address the same resource from the algorithmic side \cite{Herrman2021depth,Herrman2021}. The closest studies are those of Schmidbauer et al.\ and Rovara et al.\ Schmidbauer et al.\ studied how polynomial reductions change QAOA circuits and developed automatic quadratization \cite{Schmidbauer2024,Schmidbauer2026}. Rovara et al.\ selected auxiliaries to control the interaction-graph degree for devices with limited connectivity \cite{Rovara2026}. The cited formulations target full quadratization and use different resource objectives. Penalty scale is treated as a declared coefficient constraint, but gate count, depth and penalty scale alone do not determine hardware duration or success.

Penalty design for constrained problems remains an active topic. Existing methods include linear Ising penalty functions \cite{Mirkarimi2024}, penalty-weight selection \cite{Ayodele2022},  and unbalanced penalisation \cite{MontanezBarrera2024}, while other approaches change the variable encoding rather than the interaction locality. Examples include encoding-independent formulations \cite{Dominguez2023}, analyses of discrete-variable encoding trade-offs \cite{Sawaya2023}, domain-wall encodings \cite{Chancellor2019}, and qubit-efficient encodings \cite{Tan2021}. Section~\ref{sec:selection} selects within a pointwise-exact family using joint limits on physical width, compiled gates, depth and penalty. The comparison concerns the stated representation class and compilation protocol.

\subsection{Warm Starts and Initialisation}
\label{sec:related:warmstart}

Warm-start QAOA constructs the initial state and a compatible mixer from a classical relaxation \cite{Egger2021}. Semidefinite-programming initialisation was studied for MaxCut\cite{TateSDP2023}, warm-started QAOA with custom mixers was shown to converge for MaxCut under stated conditions \cite{TateCustom2023}, and annealing-inspired parameter initialisation offers a complementary strategy \cite{Sack2021}. Standard QAOA can fail to improve when started from a strong classical string \cite{Cain2023}, which motivates the separate measurement of the preparation gain and the evolution gain in Section~\ref{sec:experiments}. The mixer itself is a design choice within the alternating-operator framework \cite{Hadfield2019}. XY mixers preserve Hamming-weight constraints \cite{WangXY2020}, Grover mixers shift complexity from the mixer to state preparation \cite{Bartschi2020}, and multi-angle variants add parameters per term \cite{HerrmanMultiAngle2022}. These alternative mixers and ansätze are not used here, and the penalty-based treatment of auxiliaries is retained so that all representations use the same local mixer form.

An auxiliary for $x_i x_j$ requires an initial probability. This paper compares a relaxation pair moment with the product of the single-bit probabilities. Neither choice makes a product state satisfy $y_p = x_i x_j$ in every measurement outcome, so the comparison forms part of the evaluation protocol and does not by itself establish a general benefit from pair moments.

\section{Exact Selective Locality Reduction}
\label{sec:reduction}

This section develops the exact representation family positioned between native higher-order encoding and complete quadratization. It sets the notation, states a sharp penalty condition for one shared substitution, extends the condition to sequences of substitutions, and specialises the result to cubic objectives. Table~\ref{tab:notation} lists the symbols used in the main text and in the Supplementary Information.

Let $\B=\{0,1\}$, $[n]=\{1,\dots,n\}$, and $x_A=\prod_{i\in A}x_i$ with $x_\emptyset=1$.
Every function $f\colon\B^n\to\R$ has a unique multilinear expansion
$f(x)=\sum_{A\subseteq[n]}c_A x_A$. Its monomial support is
$\msupp(f)=\{A : c_A\neq 0\}$ and its degree is $\deg f=\max_{A\in\msupp(f)}|A|$
\cite{BorosHammer2002,Hadfield2021}. The diagonal encoding $H_f=f(N_1,\dots,N_n)$ with $N_i=(I-Z_i)/2$ satisfies $H_f|x\rangle=f(x)|x\rangle$, and its largest Pauli weight equals $\deg f$ \cite{Hadfield2021}. A representation $F_\sigma\colon\B^{n}\times\B^{a_\sigma}\to\R$ with $a_\sigma$ auxiliary bits is
\emph{pointwise exact} when
\begin{equation}
f(x)=\min_{y\in\B^{a_\sigma}}F_\sigma(x,y)\quad\text{for every }x\in\B^n .
\label{eq:pointwise}
\end{equation}
This identity preserves the optimal value and the projected minimisers. It requires $F_\sigma(x,y)\ge f(x)$ for every auxiliary assignment, but does not uniquely determine the full spectrum of $H_{F_\sigma}$. Classical exactness
therefore does not determine finite-depth QAOA behaviour, a distinction that
organises the experiments in Section~\ref{sec:experiments}. All constructions use the Rosenberg penalty \cite{Rosenberg1975}
\begin{equation}
\PR(a,b,y)=ab-2ay-2by+3y,
\label{eq:rosenberg}
\end{equation}
which equals zero when $y=ab$ and is at least one otherwise.

\begin{table}[!t]
\caption{Notation used in the main text and the Supplementary Information.}
\label{tab:notation}
\centering
\scriptsize
\setlength{\tabcolsep}{3pt}
\renewcommand{\arraystretch}{1.05}
\begin{tabular}{@{}l p{0.60\columnwidth}@{}}
\toprule
Symbol & Meaning \\
\midrule

$\B$, $[n]$ & $\{0,1\}$ and $\{1,\dots,n\}$ \\
$x$, $y$, $u=(x,y)$ & Original bits, auxiliary bits, combined register \\
$x_A$, $c_A$ & Monomial $\prod_{i\in A}x_i$ and its coefficient in $f$ \\
$\msupp(f)$, $\deg f$ & Monomial support of $f$ and its degree \\
$f$, $F_\sigma$ & Original objective and its pointwise-exact representation \\
$H_f$, $N_i$, $Z_i$ & Diagonal cost Hamiltonian, $(I-Z_i)/2$, Pauli $Z$ on qubit $i$ \\
$\sigma$, $a_\sigma$, $N_\sigma$ & Representation, auxiliary count, total qubits $n+a_\sigma$ \\
$\ell$, $L_\sigma$ & Target locality of reduced terms, maximum retained locality \\
$\mathcal{C}_3(f)$ & Cubic supports $\{A : |A|=3,\ c_A\neq 0\}$ \\
$D$, $R$ & Cubic terms reduced and cubic terms retained natively \\
$p$, $y_p$, $P_\rho$ & Variable pair, its product auxiliary, active pair set \\
$\rho$, $D_p$, $k_p$ & Term-to-pair assignment, terms assigned to pair $p$, and their number \\
$A_p$, $\norminf{A_p}$ & Aggregate residual of pair $p$ and its Boolean sup norm \\
$M_p$, $T_p$, $\mathcal{P}$ & Pair penalty, exactness threshold, deterministic penalty rule \\
$\PR$ & Rosenberg penalty of \eqref{eq:rosenberg} \\
$\Phi$, $\phi_\rho$ & Recursively consistent lift and its cubic form $(x_ix_j)_{p\in P_\rho}$ \\
$\partial_2 D$, $\tau_2(D)$ & Pair shadow of $D$ and its set-cover number \\
$\Delta_2(D)$, $r_{\mathrm{pair}}$ & Maximum pair multiplicity and pair-reuse score \\

$G_{2q,\sigma}$, $D_{2q,\sigma}$, $S_\sigma$ & Compiled two-qubit gates, two-qubit depth, SWAP count \\
$Q^{\Pi}_\sigma$ & Physical width of the circuit compiled under protocol $\Pi$ \\
$M_{\max,\sigma}$, $\kappa_\sigma$ & Largest active penalty and Pauli-coefficient dynamic range \\
$\Jres$, $w$, $s$ & Resource score, its weights, its normalisation scales \\
$Q_{\max}$, $B_G$, $B_D$, $B_M$ & Budgets on qubits, gates, depth, and penalty \\
$(r_0,M_0,\Delta_0,h_0)$ & Static budgets on auxiliaries, penalty, auxiliary degree, retained terms \\
$K_Q$, $R_Q$ & Maximum reducible term count and its pair-star closed form \\
$\mathcal{F}_{\mathrm{sel}}$, $\Sigma^{\mathcal{P}}_{\mathrm{ex}}$ & Feasible set and finite exact family under rule $\mathcal{P}$ \\
$\Efib$ & Consistency excess \\

$\bar\mu_i$, $\bar q_{ij}$ & Raw single and pair pseudo-moments from the relaxation \\
$\mu_i$, $q_p$, $\delta$ & Preparation probabilities limited to a fixed range and the range parameter \\
$|\Phi^{\mathrm{ws}}_\sigma\rangle$, $H^{\mathrm{ws}}_{\mathrm{M},\sigma}$ & Product warm state and its matched mixer \\
$\gamma$, $\beta$, $L$ & Cost angle, mixer angle, QAOA depth \\
$B$, $L_\sigma(B)$ & Two-qubit gate budget and the depth it permits \\
$E_f$, $P_{\mathrm{opt}}$, $R_{\mathrm{inc}}$ & Original-objective expectation, optimal-solution probability, auxiliary inconsistency \\
\bottomrule
\end{tabular}
\end{table}

\subsection{Sharp Aggregate-Residual Penalty}
\label{sec:reduction:penalty}

A selective reduction replaces one product $u_iu_j$ by a new bit $y_p$ in a chosen subset of the terms that contain it. The penalty must then control the combined effect of every term sharing $y_p$, not the coefficient of any single term. Let $q(u)=\sum_S d_S u_S$ be a collected multilinear polynomial on $\B^N$, fix a pair $p=\{i,j\}$, and choose a family $\mathcal{S}_p$ of terms of degree at least three that contain $p$. Let $v$ denote the coordinates other than $u_i$ and $u_j$. The selected residual is $A_p(v)=\sum_{S\in\mathcal{S}_p}d_S u_{S\setminus p}$, so that
\begin{equation}
q(u)=B_p(u)+u_iu_jA_p(v),
\label{eq:split}
\end{equation}
where $B_p$ collects all unselected terms. The substituted polynomial is
\begin{equation}
Q_{p,M_p}(u,y_p)=B_p(u)+y_pA_p(v)+M_p\,\PR(u_i,u_j,y_p).
\label{eq:substituted}
\end{equation}

\begin{proposition}[Sharp aggregate-residual penalty]
\label{prop:sharp}
Let $\norminf{A_p}=\max_{v\in\B^{N-2}}|A_p(v)|$ and $M_p\ge 0$. Then
\begin{equation}
q(u)=\min_{y_p\in\B}Q_{p,M_p}(u,y_p)\ \ \forall u
\ \iff\ M_p\ge\norminf{A_p}.
\label{eq:threshold}
\end{equation}
The consistent value $y_p=u_iu_j$ is the unique minimiser for every $u$ if and only if $M_p>\norminf{A_p}$. Moreover, for every $u$ and $y_p$,
\begin{equation}
Q_{p,M_p}(u,y_p)-q(u)\ge\bigl(M_p-\norminf{A_p}\bigr)\mathbf{1}\{y_p\neq u_iu_j\}.
\label{eq:margin}
\end{equation}
\end{proposition}

The proof evaluates the excess of the unique inconsistent choice of $y_p$ in the three cases $(u_i,u_j)=(0,0)$, $(u_i,u_j)\in\{(0,1),(1,0)\}$, and $(u_i,u_j)=(1,1)$. The excess equals $A_p+3M_p$, $A_p+M_p$, and $M_p-A_p$ respectively. Requiring all three values to be nonnegative for every $v$ is equivalent to \eqref{eq:threshold}, and \eqref{eq:margin} follows from $|A_p|\le\norminf{A_p}$. The threshold is sharp because the maximum defining $\norminf{A_p}$ is achieved at some point of the finite cube. Details are given in Supplementary Section~S1.

\begin{remark}
Proposition~\ref{prop:sharp} uses the classical Rosenberg mechanism. Boros and Gruber \cite{BorosGruber2014} group terms with a common product and use $M>\max|A|$ for the residual in $f=xyA+B$. Here the residual contains only the selected terms. Equality at the threshold gives pointwise exactness, while strict inequality gives a unique consistent auxiliary. The margin in \eqref{eq:margin} also supports sequential substitution.
Partial substitution itself is not new, and the contribution of this paper is the exact design space built on it together with its feasibility analysis under joint budgets.
\end{remark}

\subsection{Sequential Compositional Exactness}
\label{sec:reduction:composition}
 
Successive substitutions may act on terms that already contain earlier auxiliaries, so exactness must be shown to survive composition. Set $g^{(0)}=f$. At step $r$, choose a pair of current coordinates and compute the residual $A_r$ from the collected polynomial $g^{(r-1)}$, with all coordinates treated as independent Boolean variables. Then form $g^{(r)}$ by \eqref{eq:substituted} with a new auxiliary $y_r$ and a penalty $M_r\ge\norminf{A_r}$. Let $\Phi(x)=(\phi_1(x),\dots,\phi_R(x))$ denote the recursively consistent lift, in which $\phi_r(x)$ is the product of the two coordinates replaced at step $r$, evaluated at $(x,\phi_{<r}(x))$.

\begin{theorem}[Sequential compositional exactness]
\label{thm:composition}
After $R$ substitutions with margins $\delta_r=M_r-\norminf{A_r}\ge 0$, every $(x,y)$ satisfies
\begin{equation}
g^{(R)}(x,y)-f(x)\ge\sum_{r=1}^{R}\delta_r\,\mathbf{1}\bigl\{y_r\neq\phi_r(x,y_{<r})\bigr\},
\label{eq:telescope}
\end{equation}
with a nonnegative right-hand side, and $g^{(R)}(x,\Phi(x))=f(x)$. Hence $f(x)=\min_{y\in\B^R}g^{(R)}(x,y)$ for every $x$.
If every $\delta_r>0$, then $\Phi(x)$ is the unique minimiser.
\end{theorem}

Applying \eqref{eq:margin} at each step gives a one-step inequality, and adding these inequalities over $r$ yields \eqref{eq:telescope} because the intermediate terms cancel. Along the consistent lift, every penalty equals zero and every substituted block equals the block it replaced, which gives the equality. Uniqueness follows by considering the first index at which $y$ differs from $\Phi(x)$. The proof is given in Supplementary Section~S1.

Theorem~\ref{thm:composition} implies that any selected term of degree $k$ can be reduced to a target locality $\ell\ge 2$ with at most $\max\{k-\ell,0\}$ new auxiliaries, while every unselected term remains native. A chain construction that achieves this bound with penalties $M\ge|c_A|$ at each step is given in Supplementary Section~S1. Sharing an intermediate product between terms can lower the auxiliary count, but the shared penalty must then be recomputed from the aggregate residual. The recursion itself is the standard Rosenberg construction \cite{Rosenberg1975,BorosGruber2014}. The contribution here is its explicit selective form with a pointwise guarantee at every stage. For residuals of higher degree, computing $\norminf{A_p}$ exactly can require a search over the Boolean cube, and the sum $\sum_{S\in\mathcal{S}_p}|d_S|$ is a sufficient penalty that is easy to compute but can exceed the sharp threshold.

\subsection{Cubic Closed Form and Pair Reuse}
\label{sec:reduction:cubic}

For cubic objectives, which are used in the experiments, the threshold has an explicit form. Decompose the objective as $f=f_{\le 2}+\sum_{A\in\mathcal{C}_3(f)}c_Ax_A$, where
$\mathcal{C}_3(f)=\{A\subseteq[n] : |A|=3,\ c_A\neq 0\}$. A design $\sigma=(D,\rho)$ chooses the reduced supports $D\subseteq\mathcal{C}_3(f)$ and, for each $A\in D$, one pair $\rho(A)\in\binom{A}{2}$. Let $R=\mathcal{C}_3(f)\setminus D$, let $P_\rho=\rho(D)$ be the set of active pairs, and let $D_p=\{A\in D : \rho(A)=p\}$. With $A_p(x)=\sum_{A\in D_p}c_Ax_{A\setminus p}$, the representation is
\begin{align}
F_{\rho,M}(x,y)={}&f_{\le 2}(x)+\sum_{A\in R}c_Ax_A
+\sum_{p\in P_\rho}y_pA_p(x)\nonumber\\
&+\sum_{p=\{i,j\}\in P_\rho}M_p\,\PR(x_i,x_j,y_p).
\label{eq:cubic_rep}
\end{align}
Each $A_p$ is linear in distinct remaining coordinates, so its Boolean sup norm is the larger of the sum of its positive coefficients and the sum of the magnitudes of its negative coefficients,
\begin{equation}
T_p=\max\Bigl\{\sum_{\substack{A\in D_p\\ c_A>0}}c_A,\ \sum_{\substack{A\in D_p\\ c_A<0}}|c_A|\Bigr\}
=\norminf{A_p}.
\label{eq:cubic_threshold}
\end{equation}
 
\begin{proposition}[Exact cubic mixed-sign reduction]
\label{prop:cubic}
For nonnegative penalties, $f(x)=\min_yF_{\rho,M}(x,y)$ for every $x$ if and only if $M_p\ge T_p$ for every active pair. Strict inequalities give the unique consistent lift $\phi_\rho(x)=(x_ix_j)_{p=\{i,j\}\in P_\rho}$.
\end{proposition}

Minimisation separates over the auxiliaries, so the claim follows from Proposition~\ref{prop:sharp} applied to each pair. Let $k_p=|D_p|$ denote the number of terms assigned to pair $p=\{i,j\}$. The auxiliary $y_p$ couples to $x_i$, $x_j$, and the $k_p$ distinct third variables, so its interaction degree is $k_p+2$. Before selection, a deterministic penalty rule $\mathcal{P}$ is fixed that assigns $M_p=T_p+\delta_p^{\mathcal{P}}(\sigma)$ with $\delta_p^{\mathcal{P}}(\sigma)>0$ to every active pair of every design. This defines the finite exact family $\Sigma^{\mathcal{P}}_{\mathrm{ex}}$, in which every member is pointwise exact with a unique consistent lift. Fixed additive margins $M_p=T_p+\varepsilon$ and the multiplicative rule $M_p=1.1\,T_p$ are the two special cases used in the experiments, and every comparison applies one rule to all representations. The endpoints of the family are the native design $D=\emptyset$ and the fully reduced designs $D=\mathcal{C}_3(f)$, and a design with $0<|D|<|\mathcal{C}_3(f)|$ is strictly mixed. The auxiliary count is $a_\sigma=|P_\rho|$ rather than $|D|$, because several cubic terms sharing a pair use one auxiliary. The minimum number of auxiliaries that reduces all of $D$ is the set-cover number $\tau_2(D)$ of the pair shadow $\partial_2D=\bigcup_{A\in D}\binom{A}{2}$. It satisfies $\lceil |D|/\Delta_2(D)\rceil\le\tau_2(D)\le|D|$, where $\Delta_2(D)$ is the maximum number of terms containing a common pair
\cite{VermaLewisKochenberger2021,Schmidbauer2026}. Supplementary Section~S1 gives the integer-programming form and the overlap-graph view, and minimising the auxiliary count need not minimise penalties or compiled resources. Although the general construction permits earlier auxiliaries as inputs, the experiments use direct pairs of original variables.

\subsection{Worked Max-3SAT Example}
\label{sec:reduction:example}

A small instance illustrates the mixed-sign threshold and the three representations that the experiments compare. Consider the weighted clauses $C_1=(\neg x_1\vee\neg x_2\vee\neg x_3)$, $C_2=(x_2\vee x_3\vee x_4)$, and $C_3=(\neg x_1\vee\neg x_4\vee\neg x_5)$ with weights $5$, $4$, and $2$. The total violated weight expands and aggregates to
\begin{equation}
f^{\mathrm{ex}}(x)=g(x)+5x_1x_2x_3-4x_2x_3x_4+2x_1x_4x_5,
\label{eq:example_f}
\end{equation}
where $g(x)=4-4(x_2+x_3+x_4)+4(x_2x_3+x_2x_4+x_3x_4)$ is quadratic. The first two cubic terms share the pair $\{2,3\}$ with opposite signs. Substituting $y_{23}=x_2x_3$ gives the aggregate residual $A_{23}=5x_1-4x_4$, so \eqref{eq:cubic_threshold} yields $T_{23}=\max\{5,4\}=5$. Choosing $M_{23}=6$ produces the selective representation
\begin{align}
F_{\mathrm{sel}}(x,y_{23})={}&g(x)+5x_1y_{23}-4y_{23}x_4+2x_1x_4x_5\nonumber\\
&+6\,\PR(x_2,x_3,y_{23}),
\label{eq:example_sel}
\end{align}
which retains one cubic term natively. Full quadratization additionally substitutes $y_{14}=x_1x_4$ in the remaining term, with $T_{14}=2$ and $M_{14}=3$. Direct evaluation of all $32$ original assignments confirms that all three representations satisfy \eqref{eq:pointwise} with unique consistent minimisers.

Table~\ref{tab:example} compares the logical resources of the three representations. The selective design removes two cubic terms with one auxiliary and saves one auxiliary and four quadratic couplings relative to full reduction. Fig.~\ref{fig:example_circuits} shows the three cost layers under the reference synthesis of Section~\ref{sec:selection:model}, in which the native layer has three weight-three gadgets and eight weight-two gadgets.
The extra weight-2 strings arise from expanding cubic Boolean terms, so Boolean couplings and Pauli gate counts differ. The native, selective and full circuits use 28, 24 and 22 CNOTs on five, six and seven qubits. These counts show a resource trade-off, but they do not order solution quality. To assess quality on this same example, one must evaluate $E_{f^{\mathrm{ex}}}=\sum_{x,y}f^{\mathrm{ex}}(x)|\langle x,y|\psi\rangle|^2$ on each circuit. Section~\ref{sec:selection:warmstart} defines this separate test.

\begin{table}[!t]
\caption{Logical resources of the three representations of the worked Max-3SAT instance.}
\label{tab:example}
\centering
\scriptsize
\setlength{\tabcolsep}{4pt}
\begin{adjustbox}{max width=\columnwidth}
\begin{tabular}{@{}lccccl@{}}
\toprule
Encoding & Qubits & Aux. & Cubic & Couplings & Penalties \\
\midrule
Native & 5 & 0 & 3 & 3 & none \\
Selective & 6 & 1 & 1 & 7 & $M_{23}=6$ \\
Fully reduced & 7 & 2 & 0 & 11 & $M_{23}=6$, $M_{14}=3$ \\
\bottomrule
\end{tabular}
\end{adjustbox}
\end{table}

\Figure[t!](topskip=0pt, botskip=0pt, midskip=0pt)[width=\textwidth]{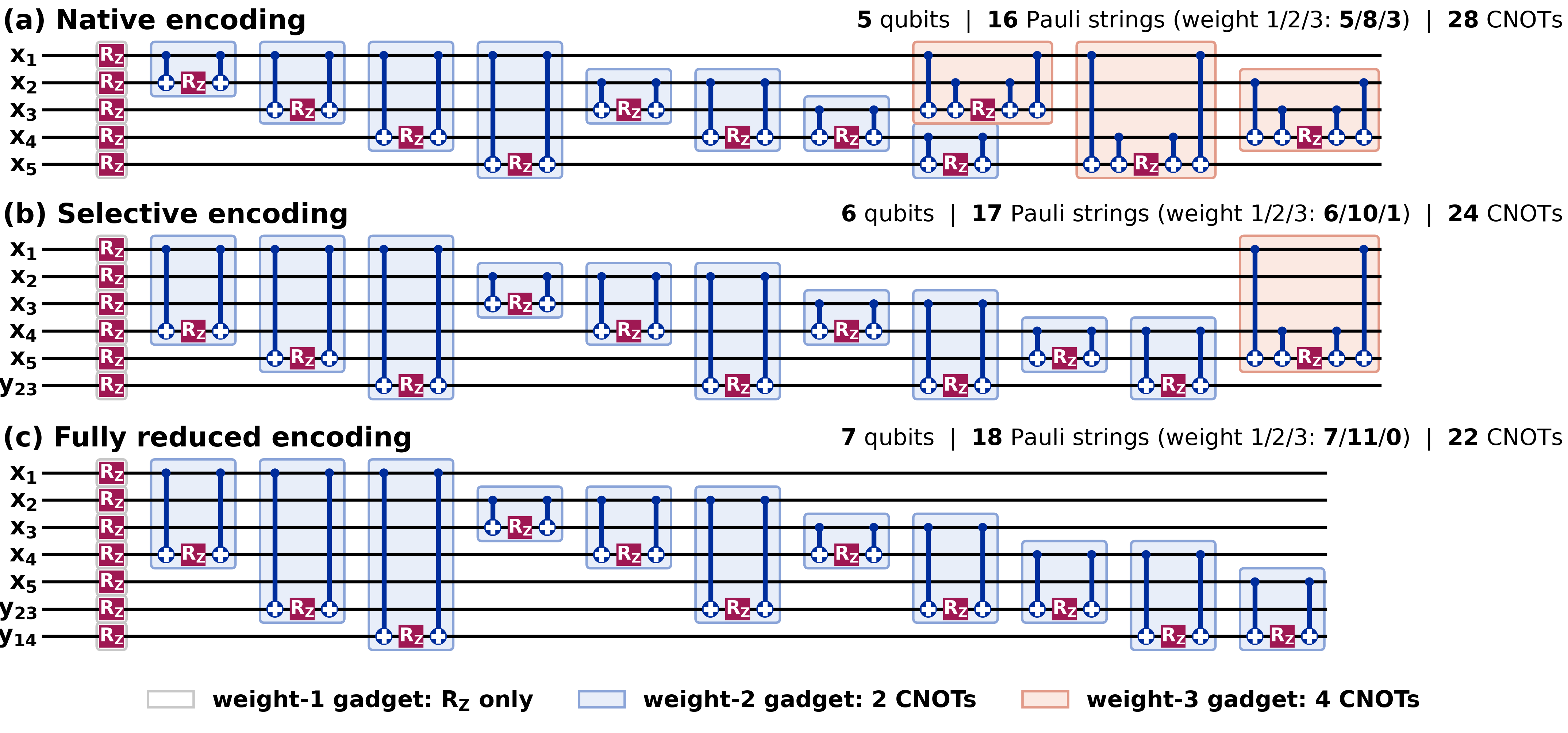}
{{One QAOA cost layer for the (a) native, (b) selective, and (c) fully reduced encodings of the worked instance \eqref{eq:example_f}.}\label{fig:example_circuits}}

\section{Resource-Aware Representation Selection}
\label{sec:selection}

Section~\ref{sec:reduction} defines a finite family of exact representations for a cubic objective. This section formulates the choice among them as a constrained design problem. A fixed compilation model assigns resources to each representation, and a scalar score ranks the representations under hard budgets. The section then characterises when the budgets permit only a mixed representation and describes the search procedures used in the experiments. The same relaxation that supplies an optional consistency diagnostic also supplies the warm start. Fig.~\ref{fig:pipeline} summarises the workflow.
 
\Figure[t!](topskip=0pt, botskip=0pt, midskip=0pt)
[width=0.98\columnwidth,draft=false]
{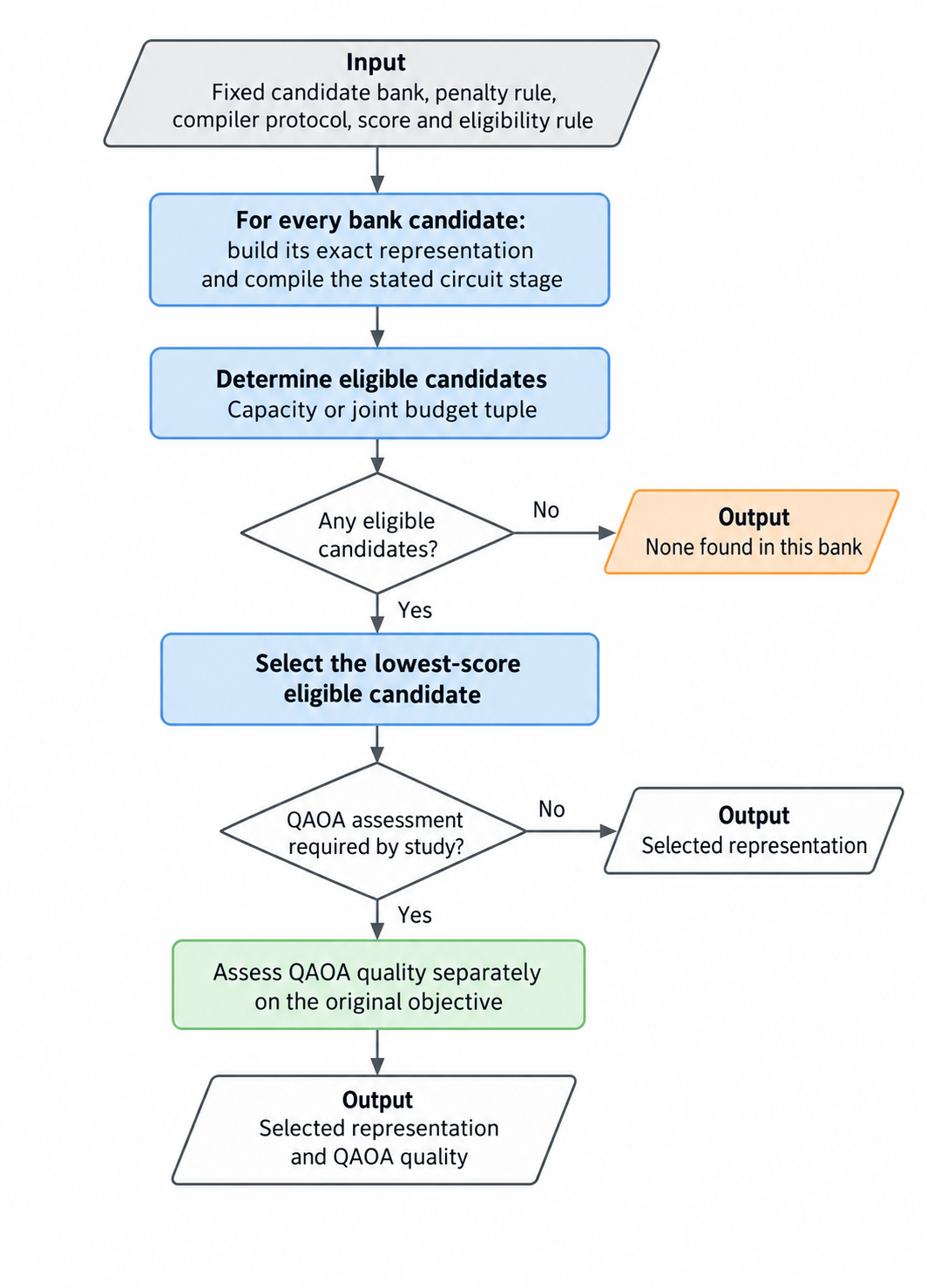}
{Fixed candidate-set evaluation used in the primary experiments.
\label{fig:pipeline}}

\subsection{Compilation and Circuit Model}
\label{sec:selection:model}

The resource protocol specifies the circuit stage that is scored. The cost-layer experiments score the single layer $U_F(\gamma) = e^{-i\gamma H_{F_\sigma}}$, whereas the joint experiment scores the complete circuit, including preparation and the matched mixer.

After the substitution $u_j\mapsto N_j$, every term of $F_\sigma$ becomes a product of commuting diagonal operators, so the cost layer factorises exactly into one phase gadget per Pauli string in the collected expansion $H_{F_\sigma}=\sum_T h_T Z_T$ \cite{Hadfield2021,CampbellDahl2022}. Equal strings are combined before synthesis and only exact zero coefficients are removed. The angle $\gamma$ is treated as a free parameter during synthesis, so no gate is removed for a particular numerical value.

The reference compiler uses an all-to-all architecture, directed CNOT gates, and arbitrary $R_z$ rotations. Original qubits occupy indices $1$ to $n$ and active auxiliaries follow in lexicographic pair order. Strings are processed in increasing weight and then lexicographically. For a string with support $T=\{q_1<\dots<q_k\}$, the gadget in Fig.~\ref{fig:gadgets} computes the parity of $T$ onto $q_k$ with a CNOT ladder, applies $R_z(2\gamma h_T)$ on $q_k$, and uncomputes the ladder \cite{CampbellDahl2022,Verchere2023}. It costs $2(k-1)$ CNOTs, so the cubic family has reference cost

\begin{equation}
G^{\mathrm{ref}}_{2q,\sigma}=2p_{2,\sigma}+4p_{3,\sigma},\qquad S^{\mathrm{ref}}_\sigma=0,
\label{eq:refcost}
\end{equation}

where $p_{k,\sigma}$ counts nonzero strings of weight $k$. Two-qubit depth $D^{\mathrm{ref}}_{2q,\sigma}$ places each emitted CNOT in the first layer after both of its qubits are free, with the gate order preserved on each wire. No gates are cancelled or reordered across gadgets, so \eqref{eq:refcost} is the cost of this synthesis rather than a lower bound. Phase-gadget optimisation and joint synthesis of
commuting exponentials can reduce both count and depth \cite{CowtanPhase2020,Amy2018,Sriluckshmy2023}. The experiments therefore use two synthesis strategies, the canonical one just described and a reuse strategy with cross-gadget optimisation, and report whether the representation ranking is preserved.
 
\Figure[t!](topskip=0pt, botskip=0pt, midskip=0pt)[width=0.98\columnwidth]{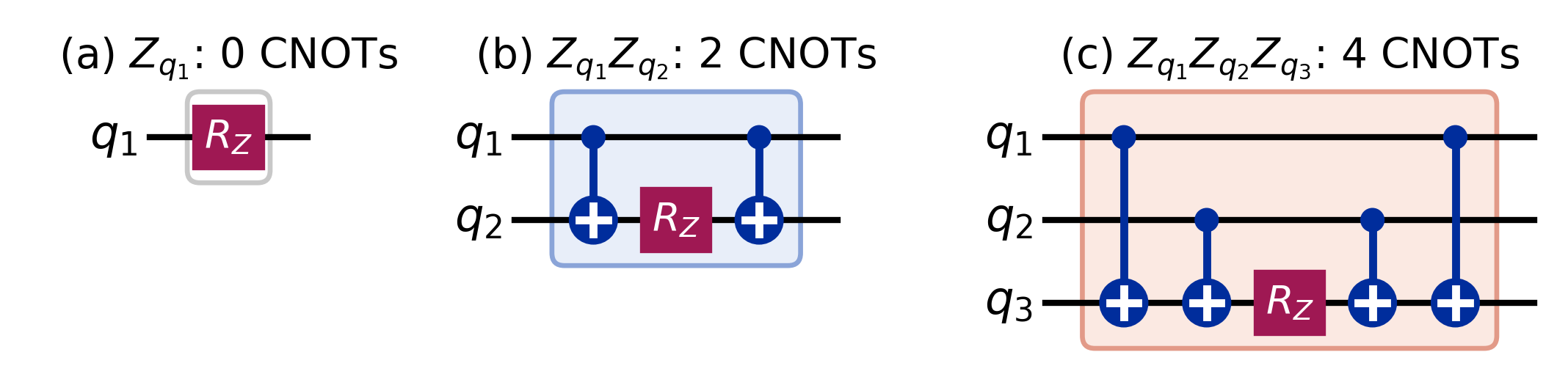}
{Reference parity gadgets for Pauli terms of weight (a) one, (b) two, and (c) three.
\label{fig:gadgets}}

Device-level resources are produced by a fixed protocol $\Pi$. It consists of the circuit construction rules above, the circuit stage, a coupling map, basis gates, compiler version, layout and routing passes, a finite seed bundle, and an aggregation rule \cite{Li2019,Zulehner2019}. Applying $\Pi$ to a representation returns the compiled two-qubit gate count $G_{2q,\sigma}$, two-qubit depth $D_{2q,\sigma}$, and inserted SWAP count $S_\sigma$ \cite{Weidenfeller2022,Schmidbauer2024}. Routing can occupy device sites beyond the logical register. Let $I^{\Pi}_\sigma$ be the input sites and $T^{\Pi}_\sigma$ the sites touched by compiled operations. The physical width is
\begin{equation}
Q^{\Pi}_\sigma=|I^{\Pi}_\sigma\cup T^{\Pi}_\sigma|\ \ge\ N_\sigma,
\label{eq:physwidth}
\end{equation}
Equality holds for the all-to-all reference, but passing the logical-width check does not guarantee that a routed circuit fits. The largest active penalty is $M_{\max,\sigma}=\max_p M_p$, with value zero for native. The dynamic range $\kappa_\sigma$ is the ratio of the largest to the smallest magnitude among nonzero, nonidentity Pauli coefficients. It is defined when this set is nonempty and is unchanged by nonzero global rescaling. Gate count and depth alone do not determine hardware duration or success.

\subsection{Resource Score and Budgets}
\label{sec:selection:score}
 
Designs are ranked by a weighted sum of normalised resources. For positive scales $s=(s_a,s_G,s_D,s_M)$ and nonnegative weights $w$ with $\sum_k w_k=1$,
\begin{equation}
\scalebox{0.95}{$
\Jres(\sigma)=w_a\frac{a_\sigma}{s_a}+w_G\frac{G_{2q,\sigma}}{s_G}
+w_D\frac{D_{2q,\sigma}}{s_D}+w_M\frac{M_{\max,\sigma}}{s_M}.
$}
\label{eq:score}
\end{equation}

The first term counts auxiliaries rather than physical sites, and the compiler supplies gate count and depth \cite{Verchere2023,Weidenfeller2022,Schmidbauer2024}. The score is one preference among feasible designs. Hard budgets on physical width, gates, depth, and penalty define the feasible set and the selection target,
\begin{align}
\mathcal{F}_{\mathrm{sel}}={}&\bigl\{\sigma\in\Sigma^{\mathcal{P}}_{\mathrm{ex}} :
Q^{\Pi}_\sigma\le Q_{\max},\ G_{2q,\sigma}\le B_G,\nonumber\\
&\hphantom{\bigl\{\sigma\in\Sigma^{\mathcal{P}}_{\mathrm{ex}} :}
D_{2q,\sigma}\le B_D,\ M_{\max,\sigma}\le B_M\bigr\},
\label{eq:feasible}\\
J^\star_{\mathrm{res}}={}&\inf_{\sigma\in\mathcal{F}_{\mathrm{sel}}}\Jres(\sigma),
\qquad \inf\emptyset=+\infty .
\label{eq:target}
\end{align}
Within each experiment, the objective scale is fixed, and the penalties $M_p$, bound $B_M$ and score scale $s_M$ use the same objective units. The penalty bound is a declared modelling constraint, not a calibrated hardware resolution limit. For $c>0$, rescaling the whole encoding gives $e^{-i(\gamma/c)H_{cF_\sigma}}=e^{-i\gamma H_{F_\sigma}}$. This scaling must include every penalty and additive margin, and scaling $B_M$ and $s_M$ by the same $c$ preserves the penalty comparisons. The angle domain and scan grid must change with the units. Scaling only $f$ while keeping a fixed additive margin gives a different encoding. Collected Pauli coefficients can add or cancel, so a larger penalty need not increase $\kappa_\sigma$. Claims about physical precision or duration require a control model, which is not used here.

For a nonempty feasible set, the infimum is a minimum, since the family is finite. The budgets and the score serve different purposes. A design can have the lowest score and still fail a budget, and a design can be the only feasible one without having the lowest score. To measure the value of retaining cubic terms, the experiments compare the optimum over full reductions with the optimum over the whole family,
\begin{equation}
\inf_{\rho:(\mathcal{C}_3(f),\rho)\in\mathcal{F}_{\mathrm{sel}}}\Jres
\quad\text{against}\quad
\inf_{(D,\rho)\in\mathcal{F}_{\mathrm{sel}}}\Jres ,
\label{eq:classsep}
\end{equation}
The ideal optima use the same score, compiler, penalty rule and budgets. The full family is a subset of the unrestricted family, so the unrestricted optimum cannot be larger. Finite searches can miss different designs and need not preserve this ordering. The experiments use matching candidate counts and report the best evaluated design.

The consistency excess is an optional diagnostic that is not used in primary resource selection. Its predictive value and effects on selection are assessed separately.

For independent $X_i\sim\mathrm{Bernoulli}(\eta_i)$ and $Y_p\sim\mathrm{Bernoulli}(\xi_p)$, the consistency excess is

   \begin{align}
   \Efib(\sigma;\eta,\xi)&=\mathbb{E}\bigl[F_\sigma(X,Y)-f(X)\bigr]\nonumber\\
   &=\sum_{p\in P_\rho}\Bigl[(\xi_p-\eta_i\eta_j)A_p(\eta)+M_p\,\ell_p\Bigr],
   \label{eq:fib}
   \end{align}

with $\ell_p=\eta_i\eta_j+\xi_p(3-2\eta_i-2\eta_j)$ the expected Rosenberg penalty and $A_p(\eta)=\sum_{A\in D_p}c_A\prod_{k\in A\setminus p}\eta_k$. Pointwise exactness makes this excess nonnegative. Under the strict penalties $M_p>T_p$ used here, it is zero exactly when the auxiliaries are consistent with probability one. At $M_p=T_p$, an inconsistent assignment can also have zero excess. Evaluation at relaxation pseudo-moments is a diagnostic, not the expectation of a prepared state. Supplementary Section~S4 tests its predictive value.

\subsection{Feasibility Boundaries and Mixed-Only Regimes}
\label{sec:selection:feasibility}
The next results concern direct cubic pair substitutions. They exclude coefficient splitting and penalties that couple different auxiliaries. 
Static feasibility, fixed-synthesis circuit counts, and empirical compiled certificates have separate scopes, and the three are labelled separately below.

Before compilation, four static resources of a design $\sigma=(D,\rho)$ are determined directly by the objective. With a common margin $\varepsilon\ge 0$ and $M_p=T_p+\varepsilon$, these are the auxiliary count $|P_\rho|$, the highest penalty $\max_p(T_p+\varepsilon)$, the highest auxiliary degree $\max_p(k_p+2)$, and the number of retained cubic terms $|\mathcal{C}_3(f)\setminus D|$, with empty maxima equal to zero. Let $(r_0,M_0,\Delta_0,h_0)$ be the budgets for these four resources, let $m=|\mathcal{C}_3(f)|$, and let $\mathcal{Q}(r_0,M_0,\Delta_0)$ be the family of reduced sets $D$ that meet the first three budgets under some pair assignment. Define
\begin{equation}
K_Q(r_0,M_0,\Delta_0)=\max_{D\in\mathcal{Q}(r_0,M_0,\Delta_0)}|D| ,
\label{eq:KQ}
\end{equation}
This maximum is the number of terms that can be reduced under the first three budgets. Computing it may require optimisation, and the characterisation does not provide a general efficient algorithm. Because removing assigned terms cannot increase the auxiliary count, signed coefficient masses or auxiliary degrees, $\mathcal{Q}$ is closed under taking subsets.
 
\begin{theorem}[Static feasibility boundary]
\label{thm:boundary}
A direct-pair representation meeting all four budgets exists if and only if $K_Q+h_0\ge m$. The native endpoint is feasible if and only if $h_0\ge m$, and a fully reduced representation is feasible if and only if $K_Q=m$. Consequently, feasible representations exist and all of them are strictly mixed exactly when
\begin{equation}
K_Q<m,\qquad h_0<m,\qquad K_Q+h_0\ge m .
\label{eq:mixedonly}
\end{equation}
\end{theorem}
 
At least $m-h_0$ terms must be reduced to meet the retained-term budget, and at most $K_Q$ can be reduced under the remaining budgets. Feasibility requires these two limits to overlap, and downward closure supplies a feasible set of any intermediate size.
The proof is given in Supplementary Section~S1.
 
For one structured family, the quantity $K_Q$ has a closed form. A pair-star with core $\{a,b\}$ and $\ell$ leaves has supports $\{a,b,w_t\}$ for $t=1,\dots,\ell$, and a pair-star forest is a disjoint union of stars.
 
\begin{theorem}[Unit-weight pair-star forests]
\label{thm:star}
Let all cubic coefficients equal $+1$, let the star sizes be $\ell_1,\dots,\ell_s\ge 1$, and let $C=\max\{0,\min\{\lfloor M_0-\varepsilon\rfloor,\lfloor\Delta_0\rfloor-2\}\}$.
If $r_0=0$ or $C=0$, then $K_Q=0$. Otherwise, let $g_i=\min\{\ell_i,C\}-1$, sort the gains so that $g_{(1)}\ge g_{(2)}\ge\cdots$, and set $\rho_0=\min\{r_0,s\}$. Then
\begin{equation}
K_Q=\min\Bigl\{m,\ r_0+\sum_{j=1}^{\rho_0}g_{(j)}\Bigr\}=:R_Q.
\label{eq:RQ}
\end{equation}
\end{theorem}
 
An auxiliary reduces one term through a private pair. Using the core of star $i$ reduces up to $\min\{\ell_i,C\}$ terms, where $C$ is the largest number of terms an auxiliary can absorb under the penalty and degree budgets. The extra gain of that choice is $g_i$, and auxiliaries are spent on the largest gains first. Supplementary Section~S1 proves the upper bound and its achievement. Combined with Theorem~\ref{thm:boundary}, the formula classifies every budget vector for this family, as Table~\ref{tab:classes} shows.
 
\begin{table}[!t]
\caption{Feasibility classes for unit-weight pair-star forests under static budgets.}
\label{tab:classes}
\centering
\footnotesize
\begin{adjustbox}{max width=\columnwidth}
\begin{tabular}{@{}ll@{}}
\toprule
Class & Necessary and sufficient condition \\
\midrule
Both endpoints feasible & $R_Q\ge m$ and $h_0\ge m$ \\
Native endpoint only & $R_Q<m$ and $h_0\ge m$ \\
Full endpoint only & $R_Q\ge m$ and $h_0<m$ \\
Strictly mixed only & $R_Q<m$, $h_0<m$, and $R_Q+h_0\ge m$ \\
None feasible & $R_Q+h_0<m$ \\
\bottomrule
\end{tabular}
\end{adjustbox}
\end{table}
 
\begin{corollary}[Infinite static mixed-only family]
\label{cor:static}
For every $L,H\ge 1$, take the disjoint union of an $L$-leaf unit pair-star and $H$ unit isolated triples. Under budgets $(r_0,M_0,\Delta_0,h_0)=(1,L+\varepsilon,L+2,H)$, neither endpoint is feasible, whereas reducing the star through its core and retaining all isolated triples is feasible.
\end{corollary}
 
The mixed design assigns its single auxiliary to the reusable core. Native fails because it retains $L+H>H$ cubic terms, and full reduction fails because each isolated triple needs its own auxiliary. Neither endpoint is uniformly more feasible than the other.
Under budgets $(1,h+1+\varepsilon,h+3,h)$ with $h\ge 2$, native is feasible for $h$ isolated triples while full reduction is not, and full reduction is feasible for a single $(h+1)$-leaf star while native is not. Interaction order alone does not decide feasibility.
 
The static result does not count gates, so the next result adds gate and depth budgets under the reference synthesis of Section~\ref{sec:selection:model}.
 
\begin{corollary}[Infinite compiled mixed-only family]
\label{cor:compiled}
Let $L\ge 2$, $H\ge 1$, and $\varepsilon>0$. On $n=L+2+3H$ original bits, consider
\begin{equation}
f_{L,H}(x)=x_ax_b\sum_{t=1}^{L}x_{w_t}+\sum_{j=1}^{H}x_{u_j}x_{v_j}x_{z_j},
\label{eq:family}
\end{equation}
with variables ordered $a,b,w_1,\dots,w_L$, then each triple in order, then the auxiliaries in pair order. Under $M_p=T_p+\varepsilon$ and budgets
\begin{align}
Q_{\max}&=n+1, & B_G&=2L+6+10H,\nonumber\\
B_D&=2L+6, & B_M&=L+\varepsilon,
\label{eq:familybudgets}
\end{align}
the native circuit violates both the gate and the depth budget. Every fully reduced direct-pair representation violates the width budget, whereas reducing only the star through its core meets all four budgets with a unique consistent lift.
\end{corollary}
 
Fig.~\ref{fig:family} illustrates the case $L=3$, $H=2$. The native cost layer needs $8L+2+10H$ CNOTs because each cubic monomial also generates weight-two strings, whereas the mixed layer needs $2L+6+10H$, namely $2L$ for the couplings of the auxiliary, $6$ for its penalty, and $10$ for each retained triple. Every full reduction needs at least $H+1$ auxiliaries and therefore exceeds the width $n+1$. Supplementary Section~S1 counts the Pauli supports, CNOT gates, and layers explicitly, and the corollary concerns only this synthesis and the direct-pair class. Gate cancellation, other gadget orderings, and routing change the counts, and the corollary makes no claim about solution quality. Section~\ref{sec:experiments} tests whether such mixed-only regions occur for random instances under a real compiler.
 
\Figure[t!](topskip=0pt, botskip=0pt, midskip=0pt)[width=0.98\columnwidth]{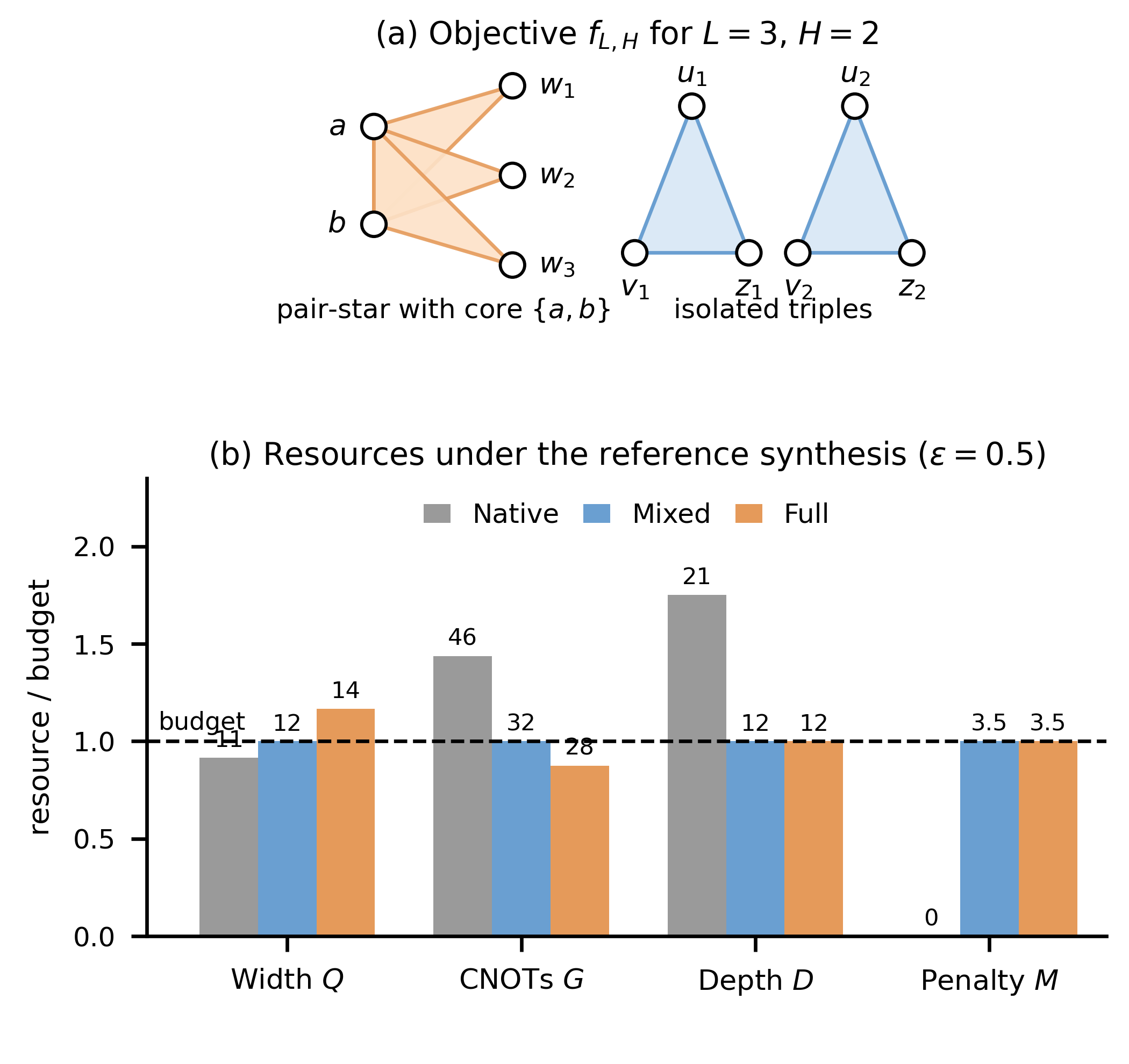}
{The compiled mixed-only family of Corollary~\ref{cor:compiled}. (a) Hypergraph of $f_{3,2}$. (b) Resources of the native, mixed, and full encodings relative to the budgets \eqref{eq:familybudgets}.\label{fig:family}}

\subsection{Search Procedures}
\label{sec:selection:search}
The primary experiments evaluate fixed, seeded candidate sets and report the best evaluated design. These sets include minimum-auxiliary, maximum-reuse, resource-optimised, and distance-aware reductions, together with random controls that match the auxiliary and reduced-term counts. Each candidate is compiled under the fixed protocol before selection.

Each experiment has its own eligibility rule. The cost-layer experiment selects by score within device capacity and tests every candidate in its budget maps. Preparation/evolution uses the reference-synthesis score without the four-budget filter. The joint experiment filters full and strictly mixed candidate sets by each budget tuple, then selects by score. The routing follow-up reuses these saved sets. No primary choice uses QAOA outcomes or a threshold on the consistency excess. 
The distance-aware rule is a comparator defined for this paper and not a reproduction of an external algorithm.

A \textit{candidate-set mixed-only} cell has a feasible strictly mixed candidate and no feasible endpoint in the tested candidate set. A \textit{certified mixed-only} cell also excludes every full assignment in the stated direct-pair class under the fixed penalty rule. A \textit{joint mixed-only success} requires the selected mixed circuit to meet both resource and quality targets, while neither compared endpoint meets both. An endpoint may fail either test.

Complete endpoint exclusion is reported separately. Since $Q_\sigma^\Pi\ge N_\sigma$, failing the logical-width check rules out compiled feasibility, although passing this check does not prove it. Failure in a finite candidate set means only that no feasible candidate was found.

A Pareto search is an exploratory comparator under the reference synthesis. It searches the $4^m$ action tree with scalar and non-dominated rankings \cite{DebEtAl2002}. Optional branch-and-bound can certify the score using admissible bounds \cite{LandDoig1960,LawlerWood1966}. The compiled part of these bounds is zero, so small cases require about as many nodes as a complete search. These tests check the implementation, but they do not show a faster solver or global optimality of the primary fixed candidate sets. Supplementary Section~S5 gives the algorithms and validation.

\subsection{Warm-Start Initialisation on the Enlarged Register}
\label{sec:selection:warmstart}
 
A sparse linear relaxation with single, pair and supported triple pseudo-moments in $[0,1]$, linked by McCormick and reformulation-linearisation constraints \cite{SheraliAdams1990,McCormick1976}, provides a product warm start \cite{Egger2021,TateSDP2023,TateCustom2023}. Only the single and pair coordinates $(\bar\mu_i,\bar q_{ij})$ are exported. A reformulation that adds a variable for $x_ix_j$ must also specify its initial marginal, and the pair pseudo-moment $\bar q_{ij}$ can differ from the product $\bar\mu_i\bar\mu_j$. This defines a specific initial state for the enlarged register, which Section~\ref{sec:experiments} tests.
 
With $\mathrm{clip}_\delta(t)=\min\{1-\delta,\max\{\delta,t\}\}$ for $0<\delta<1/2$, set $\mu_i=\mathrm{clip}_\delta(\bar\mu_i)$. Pair warming uses $q_p=\mathrm{clip}_\delta(\bar q_{ij})$. 
The primary preparation and evolution experiment uses $q_p=\mathrm{clip}_\delta(\mu_i\mu_j)$ for independence warming, so the original marginals are bounded, multiplied, and bounded again.
The earlier exploratory comparison uses $q_p=\mu_i\mu_j$ without this final adjustment. Neither rule enforces the auxiliary relation in every product-state outcome. Let $|\phi(t)\rangle=\sqrt{1-t}\,|0\rangle+\sqrt{t}\,|1\rangle$. The warm state is
\begin{equation}
|\Phi^{\mathrm{ws}}_\sigma\rangle=\bigotimes_{i=1}^{n}|\phi(\mu_i)\rangle\otimes\bigotimes_{p\in P_\rho}|\phi(q_p)\rangle .
\label{eq:warmstate}
\end{equation}
It is the unique ground state of the matched mixer
\begin{align}
\scalebox{0.77}{$
H^{\mathrm{ws}}_{\mathrm{M},\sigma}
=\sum_{j=1}^{N_\sigma}h_{\mathrm{M}}(\omega_j),
\qquad
h_{\mathrm{M}}(t)
=-2\sqrt{t(1-t)}\,X-(1-2t)\,Z
$}
\label{eq:warmmixer}
\end{align}
where $\omega=(\mu,q)$ lists the preparation probabilities in register order. With $U_F(\gamma)=e^{-i\gamma H_{F_\sigma}}$ and $U_{\mathrm{M}}(\beta)=e^{+i\beta H^{\mathrm{ws}}_{\mathrm{M},\sigma}}$, the depth-$L$ ansatz is
\begin{equation}
|\psi_{L,\sigma}\rangle=U_{\mathrm{M}}(\beta_L)U_F(\gamma_L)\cdots
U_{\mathrm{M}}(\beta_1)U_F(\gamma_1)\,|\Phi^{\mathrm{ws}}_\sigma\rangle ,
\label{eq:ansatz}
\end{equation}
and $L=0$ denotes the initial state. Uniform probabilities on every qubit recover the cold start $|+\rangle^{\otimes N_\sigma}$ with the standard $X$ mixer at the same optimiser coordinates, which serves as a null control. Each mixer term factorises into single-qubit gates as $R_y(\theta_j)R_z(2\beta)R_y(-\theta_j)$ with $\theta_j=2\arcsin\sqrt{\omega_j}$.
 
Solution quality is always measured on the original objective through $E_f(\psi)=\sum_{x,y}f(x)\,|\langle x,y|\psi\rangle|^2$, which sums over all auxiliary outcomes, and the classical optimiser minimises $E_f$ directly. The improvement of a warm-started circuit over the cold initial state separates into two parts,
\begin{equation}
\scalebox{0.9}{$
f_{\mathrm{cold},0}-f_{\mathrm{warm},1}
=\underbrace{f_{\mathrm{cold},0}-f_{\mathrm{warm},0}}_{\text{preparation gain}}
+\underbrace{f_{\mathrm{warm},0}-f_{\mathrm{warm},1}}_{\text{evolution gain}}
$}
\label{eq:gains}
\end{equation}
where subscripts $0$ and $1$ denote the initial state and the optimised depth-one circuit. All representations share the same original-bit distribution at warm depth zero. Their evolution gains compare the tested representations under the stated circuit and optimisation protocol. Classical rounding of the relaxation and independent sampling from it provide classical controls for the preparation gain. The local mixer does not preserve the consistent subspace $y_p=x_ix_j$, and the penalties do not enforce consistency on individual measurement outcomes, so auxiliary inconsistency is recorded as a separate diagnostic.
Constraint-preserving mixers are a different design choice within the alternating-operator framework \cite{Hadfield2019} and are not studied here. The encoded energy $\langle H_{F_\sigma}\rangle$ is representation specific and is reported only as a diagnostic.

\section{Experimental Evaluation}
\label{sec:experiments}

This section evaluates the framework in four experiments. The first two concern compiled resources and test whether strictly mixed representations meet joint resource budgets that exclude both endpoints on sparse 12-qubit targets, and whether such regions persist under routing variation. The third decomposes the improvement of a warm-started depth-one QAOA circuit into the classical preparation gain and the gain of the subsequent evolution. The fourth tests whether a mixed representation can be the only design that meets a resource budget and a solution-quality target on the same compiled circuit. 
The routing follow-up reuses 60 of the cost-layer objectives, whereas the two QAOA experiments use separate new objectives, and no cohort or protocol is pooled across experiments.

Table~\ref{tab:studies} summarises the four experiments. The theoretical results of Sections~\ref{sec:reduction} and~\ref{sec:selection} hold independently of these data, and every experimental finding is restricted to the tested instances, compiler, and protocols.

\subsection{Benchmark Families and Instance Tiers}
\label{sec:experiments:benchmarks}

Weighted Max-3SAT minimises the total weight of violated clauses\cite{Nusslein2023,Nagies2025}. The pure cubic spin objective is $E(s)=\sum_{\{i,j,k\}}K_{ijk}s_is_js_k$, with $K_{ijk}\in\{-1,+1\}$ and $s_i=1-2x_i$\cite{Basso2022,PelofskeShortDepth2024}. Equal Boolean monomials are combined before penalties are set.

The archived code uses $E/8$ in the cost-layer compilation experiment, whereas the preparation and evolution and the joint QAOA experiments use the unscaled $E$, so their reported spin objectives and gains are in unscaled units.
The six-interaction QAOA objective lies in $[-6,6]$, so the values in Tables~\ref{tab:gains} and~\ref{tab:optimisers} are within its range. Each penalty uses the objective units of its own experiment. Penalty bounds at different scales are not compared directly.

Each random instance in the cost-layer experiment has six original variables and samples five distinct cubic supports, giving $3^5$ full pair assignments. Its Max-3SAT clauses have unit weight, and the generator does not require every variable to occur. The two random QAOA cohorts instead use six clauses or spin interactions, require all six variables to occur, and retain at least two nonzero cubic supports after aggregation. Their Max-3SAT clause weights are sampled from \(\{1,2,3,4,5\}\). In general, the full assignment count is $3^{|\mathcal C_3(f)|}$ after aggregation, rather than $3$ raised to the raw clause count. Each constructed compilation instance has nine variables, a four-leaf pair-star and one isolated triple, with weights in $\{0.75,1,1.25\}$. The joint mechanism cohort uses cubic weights in $\{1,2,3\}$ and affine coefficients in $\{-2,-1,0,1,2\}$. Each experiment follows its own frozen protocol.

The compilation and joint experiments use Qiskit~2.5.2~\cite{Qiskit2024} with basis $\{R_z, X, \sqrt{X}, \mathrm{CX}\}$, optimisation level~3, SABRE routing~\cite{Li2019}, identity initial layout, and seed 1729 unless stated otherwise. The targets are a 12-qubit line, a 12-qubit ring, and a $3\times 4$ grid of equal capacity. Both synthesis strategies of Section~\ref{sec:selection:model} are applied, and the compiler may optimise across gadget boundaries. The recorded resources are the physical width $Q$ of \eqref{eq:physwidth}, the CX count $G$, the CX-only depth $D$, and the largest penalty $M$. The score \eqref{eq:score} uses weights $w=(1/4,1/4,1/4,1/4)$ and scales $s=(4,160,120,8)$. The compilation experiments apply the penalty rule $M_p = 1.1\,T_p$ and the QAOA experiments apply $M_p = T_p + 1$, in each case to all representations within one experiment.

The cost-layer experiment uses equally sized full and unrestricted candidate sets. Each set has 32 candidates for a random instance and 27 for a constructed instance. 
The constructed-instance set includes every full assignment that fits the device, and the full set includes minimum-auxiliary, maximum-reuse, and distance-aware choices.
The distance-aware rule chooses the shortest original-bit pair and breaks ties by reuse. The unrestricted set also includes native, shared-pair groups and reuse prefixes. All $3^5$ full assignments of each random instance are checked for logical width and penalty. All capacity-feasible full assignments of each constructed instance are compiled. Random controls match auxiliary and reduced-term counts, and the largest penalty where possible.

Comparisons are paired by instance. Seeds, restarts, and nearby budget cells are not independent samples. Unless stated otherwise, uncertainty intervals in the primary experiments use $10{,}000$ paired instance-bootstrap resamples.

\begin{table}[!t]
\caption{The four experiments and the evidence each provides.}
\label{tab:studies}
\centering
\scriptsize
\setlength{\tabcolsep}{3pt}
\begin{tabular}{@{}p{0.25\columnwidth}p{0.30\columnwidth}p{0.40\columnwidth}@{}}
\toprule
Experiment & Instances & Circuit scope and purpose \\
\midrule
Cost-layer compilation & 90 new: 30 Max-3SAT, 30 cubic spin, 30 constructed & One cost layer on three 12-qubit targets; feasibility of exact representations under joint budgets \\
Routing-seed follow-up & Reuses 60 of the above (Max-3SAT, constructed) & Same candidate sets under five routing seeds; stability of mixed-only regions \\
Preparation and evolution & 60 new: 30 Max-3SAT, 30 cubic spin & Reference-compiler selection, noiseless depth one; preparation gain against evolution gain \\
Joint resource and quality & 60 new random plus 12 mechanism & Complete depth-one circuit on ring-12; simultaneous resource and quality feasibility \\
\bottomrule
\end{tabular}
\end{table}

\subsection{Joint-Budget Compilation and Mixed-Only Regions}
\label{sec:experiments:resources}

For every instance, target, and synthesis strategy, all candidates are compiled once and evaluated against a frozen four-dimensional budget grid. The grid has $Q_{\max}\in\{6,\dots,12\}$, $B_G\in\{0,4,\dots,240\}$,
$B_D\in\{24,48,72,96,120,160,240\}$, and \\
$B_M\in\{0,1.1,2.2,3.3,4.4,6.6,8.8\}$. Each of the $20{,}923$ cells is labelled as both endpoints, native only, full only, strictly mixed only, or none found, based on its feasible candidate classes. Endpoint labels may also contain feasible mixed designs. ``None found'' refers to the finite candidate set. Certification requires a compiled mixed candidate to pass all budgets, while native must fit width and fail gates or depth. A check of all full assignments must exclude each one by logical width or penalty within the stated direct-pair class and fixed penalty rule.

Table~\ref{tab:mixedonly} counts instances with a certified mixed-only cell. At least one setting has such a cell for 23 of 30 Max-3SAT and all 30 constructed instances. Every setting has one for 3 and 21 instances, respectively. Only 2 of 30 spin instances have one in any setting, and none does under reuse synthesis. These are observations for the tested families.
The weak incidence for spin glasses has a partial structural explanation under the reference synthesis. A cubic spin interaction is a single weight-three Pauli string costing four CNOT gates, because the quadratic terms of its Boolean expansion cancel in Pauli form. Substituting an auxiliary for the pair of an isolated pair-star with $\ell$ leaves removes this cancellation. The reduced layer then contains $3\ell+3$ weight-two strings and costs $6\ell+6$ CNOT gates in place of $4\ell$, so reduction of an isolated star never lowers the reference gate count. Signed terms can cancel the original-to-leaf strings, and Supplementary Section~S1 gives an eight-variable objective whose four-auxiliary reduction lowers the reference count from 64 to 56 CNOT gates. Although the isolated-star count does not apply to every signed objective, it illustrates how direct-pair reduction can increase the reference gate count. The low incidence of certified mixed-only cells in Table~6 is consistent with this mechanism, but the calculation alone does not explain why such cells are rare in the tested instances.
Fig.~\ref{fig:phasemaps} shows the phase maps of one constructed instance on the three targets. The maps are drawn at $B_D = 96$ and $B_M = 4.4$. The mixed candidate library is finite, and for this constructed family every capacity-feasible full assignment is evaluated.

\begin{table}[!t]
\caption{Instances with at least one certified mixed-only cell (canonical / reuse synthesis).}
\label{tab:mixedonly}
\centering
\footnotesize
\begin{tabular}{@{}lccc@{}}
\toprule
Family ($N=30$ each) & Line-12 & Ring-12 & Grid-12 \\
\midrule
Weighted Max-3SAT & 16 / 8 & 16 / 8 & 18 / 8 \\
Cubic spin glass & 1 / 0 & 1 / 0 & 1 / 0 \\
Constructed pair-star & 30 / 29 & 30 / 22 & 30 / 28 \\
\bottomrule
\end{tabular}
\end{table}

\Figure[t!](topskip=0pt, botskip=0pt, midskip=0pt)[width=0.85\textwidth]{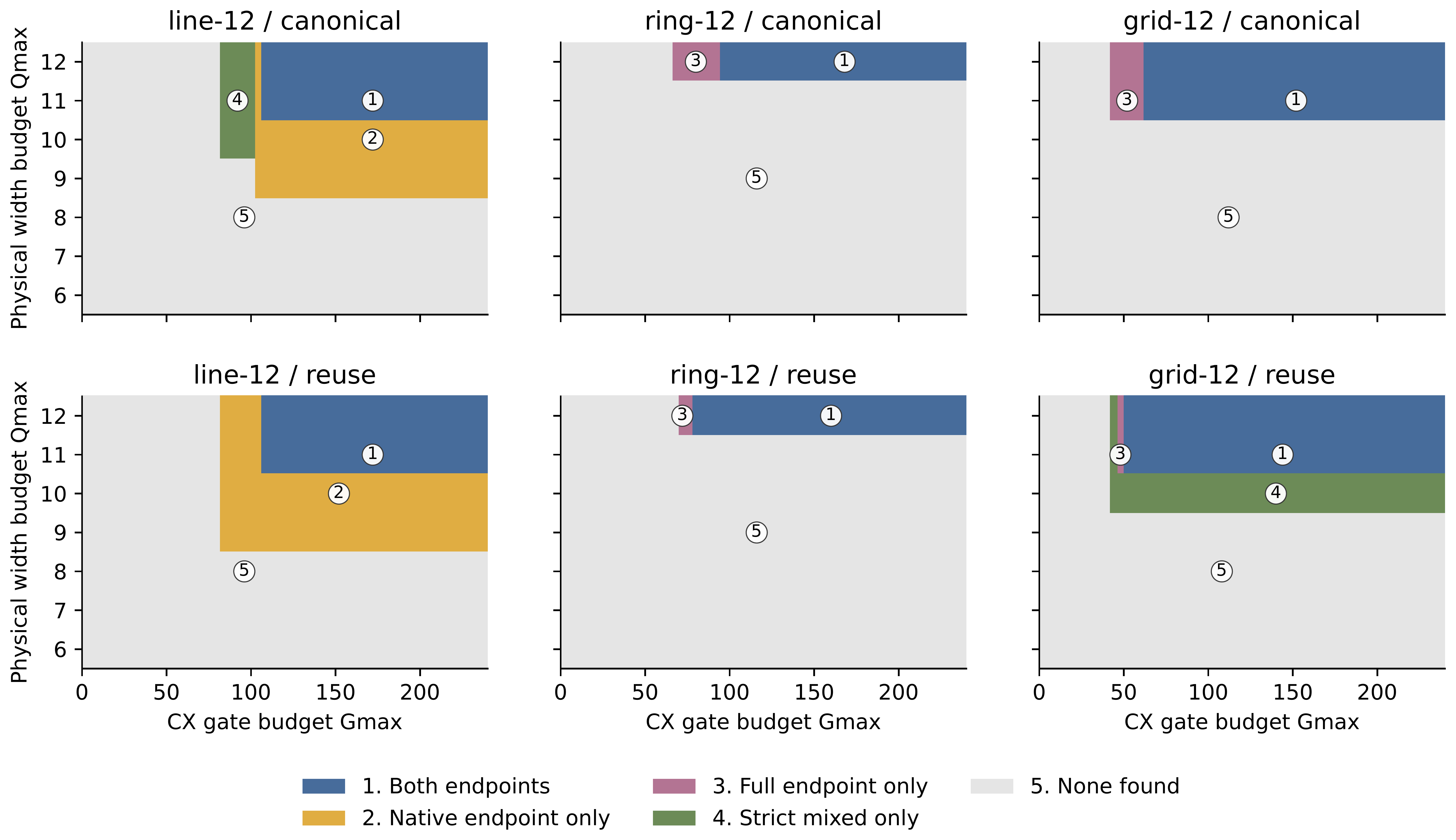}
{{Joint-budget feasibility classes for one constructed instance on three 12-qubit targets under canonical (top) and reuse (bottom) synthesis.
}\label{fig:phasemaps}}

One ring-12 cell illustrates the certificate under canonical synthesis at $(Q_{\max},B_G,B_D,B_M)=(12,92,48,2.2)$. Native has resources $(12,96,55,0)$, so it fails gates and depth but passes width and penalty. Every full assignment with logical width at most 12 has $M_{\max}\ge2.475$ under the fixed penalty rule, and all other full assignments fail the width. A strictly mixed candidate has resources $(12,92,43,2.2)$ and passes every budget. This cell was selected after observing the results, within the unchanged grid.

Table~\ref{tab:witness-margins} gives the budget margins. Native exceeds gates by four and depth by seven. Mixed has five depth units to spare, but no spare width, gates or penalty. Every width-feasible full assignment exceeds the penalty budget by at least $0.275$. These margins apply to the reported compilation but do not establish robustness to other compiler settings.

\begin{table}[!t]
\caption{Signed resource margins for the two reported witnesses.}
\label{tab:witness-margins}
\centering
\footnotesize
\setlength{\tabcolsep}{5pt}
\begin{tabular}{@{}llrrrr@{}}
\toprule
Witness & Design & $m_Q$ & $m_G$ & $m_D$ & $m_M$ \\
\midrule
Cost layer & Native & 0 & $-4$ & $-7$ & 2.2 \\
 & Mixed & 0 & 0 & 5 & 0 \\
Joint secondary & Native & 2 & $-1$ & 14 & 8 \\
 & Mixed & 1 & 6 & 28 & 1 \\
\bottomrule
\end{tabular}
\par\smallskip
\parbox{\columnwidth}{\footnotesize The cost-layer budget is $(12,92,48,2.2)$. The secondary joint budget is $(8,64,64,8)$. The entries are arithmetic differences from the reported vectors.}
\end{table}

The score-based comparison of \eqref{eq:classsep} provides a complementary view. The best strictly mixed candidate has a lower score than the resource-optimised full candidate on every instance in every setting. When the native design is included in the comparison, however, it achieves the lowest score in every random-instance setting and in 173 of 180 constructed settings. The matched random control can then equal or exceed the selected mixed design. Mixed-only cells therefore arise from hard budgets that exclude the native design, not from a score advantage. The two synthesis strategies rank candidates similarly, with median Spearman correlations ranging from $0.905$ to $0.955$ for the random families and from $0.729$ to $0.812$ for the constructed family. They nevertheless change the selected full assignment on 6 to 14 of 30 Max-3SAT instances and 10 to 13 of 30 spin-glass instances, depending on the target. The median Jaccard index of the Pareto sets in $(Q,G,D,M)$ ranges from $0.250$ to $0.333$ for Max-3SAT. Similar rankings therefore do not imply identical feasible regions, and Supplementary Section~S2 contains the correlation, frontier, and control tables.

\subsection{Sensitivity to Routing Seeds}
\label{sec:experiments:seeds}

The 30 Max-3SAT and 30 constructed instances were recompiled with their saved candidate sets on three targets and under both synthesis strategies at five routing seeds. The explicit SABRE settings differ from the earlier run, so this study uses a new seed-1729 baseline. Its $149{,}439$ candidate-set mixed-only cells have a feasible mixed candidate and no feasible endpoint candidate in the candidate set.
Of these cells, 79,069 (52.91\%) remain mixed-only at every seed, although the feasible mixed design may change with the seed.
This is a candidate-set result, not class-wide endpoint exclusion.

A post-hoc check now fixes the design. At each baseline mixed-only cell, the feasible mixed candidate with the lowest resource score is selected, with ties broken by candidate ID. This representation and the budget are kept fixed across seeds. It remains feasible at all five seeds in $100{,}967$ cells ($67.56\%$). If each seed may choose another mixed candidate, $120{,}838$ cells ($80.86\%$), retain mixed feasibility. Mixed feasibility does not imply mixed-only status because an endpoint may become feasible too. The cells share instances, so these are descriptive fractions, not independent trials.

Mixed-only retention ranges from $35.6\%$ for constructed instances on the grid to $84.1\%$ on the ring. Between $254$ and $262$ of the $360$ settings contain a mixed-only cell at each seed. Selection should therefore use the intended compiler settings, and Supplementary Section~S2 gives the seed-specific regions.

\subsection{Preparation Gain Versus QAOA Evolution}
\label{sec:experiments:qaoa}

The decomposition \eqref{eq:gains} separates the classical preparation gain from the gain of one QAOA layer. Thirty new six-variable instances per family are compared under native, strictly mixed with $0<|D|<|\mathcal{C}_3(f)|$, and full representations in noiseless depth-one simulation. Mixed and full selection use the all-to-all reference score with up to 128 seeded candidates each and no budget filter, and the native design is evaluated separately. When the native design is included, it has the lowest score on all 60 instances, so the mixed comparison is conditional on requiring a strictly mixed design. Each arm uses three common parameter starts and $3\times120$ COBYLA evaluations\cite{Powell1994}. 
Seeded uniform samples fill any unused evaluations. Every trace is retained, and the reported value is the lowest expectation across all 360 evaluations rather than an average over restart optima, which the earlier exploratory experiments used.
Pair warming is used throughout, with original and pair probabilities bounded to $[0.05,0.95]$.

\begin{table}[!t]
\caption{Preparation and depth-one evolution gains, 30 instances per family, with paired instance-bootstrap intervals.}
\label{tab:gains}
\centering
\footnotesize
\setlength{\tabcolsep}{4pt}
\begin{adjustbox}{max width=\columnwidth}
\begin{tabular}{@{}llcc@{}}
\toprule
Family & Gain & Mean [95\% interval] & Positive / 30 \\
\midrule
Max-3SAT & Preparation & 0.484 [0.279, 0.703] & 17 \\
 & Native evolution & 1.152 [0.959, 1.344] & 30 \\
 & Mixed evolution & 1.086 [0.900, 1.277] & 30 \\
 & Full evolution & 0.778 [0.661, 0.891] & 30 \\
\addlinespace[2pt]
Cubic spin & Preparation & 0.000 [0.000, 0.000] & 0 \\
 & Native evolution & 2.133 [2.004, 2.238] & 30 \\
 & Mixed evolution & 1.222 [1.063, 1.376] & 30 \\
 & Full evolution & 0.648 [0.522, 0.780] & 30 \\
\bottomrule
\end{tabular}
\end{adjustbox}
\end{table}

\Figure[t!](topskip=0pt, botskip=0pt, midskip=0pt)[width=0.9\textwidth]{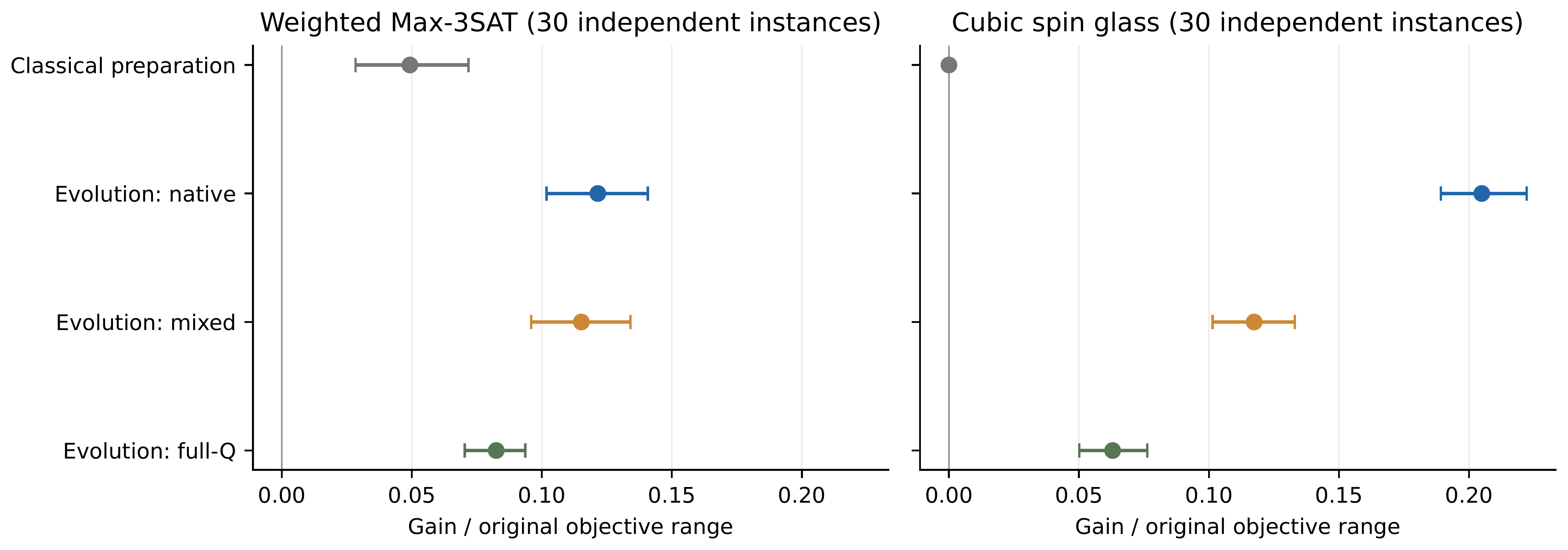} 
{{Preparation and depth-one evolution gains by representation, normalised by the objective range and averaged over 30 instances per family.}\label{fig:gains}}

Table~\ref{tab:gains} uses the original objective units. Fig.~\ref{fig:gains} divides gains by the objective range. For Max-3SAT, the mean preparation gain is $0.484$ and native evolution adds $1.152$. For spin glasses, the relaxation returns uniform marginals, the preparation gain is zero, and native evolution adds $2.133$. Native evolution improves on the warm initial state on all 30 instances of each family. Mixed evolution is close to native for Max-3SAT and about half of native for spin glasses, and full evolution is smallest in both families. The classical controls, deterministic rounding of the relaxation and independent sampling from it, quantify the preparation alone and are not a comparison with an optimised classical solver.

The QAOA Max-3SAT weights are integers in \(\{1,2,3,4,5\}\), and the unscaled spin coefficients are \(\pm 1\). Their Boolean expansions have integer coefficients, so the cubic thresholds \(T_p\) and penalties \(M_p=T_p+1\) are integers. Every encoded computational-basis energy is therefore an integer, making \(2\pi\) a valid cost-angle period. Each matched single-qubit mixer has eigenvalues \(\pm1\), so \(\pi\) is a mixer period up to a global phase. The scan covers $\gamma\in[0,2\pi)$ and $\beta\in[0,\pi)$ on a $65\times65$ grid, followed by four refinements of 240 evaluations each. These periods use the QAOA energy scale, and need not hold after rescaling. 

A post-hoc comparison pairs EE outcomes within each instance. Define $\Delta=E_{\rm other}-E_{\rm native}$, so positive values favour native. For Max-3SAT, mixed has mean difference $0.0659$ with 95\% interval $[-0.0083,0.1379]$.
The full representation has $0.3741$ with interval $[0.2388,0.5074]$.
For spin, the corresponding differences are $0.9115$ with $[0.7414,1.0866]$ and $1.4851$ with $[1.3106,1.6453]$. 
All values are in unscaled objective units, and the intervals use $10{,}000$ paired instance-bootstrap resamples with seed 3619. The mixed representation beats native on 7 Max-3SAT instances and on 1 spin instance.
The Max-3SAT mixed interval includes zero.

Table~\ref{tab:optimisers} compares equal-effort COBYLA (EE), equal-effort Nelder--Mead (NM), and scan refinement (Scan). Native has the lowest mean original objective under all three. Moving from EE to Scan changes the winner on 7 Max-3SAT and 1 spin instance, while moving from EE to NM changes it on 9 and 2 instances. These results show no consistent mean advantage for mixed or full under these protocols. Individual winners can change, and the scans do not prove global optimality. Supplementary Section~S3 gives the best-so-far gaps.

\begin{table}[!t]
\caption{Mean best original objective under three optimisation protocols. Lower is better.}
\label{tab:optimisers}
\centering
\footnotesize
\setlength{\tabcolsep}{5pt}
\begin{tabular}{@{}llrrr@{}}
\toprule
Family & Representation & EE & NM & Scan \\
\midrule
Max-3SAT & Native & 0.631 & 0.772 & 0.405 \\
 & strictly mixed & 0.697 & 0.799 & 0.533 \\
 & Full & 1.005 & 1.057 & 0.891 \\
\addlinespace[2pt]
Cubic spin & Native & $-2.133$ & $-1.979$ & $-2.257$ \\
 & strictly mixed & $-1.222$ & $-1.228$ & $-1.495$ \\
 & Full & $-0.648$ & $-0.535$ & $-0.888$ \\
\bottomrule
\end{tabular}
\end{table}

\subsection{Joint Resource and Solution-Quality Test}
\label{sec:experiments:joint}

The final experiment tests whether a mixed representation can be the only design that meets both a resource budget and a quality target on the same circuit. Thirty new Max-3SAT, 30 new spin-glass, and 12 nine-variable mechanism instances are used. The complete depth-one circuit, consisting of preparation, cost evolution, and matched mixer, is compiled on ring-12 with reuse synthesis, so the physical width includes all routing sites of the whole circuit. Full and strictly-mixed candidate sets of equal size are filtered by the budgets and ranked by score without reference to QAOA outcomes. Quality is measured on the original bits by the normalised gap
\begin{equation}
g=\frac{E_f-\min f}{\max f-\min f}\le 0.2,
\label{eq:gap}
\end{equation}
The extrema are obtained by evaluating $f$ on all $2^n$ assignments.
Each evaluated design receives three starts with 120 Nelder--Mead evaluations each, and the larger scan set is fixed before any quality value is observed. It contains native, full and three guardrail choices under capacity alone, plus the full and mixed choices at the primary budget. Duplicate designs share one evaluation, and the primary budget is $(Q,G,D,M)\le(12,96,64,16)$ and uses scan-refined results. Secondary cells use equal-effort results, including designs that also received the scan.

Table~\ref{tab:joint} separates resource feasibility from joint success.
Each entry gives the number of resource-feasible instances followed by the number that also meet the quality target. Native and mixed designs both pass jointly on all 30 Max-3SAT instances, and full and mixed designs both pass jointly on all 12 mechanism instances. All 30 mixed spin designs pass the resource budget, but none meets the quality target.
No primary instance has both endpoints excluded by resources because native fits every random instance and full fits every mechanism instance. Therefore, zero of 72 instances offer a resource-only exclusion of both endpoints.

There is no exclusive mixed joint success at the primary budget. The one-sided 95\% binomial upper limits are $9.5\%$ for $0/30$ and $22.1\%$ for $0/12$ under an independent-instance sampling model. These limits do not imply a general rate across other families or budgets. A fixed secondary grid has 432 budgets, with $Q_{\max}\in\{8,10,12\}$, $B_G\in\{64,\dots,256\}$, $B_D\in\{32,\dots,128\}$ and $B_M\in\{2,\dots,64\}$. It is tested at $g\le0.1$, $0.2$ and $0.3$. At $g\le0.2$, only the selected mixed design succeeds in at least one cell on 7 Max-3SAT, 0 spin and 5 mechanism instances. Of these, 3, 0 and 4 instances also have a cell with native resource failure and exclusion of all full assignments. These are conditional examples under the fixed direct-pair class, penalty rule and compiler. The cells are correlated sensitivity checks and do not change the primary result.

The secondary case \texttt{max3sat\_07}, identified after the results were available, exhibits the stronger certificate at budget $(8,64,64,8)$. The native design uses $(6,65,50,0)$ and exceeds the CX budget by one gate. This instance has six nonzero cubic supports after aggregation, so there are $3^6=729$ full pair assignments. Every one requires at least nine logical qubits and exceeds the width budget. The selected mixed circuit uses $(7,58,36,7)$, meets every limit, and achieves an equal-effort gap of $0.185$ compared with a warm initial gap of $0.25$. An independent statevector evaluation of the transpiled circuit reproduces the saved expectation. Table~\ref{tab:witness-margins} shows its margins, and native misses feasibility by only one CX gate, so the certificate applies to this reported compilation. The best of $2{,}048$ classical samples from the relaxation also meets $g\le0.2$ on every instance. A best sample and a circuit expectation are different quantities. These tests establish no quantum advantage.

The three selection rules in Supplementary Section S4 choose the same design on all 72 instances, so no selection benefit is observed. In the separate held-out test, the prepared-state diagnostic reduces inconsistency-prediction MSE by $27.7\%$. Its quality-prediction improvement has an interval containing zero, so no causal benefit from diagnostic filtering is established, and Supplementary Section~S4 gives the prediction results.

\begin{table}[!t]
\caption{Primary joint test at budget $(12,96,64,16)$ with $g\le 0.2$. }
\label{tab:joint}
\centering
\footnotesize
\setlength{\tabcolsep}{6pt}
\begin{tabular}{@{}lcccc@{}}
\toprule
Cohort & $N$ & Native & Full & strictly mixed \\
\midrule
Max-3SAT & 30 & $30\to 30$ & $8\to 7$ & $30\to 30$ \\
Cubic spin & 30 & $30\to 1$ & $0\to 0$ & $30\to 0$ \\
Mechanism & 12 & $9\to 9$ & $12\to 12$ & $12\to 12$ \\
\bottomrule
\end{tabular}
\end{table}

\subsection{Additional Comparisons}
\label{sec:experiments:additional}

The exploratory Pareto studies use different selectors and protocols from the primary experiments with fixed candidate sets. They cover 180 compilation instances with $n\in\{8,12,16,20\}$. On all-to-all coupling, selective Max-3SAT reduction saves a median of 25 two-qubit gates and 60.5 depth units relative to native. Spin glass shows a trade-off between gate count and depth. On the 12-qubit grid, 108 native, 57 selective and 21 full designs fit. These data do not extend the primary certified mixed-only incidence to larger sizes.

Table~\ref{tab:exploratory-size} groups the existing records by size. Each all-to-all entry is a median triple $(\Delta N_\sigma, \Delta G, \Delta D)$ of paired changes from native, so a negative gate or depth change is a saving, and the grid columns count designs within the 12-qubit capacity. For each fixed design, the median over five compiler seeds is taken first. For matched random controls, these medians are averaged over the five draws within each instance. Native on the same instance is then subtracted, and the median paired difference is reported. The width difference is logical $\Delta N_\sigma$, not physical width.

All 180 instances enter the all-to-all comparison. A common grid instance must fit native, selective, full and all five random controls. This leaves 9 spin and 12 Max-3SAT instances, all with $n=8$. Table~\ref{tab:exploratory-grid-changes} reports their paired changes. No full design fits at $n=12$, and even native exceeds capacity at $n\ge16$. Excluded designs are not assigned zero cost, and no common-cohort difference is available at these sizes.

\begin{table*}[!t]
\centering
\footnotesize
\setlength{\tabcolsep}{4pt}
\caption{Exploratory compilation results by original problem size. All-to-all entries give paired median changes from native.}
\label{tab:exploratory-size}
\begin{tabular}{@{}llrccrrrr@{}}
\toprule
 & & & \multicolumn{2}{c}{All-to-all: all $N$ instances fit} & \multicolumn{4}{c}{12-qubit grid: counts within capacity} \\
\cmidrule(lr){4-5}\cmidrule(l){6-9}
Family & $n$ & $N$ & Selective change & Full change & Native & Selective & Full & Common \\
\midrule
Spin glass & 8 & 36 & $(0,0,0)$ & $(+6,+24,-21)$ & 36 & 30 & 9 & 9 \\
Spin glass & 12 & 18 & $(+1,+6,-10.5)$ & $(+6.5,+50,-28)$ & 18 & 9 & 0 & 0 \\
Spin glass & 16 & 18 & $(+1,+4,-15)$ & $(+7.5,+75,-35)$ & 0 & 0 & 0 & 0 \\
Spin glass & 20 & 18 & $(+1.5,+16,-27)$ & $(+9.5,+111,-43.5)$ & 0 & 0 & 0 & 0 \\
Max-3SAT & 8 & 33 & $(+4,-12,-31)$ & $(+7,-16,-37)$ & 33 & 18 & 12 & 12 \\
Max-3SAT & 12 & 21 & $(+5,-26,-50)$ & $(+10,-38,-58)$ & 21 & 0 & 0 & 0 \\
Max-3SAT & 16 & 18 & $(+6.5,-42,-77.5)$ & $(+9.5,-52,-81)$ & 0 & 0 & 0 & 0 \\
Max-3SAT & 20 & 18 & $(+8.5,-56,-102.5)$ & $(+10.5,-77,-103.5)$ & 0 & 0 & 0 & 0 \\
\bottomrule
\end{tabular}
\par\smallskip
\parbox{0.97\textwidth}{\footnotesize Common requires native, selective, full and every matched-random control to fit. 
Selective returns native on 43 of the 90 spin instances, and the matched-random fit counts equal the selective counts.
}
\end{table*}

\begin{table}[!t]
\centering
\footnotesize
\setlength{\tabcolsep}{3pt}
\caption{Paired changes on the common grid cohort with $n=8$. 
}
\label{tab:exploratory-grid-changes}
\begin{tabular}{@{}lrcc@{}}
\toprule
Family & $N$ & Selective change & Full change \\
\midrule
Spin glass & 9 & $(0,0,0)$ & $(+3,+45,+13)$ \\
Max-3SAT & 12 & $(+2,-10,-21)$ & $(+2,-16,-29)$ \\
\bottomrule
\end{tabular}
\end{table}

Table~\ref{tab:exploratory-size-cost} gives the recorded selector cost. 
The selector uses precomputed relaxation moments, and the times exclude relaxation solving, device transpilation, and QAOA.
They depend on the platform and instance structure, so they are not total workflow costs.

\begin{table}[!t]
\centering
\footnotesize
\caption{Exploratory selector cost by original problem size. }
\label{tab:exploratory-size-cost}
\begin{tabular}{@{}llrr@{}}
\toprule
Family & $n$ & Candidate evaluations & Time (s) \\
\midrule
Spin glass & 8 & 2560 & 0.74 \\
Spin glass & 12 & 4096 & 1.91 \\
Spin glass & 16 & 5632 & 2.69 \\
Spin glass & 20 & 7168 & 6.21 \\
Max-3SAT & 8 & 2368 & 1.19 \\
Max-3SAT & 12 & 3520 & 1.48 \\
Max-3SAT & 16 & 4864 & 2.82 \\
Max-3SAT & 20 & 6688 & 5.02 \\
\bottomrule
\end{tabular}
\par\smallskip
\parbox{\columnwidth}{\footnotesize The sample counts are the $N$ values in Table~\ref{tab:exploratory-size}. Costs include instances that later fail the grid capacity check. Source: corrected exploratory selector records.}
\end{table}

A 14-instance noiseless QAOA experiment at depths one and two finds native best on average under its protocol. A separate 20-instance signal-screened experiment does not establish a general pair-moment benefit. The 18-instance simulated-noise experiment shows no clear selective advantage at its nonzero noise levels. Its noise parameters are representative rather than hardware calibrated, and all of these experiments remain separate from the primary cohorts.

\section{Discussion}
\label{sec:discussion}

The results address classical exactness, resource feasibility, and finite-depth QAOA quality as three distinct properties, each of which needs a separate assessment.

\subsection{Feasibility as a Representation Criterion}
\label{sec:discussion:feasibility}
Mixed representations can be necessary under joint resource limits. Within the stated direct-pair class and fixed penalty rule, both endpoints can fail while a mixed design fits. Theorem~\ref{thm:boundary} characterises static feasibility, and Corollary~\ref{cor:static} gives an infinite static mixed-only family. Corollary~\ref{cor:compiled} extends the separation to gates and depth for the stated reference synthesis. Section~\ref{sec:experiments:resources} gives separate compilation evidence under the tested toolchain. These results do not give an efficient feasibility test for arbitrary devices.

Hard budgets and weighted scores answer different questions. Native has the lowest unrestricted score in every random-instance setting of the primary compilation experiment, yet a mixed design can still be the only feasible choice.

\subsection{Dependence on Class, Structure, and Compiler}
\label{sec:discussion:dependence}

The findings depend on the representation class, the instance structure, and the compiler. The static theorem needs no compiler, the reference gate and depth bounds apply only to the stated synthesis, and neither result excludes other quadratization methods. In compilation the effect is strongest for the constructed pair-star family, occurs for some Max-3SAT instances, and disappears for the tested spin instances under reuse synthesis. For spin glasses, the native objective is already compact in Pauli form, and the isolated-star calculation in Section~V-B gives one mechanism by which direct-pair reduction can increase the reference gate count, although signed overlapping terms can cancel and reverse the ordering. 

Different synthesis and routing settings can change the feasible set, so selection should use the settings of the planned circuit. The fixed-design check in Section~\ref{sec:experiments:seeds} separates persistence of one mixed design from the existence of some feasible mixed design at each seed. Neither alone guarantees that endpoints remain infeasible.

Gate count and depth do not fix hardware duration or success probability, and the penalty budget is a declared modelling constraint at a fixed objective scale rather than a calibrated hardware limit. Increasing penalties need not increase the coefficient dynamic range. The score weights state a preference between resources, and reporting the feasible resource frontier makes that preference visible.

\subsection{Resources Versus Solution Quality}
\label{sec:discussion:quality}

Resource feasibility does not guarantee solution quality. Native has the best mean original objective under equal-effort COBYLA, Nelder-Mead, and the larger scan. Although the winning design changes for some instances, the tested protocols show no consistent mean advantage for mixed or full reduction. The primary joint test finds no mixed-only success at its primary budget, and secondary examples establish joint success only at particular budgets. The best classical sample meets the quality target on every tested instance, although a best sample and a circuit expectation are different quantities, so these tests do not establish quantum advantage.

The gain decomposition separates classical preparation from later evolution. Native evolution improves on the warm initial state for all 30 instances in each family. Auxiliary inconsistency is a possible mechanism for the smaller gains after reduction, although its observed weight does not show that variational effort is wasted. The tests do not establish a general benefit from pair pseudo-moments. The prepared-state diagnostic improves held-out prediction of auxiliary inconsistency in the tested setting. Reliable quality prediction is not established, and the tested guardrails change no selection.

Earlier work also shows that finite-depth QAOA depends on Hamiltonian structure and constraint density \cite{WangMaxCut2018,FarhiLocal2020,Akshay2020}. A good classical starting point can also leave little room for improvement \cite{Cain2023}. The exactness and feasibility results of this paper therefore do not certify solution quality.

\subsection{Implications for Practice}
\label{sec:discussion:practice}

First fix the device, circuit stage, synthesis, routing protocol, objective scale, and penalty rule, then state all budgets and score weights. Construct exact native, mixed, and full candidates and compile them under this protocol. 
Every active budget should be checked before selection by score, and the resource frontier should be kept even when both endpoints fit.
A minimum-auxiliary full reduction need not minimise penalties or compiled resources, so its failure cannot exclude every full reduction. Such exclusion needs a complete check or a valid bound over the stated class. Passing a logical check does not establish compiled feasibility.

A finite candidate set returns the best evaluated design, and the outcome is reported as ``none found'' when no candidate fits. Pair reuse can guide candidate construction, but its absence is not a reason to omit mixed designs. Selected designs should then be checked across routing seeds, and solution quality should be assessed separately.

\subsection{Limitations and Future Directions}
\label{sec:discussion:limitations}

The primary QAOA data are small, noiseless, and restricted to one layer, and no selected circuit has been run on a quantum device. Finite candidate sets can miss feasible designs. The optional compiled certification uses a trivial lower bound and is not a scalable solver. Constructed instances test a mechanism and are separate from the random families. Protocols were frozen locally rather than externally preregistered, and the primary random instances have six variables. Separate exploratory compilation tests reach 20 variables, and a separate experiment uses simulated noise. Because their protocols differ from the primary study, they do not establish how primary certified mixed-only incidence scales with size.

The framework supplies exact circuit candidates for further tests. Future work could execute complete circuits on a calibrated device, including preparation, cost evolution, and mixers. These tests should compare quality and total computational cost, including relaxation and parameter search, with strong classical methods. Larger instances, deeper circuits, other compilers, and stronger compiled bounds would test the limits of the framework. Mixers that preserve auxiliary consistency would test the proposed inconsistency mechanism.

\section{Conclusion}
\label{sec:conclusion}

Selective reduction supplies an exact design space between native encoding and full quadratization, and this paper shows how to use it under resource constraints. The aggregate-residual penalty condition preserves the original objective pointwise, is sharp, and survives sequential composition, so every member of the family is certified exact before any circuit is built. Within the stated direct-pair class and penalty rule, the static theorem characterises exactly when joint limits exclude both endpoints yet permit a mixed design, and the reference-synthesis construction extends that separation to gate and depth budgets. Compilation with an industry-standard toolchain then confirms the behaviour on 12-qubit targets for structured instances and for a substantial fraction of weighted maximum three-satisfiability instances, so the enlarged design space increases the number of instances that fit a given device without approximating the objective. These results establish resource feasibility as a legitimate representation criterion. They do not establish better finite-depth QAOA quality, since the native encoding gives the best mean objective in the primary comparisons and the joint test finds no mixed-only success at the primary budget, with secondary examples succeeding only at particular budgets. Exactness, resource feasibility, and solution quality are therefore distinct properties that must be assessed separately, and all findings remain limited to the stated class, compiler protocols, and instance sets.

\section*{Acknowledgment}
The authors thank the University of Warwick Undergraduate Research Support Scheme
for supporting this project.

\section*{Data and Code Availability}
\begingroup
\Urlmuskip=0mu plus 1mu
\def\UrlBreaks{\do\/\do\-\do\_\do\.}
Processed data, source manifests and analysis code are available at \url{https://github.com/rrocy6/Exact-mixed-locality-representation-selection-for-resource-constrained-QAOA}. The results here refer to the frozen 5 October 2026 reproduction archive. They do not refer to a changing default branch.
Primary compilation and preparation results use \path{archives/revision_20260917/revision/}. Joint tests use \path{archives/feedback_iteration_v3/}. Routing results use \path{corrections/review_v2_20260920/final_analysis/}. The 180-instance compilation results and selector costs use \path{delivery/20260913/archives/URSS_E2_E5_LOCAL_RESULTS_20260913_020457_389.zip}. Historical records are retained separately.

\endgroup

\section*{Use of AI Tools}
AI tools assisted with the construction of Figure~\ref{fig:pipeline}, suggested some illustrative examples in Section~\ref{sec:reduction:example}, generated the initial code for numerical validation, and contributed the proof idea for the necessity part of Proposition~\ref{prop:sharp}. The third author subsequently checked and revised the code and standardised its style. All AI-generated material was reviewed and verified by the authors for accuracy and correctness. The authors take full responsibility for the final content.

\begingroup

\endgroup
%

\begin{IEEEbiography}[{\includegraphics[width=1in,height=1.25in,clip,keepaspectratio]{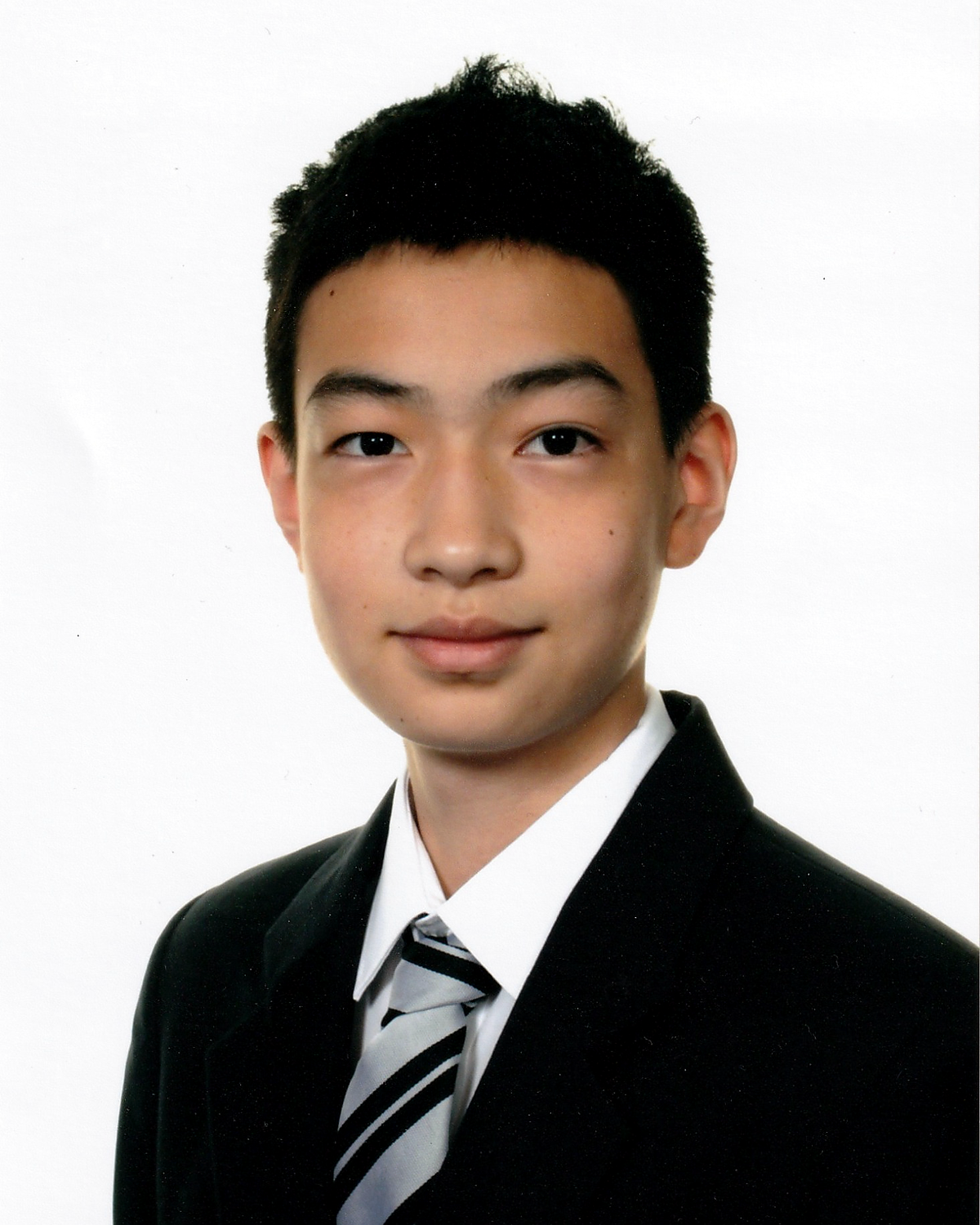}}]{Jingcong Liu}
is currently pursuing the MMath degree with the Department of Mathematics, University of Warwick, Coventry, U.K. He is also a Teaching Assistant with the Department of Mathematics and a Research Assistant with the Secure Cyber Systems Research Group (SCSRG), Warwick Manufacturing Group (WMG), University of Warwick. His research interests include quantum algorithms, quantum optimization, and the geometry of quantum dynamics.
\end{IEEEbiography}

\begin{IEEEbiography}[{\includegraphics[width=1in,height=1.25in,clip,keepaspectratio]{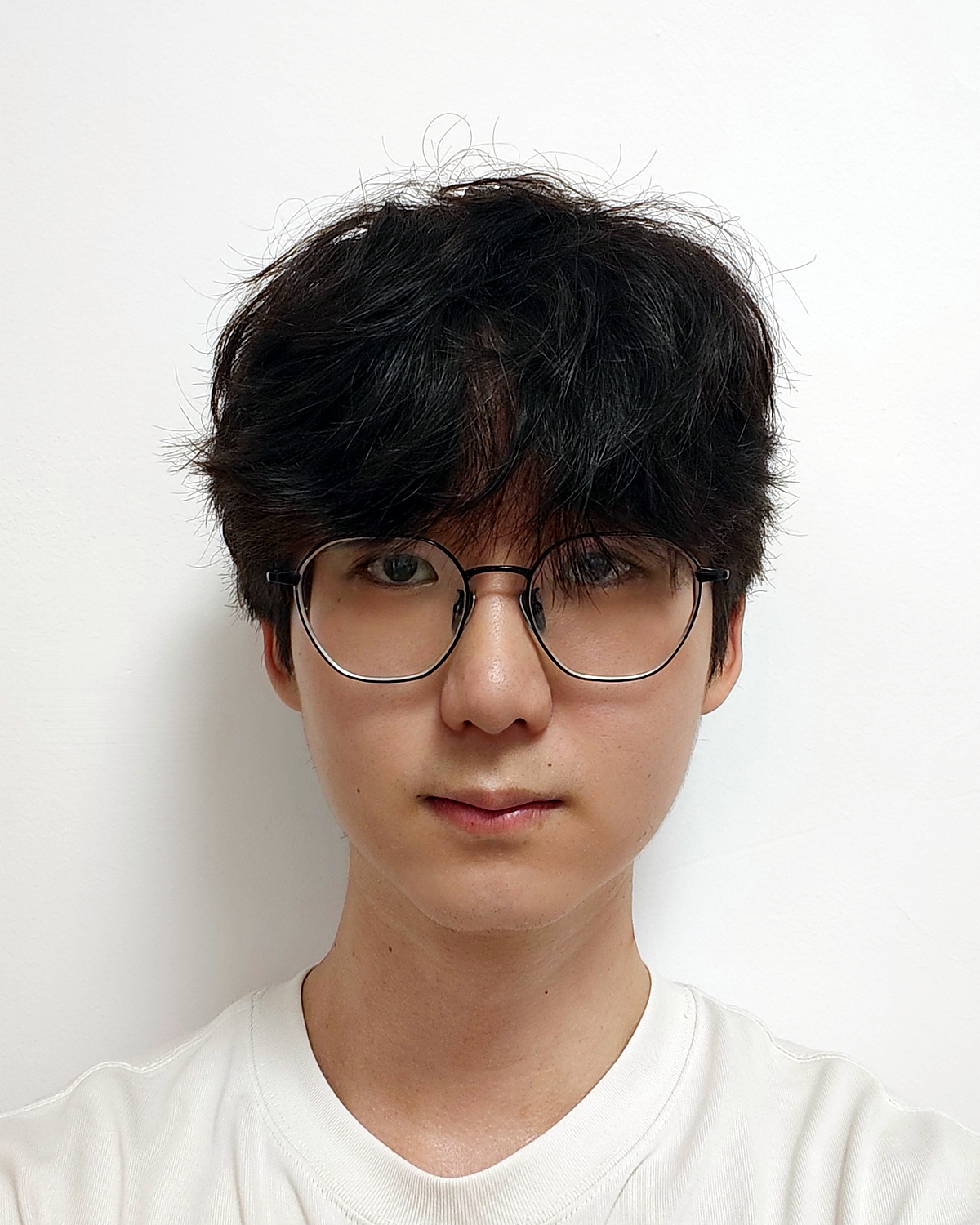}}]{Yiding Su}
is currently pursuing the B.Sc. degree in Mathematics at the University of Warwick, Coventry, U.K. His research interests include statistical modelling, computational statistics, and combinatorial optimization.
\end{IEEEbiography}

\begin{IEEEbiography}[{\includegraphics[width=1in,height=1.25in,clip,keepaspectratio]{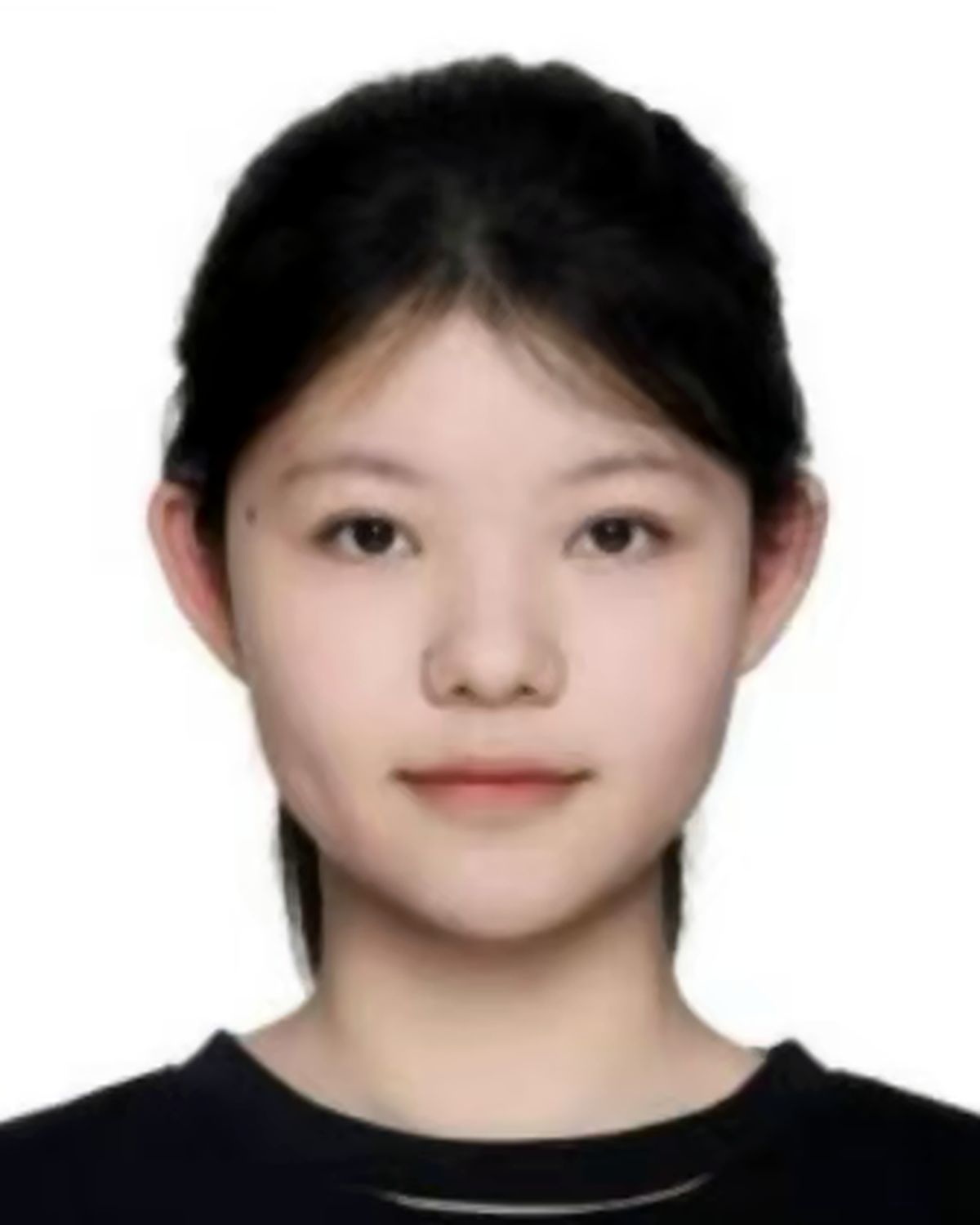}}]{Ruojun Zhang}
is currently pursuing the B.Sc. degree in Computer Science
with the School of Electronics and Computer Science,
University of Southampton, Southampton, U.K.
Her research interests include machine learning, AI agents,
and combinatorial optimization.
\end{IEEEbiography}

\begin{IEEEbiography}[{\includegraphics[width=1in,height=1.25in,clip,keepaspectratio]{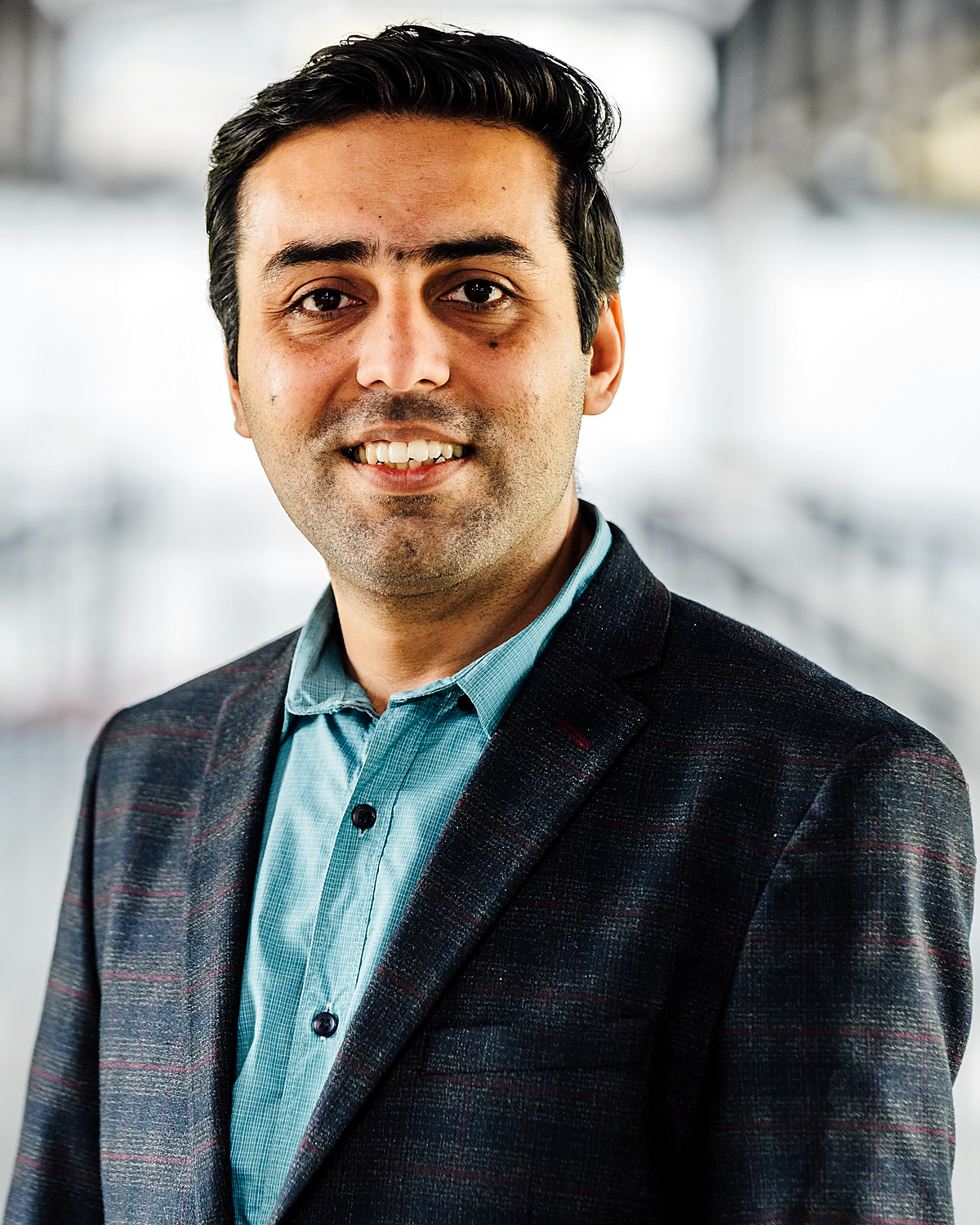}}]{Hafiz Muhammad Waseem}
received his Ph.D. in Computer Engineering from Gachon University, Seongnam, South Korea, in 2023. He served as an Assistant Professor of Quantum Computing and Information Security at Gachon University from 2023 to 2024. He is currently a Research Fellow at the Warwick Manufacturing Group, University of Warwick. His research interests include quantum computing, cryptography, quantum artificial intelligence, and cybersecurity.

\end{IEEEbiography}

\begin{IEEEbiography}[{\includegraphics[width=1in,height=1.25in,clip,keepaspectratio]{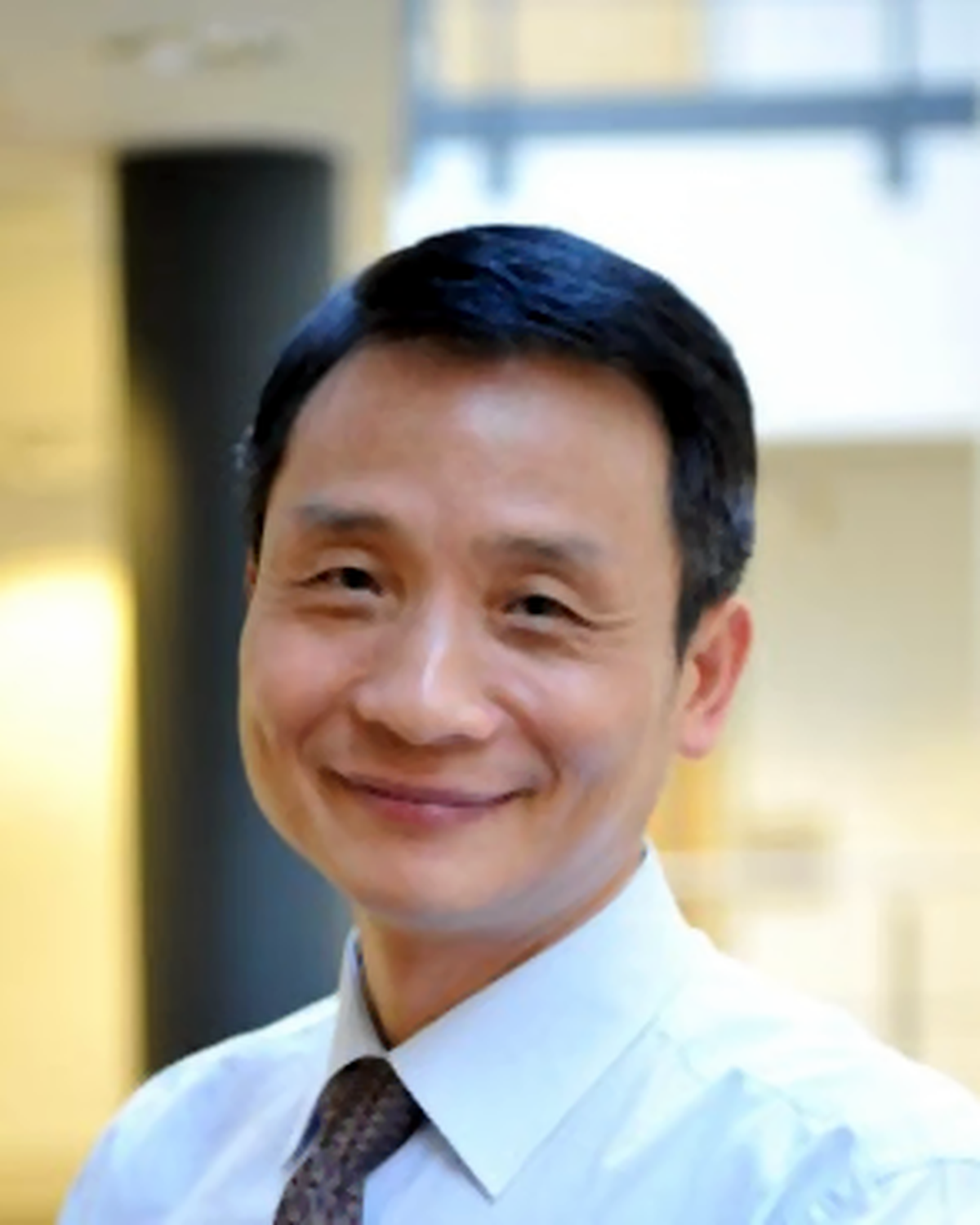}}]{Bo Chen} received the Ph.D. degree in operations research and econometrics from Erasmus University, The Netherlands, in 1994. He is Professor of Operational Research, Analytics and Management Science at Warwick Business School, University of Warwick, U.K. He was a co-principal investigator on the EPSRC Science and Innovation Award that established the Centre for Discrete Mathematics and its Applications. His research interests include combinatorial optimization, algorithm design and analysis, game theory, and the mathematical and computational foundations of artificial intelligence. Prof. Chen was awarded the higher doctorate of Doctor of Science (D.Sc.) by the University of Warwick in 2012. He is a Fellow of the Academy of Social Sciences, the Operational Research Society, and the Institute of Mathematics and its Applications.
\end{IEEEbiography}

\begin{IEEEbiography}[{\includegraphics[width=1in,height=1.25in,clip,keepaspectratio]{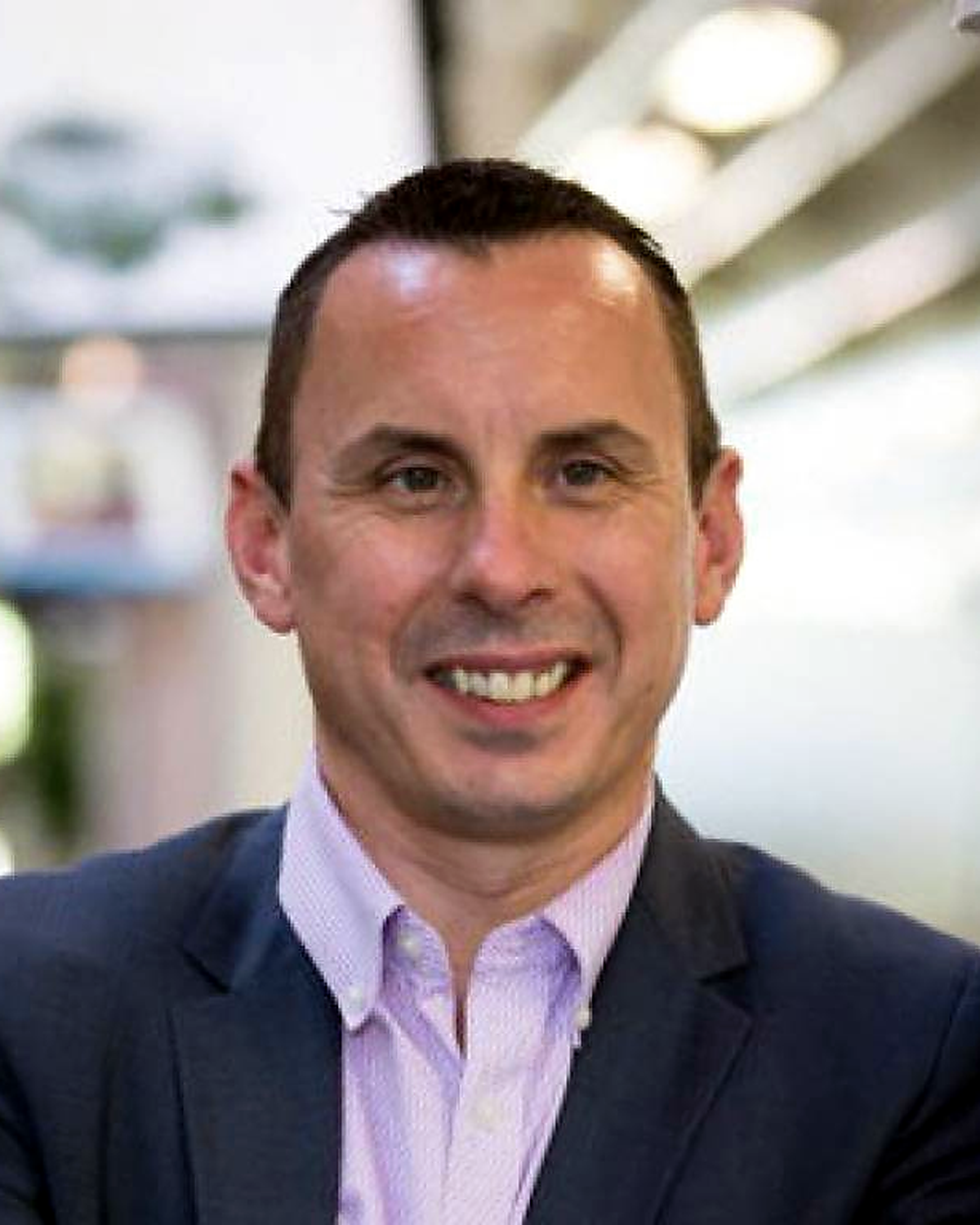}}]{Carsten Maple}
is the Director of the NCSC-EPSRC Academic Centre of Excellence in Cyber Security Research and Professor of Cyber Systems Engineering at the University of Warwick. He is also Director for Research Innovation at EDGE-AI, the National Edge Artificial Intelligence Hub, and a Professor and Fellow of the Alan Turing Institute, where he is a principal investigator on a \$9 million project developing trustworthy digital infrastructure. Carsten is also co-investigator on three further UK research and Innovation grants, leading work on trustworthy systems. He has published over 450 peer-reviewed papers, and his research has been widely reported through the media.
\end{IEEEbiography}

\SupplementSetup
\title{Supplementary Information for\\
``Exact Mixed-Locality Encodings for Resource-Constrained QAOA''}

\author{Jingcong Liu, Yiding Su, Ruojun Zhang, Hafiz Muhammad Waseem, Bo Chen, and Carsten Maple\\[0.6ex]
{\small Corresponding author: Hafiz Muhammad Waseem (e-mail: HM.Waseem@warwick.ac.uk)}}

\maketitle

\begin{abstract}
This document provides the proofs, protocols, extended results, and reproducibility records that support the main text. Equation, theorem, figure, and table numbers without the prefix S refer to the main text.
\end{abstract}

\tableofcontents

\clearpage
\section{Proofs of the Theoretical Results}
\label{sup:proofs}

This section proves the results stated in Sections~III and IV of the main text. Notation follows Table~2 of the main text. Equation numbers without the prefix S refer to the main text. Throughout, $q$ denotes a collected multilinear polynomial on $\B^N$, $p=\{i,j\}$ a pair of its coordinates, $\mathcal{S}_p$ the selected family of terms of degree at least three that contain $p$, and $A_p(v)$ the aggregate residual of (3) in the main text, which does not depend on $u_i$ or $u_j$.

\subsection{Sharp Aggregate-Residual Penalty (Proposition 1)}
\label{sup:proofs:sharp}

\begin{proof}
Fix $u$ and define $a=u_i$, $b=u_j$, and $A=A_p(v)$. With $u$ fixed, $B_p(u)$ and $A$ are constants and only $y_p$ varies. Subtracting the original polynomial gives
\begin{equation}
Q_{p,M_p}(u,y_p)-q(u)=(y_p-ab)\,A+M_p\,\PR(a,b,y_p).
\label{eq:S:excess}
\end{equation}
If $y_p=ab$, both terms are zero, so $\min_{y_p}Q_{p,M_p}(u,y_p)\le q(u)$. Equality holds if and only if the incorrect value of $y_p$ is never lower. For $(a,b)=(0,0)$, the incorrect value is $y_p=1$, with objective change $A$ and penalty $3M_p$. For $(a,b)\in\{(0,1),(1,0)\}$, the incorrect value is $y_p=1$, with objective change $A$ and penalty $M_p$. For $(a,b)=(1,1)$, the incorrect value is $y_p=0$, with objective change $-A$ and penalty $M_p$. The excess of the incorrect value is therefore $A+3M_p$, $A+M_p$, or $M_p-A$ respectively.

The second row is nonnegative for every $v$ exactly when $M_p\ge-\min_vA_p(v)$, and the third exactly when $M_p\ge\max_vA_p(v)$. The first row adds no condition, because $A+3M_p=(A+M_p)+2M_p$ and $M_p\ge 0$. The two conditions together state that $M_p\ge\max_v|A_p(v)|=\norminf{A_p}$, which proves sufficiency.

For necessity, suppose $M_p<\norminf{A_p}$ and choose $v^\star$ with $|A_p(v^\star)|=\norminf{A_p}$, which exists because the domain is finite. If $A_p(v^\star)>M_p$, take $(a,b)=(1,1)$, so that $y_p=0$ has excess $M_p-A_p(v^\star)<0$. If $A_p(v^\star)<-M_p$, take $(a,b)=(0,1)$, so that $y_p=1$ has excess $A_p(v^\star)+M_p<0$. In either case an incorrect value is strictly better for some $u$, and pointwise exactness fails.

The three rows and $M_p\ge 0$ give, for every incorrect value, $Q_{p,M_p}(u,y_p)-q(u)\ge M_p-|A_p(v)|\ge M_p-\norminf{A_p}$, while the correct value has zero excess. This proves the margin bound (6) of the main text.

If $M_p>\norminf{A_p}$, every incorrect value has strictly positive excess, so $y_p=u_iu_j$ is the unique minimiser for every $u$. If $M_p=\norminf{A_p}>0$, choose $v^\star$ as above. When $A_p(v^\star)=\norminf{A_p}$, the choice $(a,b)=(1,1)$ makes the third row zero, and when $A_p(v^\star)=-\norminf{A_p}$, the choice $(a,b)=(0,1)$ makes the second row zero. If $\norminf{A_p}=0$, then $M_p=0$ and the second row is again zero. Equality at the threshold therefore produces a tie for at least one $u$. Together with the below-threshold case, this shows that the correct value is the unique minimiser for every $u$ if and only if $M_p>\norminf{A_p}$.
\end{proof}

A small example shows why the combined residual matters. Take the selected block $2x_1x_2x_3-3x_1x_2x_4$ and replace $x_1x_2$. The residual $A=2x_3-3x_4$ takes the values $0$, $2$, $-3$, and $-1$ on the four assignments of $(x_3,x_4)$, so $\norminf{A}=3$. A penalty $M=3$ preserves the minimum value but allows a tie. At $x=(0,1,0,1)$ the incorrect value $y=1$ changes the objective by $-3+M=0$. With $M>3$ the change is positive, and with $M<3$ it is negative, so exactness fails. If both coefficients were positive, the residual $2x_3+3x_4$ would reach $5$, and a penalty of $3$ would be insufficient even though it exceeds each coefficient separately.

\subsection{Sequential Compositional Exactness (Theorem 1)}
\label{sup:proofs:composition}

Before step $r$ the available bits are $u^{(r-1)}=(x,y_{<r})$. Two distinct current
coordinates $i_r$ and $j_r$ are chosen, either of which may be an original bit or an
earlier auxiliary, and their values are written $a_r(u)=u_{i_r}$ and $b_r(u)=u_{j_r}$.
Their product is factored out of the selected terms and replaced by a new auxiliary
$y_r$ with penalty $M_r$,
\begin{equation}
g^{(r-1)}(u^{(r-1)})=B_r+a_rb_rA_r,\qquad
g^{(r)}(u^{(r-1)},y_r)=B_r+y_rA_r+M_r\,\PR(a_r,b_r,y_r),
\label{eq:S:step}
\end{equation}
where $A_r$ is computed from the current polynomial with all coordinates, including
earlier auxiliaries, treated as independent Boolean variables. The margin is
$\delta_r=M_r-\norminf{A_r}$. The recursively consistent lift is defined in order by
$\phi_r(x)=a_r(x,\phi_{<r}(x))\,b_r(x,\phi_{<r}(x))$ and $\Phi(x)=(\phi_1(x),\dots,\phi_R(x))$.

\begin{proof}
Fix $x$ and an arbitrary auxiliary vector $y$, whose earlier entries need not be
consistent. Applying the margin bound of Proposition~1 at step $r$ to the current
assignment $(x,y_{<r})$ gives
\begin{equation}
g^{(r)}(x,y_{\le r})-g^{(r-1)}(x,y_{<r})\ \ge\ \delta_r\,\mathbf{1}\bigl\{y_r\neq a_r(x,y_{<r})\,b_r(x,y_{<r})\bigr\}.
\label{eq:S:onestep}
\end{equation}
Adding \eqref{eq:S:onestep} over $r=1,\dots,R$, the intermediate polynomials cancel in
pairs and the left side reduces to $g^{(R)}(x,y)-g^{(0)}(x)=g^{(R)}(x,y)-f(x)$. The right
side is the sum in (7) of the main text, which is nonnegative because
every margin is nonnegative. Hence $\min_yg^{(R)}(x,y)\ge f(x)$.

For the consistent lift $y=\Phi(x)$, every substituted product equals its auxiliary
and every penalty is zero, so each step leaves the value unchanged and
$g^{(R)}(x,\Phi(x))=f(x)$. This gives $\min_yg^{(R)}(x,y)\le f(x)$, and the two
inequalities prove pointwise exactness.

Suppose every margin is positive and $y\neq\Phi(x)$. Let $r^\star$ be the first index
at which they differ. Then $y_{<r^\star}=\phi_{<r^\star}(x)$, so the intended product at
step $r^\star$ equals $\phi_{r^\star}(x)$ and the indicator for that step equals one.
All other terms are nonnegative, so $g^{(R)}(x,y)-f(x)\ge\delta_{r^\star}>0$. Every
other assignment therefore exceeds the value at $\Phi(x)$, and $\Phi(x)$ is the unique
minimiser.
\end{proof}

\subsection{Target-Locality Reduction and Auxiliary Bound}
\label{sup:proofs:grammar}

\begin{corollary}[Selective target-$\ell$ reduction]
\label{cor:S:target}
Fix $\ell\ge 2$ and $D\subseteq\msupp(f)$. There exists an acyclic Rosenberg reduction
$F_{D,\ell}$ with
\begin{equation}
f(x)=\min_{y\in\B^{m}}F_{D,\ell}(x,y),\qquad m\le\sum_{A\in D}\max\{|A|-\ell,0\},
\label{eq:S:target}
\end{equation}
in which every selected monomial has terminal degree at most $\ell$. A construction with
disjoint chains of new auxiliaries permits penalties $M_{A,s}\ge|c_A|$ at each step, and
strict inequalities make the consistent lift unique. If every term of degree above
$\ell$ belongs to $D$, then $\deg F_{D,\ell}\le\ell$.
\end{corollary}

A substitution records $y_S=v_{B_S}v_{C_S}$ with
$S=B_S\,\dot\cup\,C_S$ and $v_{\{i\}}=x_i$. A block with more than one variable must have
been created at an earlier step, and processing blocks in order of size prevents
circular dependence. A support written as a product of $t$ available blocks has locality
$t$ and equals the original monomial when the auxiliaries are consistent. This is the
standard recursive construction of Rosenberg \cite{Rosenberg1975} in the form surveyed
by Boros and Gruber \cite{BorosGruber2014}. The proof below gives the auxiliary bound
used in the main text.

\begin{proof}
Let $A\in D$ with $k=|A|$. If $k\le\ell$ the term is left unchanged. Otherwise order the
variables as $A=\{i_1,\dots,i_k\}$ and introduce the chain
\begin{equation}
y_{A,1}=x_{i_1}x_{i_2},\qquad y_{A,s}=y_{A,s-1}x_{i_{s+1}}\quad(2\le s\le k-\ell).
\label{eq:S:chain}
\end{equation}
After step $s$ the descendant of $c_Ax_A$ is $c_A\,y_{A,s}\prod_{t=s+2}^{k}x_{i_t}$, and
the selected residual is $c_A\prod_{t=s+2}^{k}x_{i_t}$, whose Boolean sup norm equals
$|c_A|$. Proposition~1 therefore permits $M_{A,s}\ge|c_A|$ at every step, and the final
descendant has degree $\ell$. Using disjoint chains for all selected supports and
concatenating them in any topological order, Theorem~1 gives pointwise exactness with
exactly $\sum_{A\in D}\max\{|A|-\ell,0\}$ auxiliaries, and strict penalties give
uniqueness. If every term of degree above $\ell$ is selected, all remaining objective
terms have degree at most $\ell$, because the added penalties are quadratic.
\end{proof}

\subsection{Cubic Mixed-Sign Exactness and Pair Cover (Proposition 2)}
\label{sup:proofs:cubic}

\begin{proof}
For fixed $x$, the representation (8) of the main text satisfies
\begin{equation}
F_{\rho,M}(x,y)-f(x)=\sum_{p\in P_\rho}\Bigl[(y_p-x_ix_j)A_p(x)+M_p\,\PR(x_i,x_j,y_p)\Bigr].
\label{eq:S:cubicdiff}
\end{equation}
Each summand depends on a single auxiliary, and $A_p$ is linear in distinct remaining
coordinates, so $\norminf{A_p}=T_p$ as in (9) of the main text.
If $M_p\ge T_p$ for every $p$, Proposition~1 makes every summand nonnegative and the
consistent assignment makes every summand zero, which proves sufficiency. Conversely,
if $M_p<T_p$ for some $p$, Proposition~1 supplies an original assignment and an
incorrect value of $y_p$ for which the $p$-summand is negative. Setting every other
auxiliary to its consistent product makes the remaining summands zero, including those
whose pairs share original variables with $p$, so $\min_yF_{\rho,M}(x,y)<f(x)$ and
exactness fails. The uniqueness statement follows in the same way from the strict part
of Proposition~1.
\end{proof}

For $D\subseteq\mathcal{C}_3(f)$ the pair shadow is $\partial_2D=\bigcup_{A\in D}\binom{A}{2}$.
Within the direct-pair construction, the minimum number of auxiliaries that reduces
every term of $D$ is the set-cover number
\begin{equation}
\tau_2(D)=\min\Bigl\{\sum_{p\in\partial_2D}z_p\ :\ \sum_{p\in\binom{A}{2}}z_p\ge 1\ \ \forall A\in D,\ z_p\in\{0,1\}\Bigr\}.
\label{eq:S:cover}
\end{equation}
Any pair assignment activates a set of pairs that covers every selected support, so
its size is at least $\tau_2(D)$. Conversely, any cover yields an exact reduction with
one auxiliary per covering pair and penalties above their thresholds. Minimising the
auxiliary count is the objective of several substitution-selection methods
\cite{VermaLewis2020,VermaLewisKochenberger2021,Schmidbauer2026}, and set cover is
NP-hard in general \cite{Karp1972}. The count does not minimise penalties or compiled
resources. With $\Delta_2(D)$ the largest number of selected supports containing a
common pair, $\lceil|D|/\Delta_2(D)\rceil\le\tau_2(D)\le|D|$, since each pair covers at
most $\Delta_2(D)$ supports and one pair per support always suffices. Pair reuse can be
visualised on the overlap graph whose vertices are cubic supports and whose edges join
supports sharing two variables. All terms assigned to one auxiliary $y_p$ belong to the
clique $\{A\in\mathcal{C}_3(f) : p\subset A\}$.

\subsection{Static Feasibility Boundary (Theorem 2)}
\label{sup:proofs:boundary}

\begin{proof}
Let $m=|\mathcal{C}_3(f)|$, let $K_Q$ be the largest number of terms reducible under the
auxiliary, penalty, and degree budgets, and let $h_0$ be the retained-term budget.

Suppose a feasible representation exists. It retains at most $h_0$ terms, so it reduces
at least $m-h_0$ terms, and its reduced set satisfies the first three budgets, so it
reduces at most $K_Q$ terms. Hence $m-h_0\le K_Q$. Conversely, suppose $K_Q+h_0\ge m$.
The design space is finite, so some assignment reduces exactly $K_Q$ terms within the
first three budgets and retains $m-K_Q\le h_0$ terms, which satisfies the fourth budget.

The native design uses no auxiliaries and no penalties and retains all $m$ terms, so it
is feasible exactly when $h_0\ge m$. A full reduction reduces all $m$ terms, which is
possible exactly when $K_Q=m$, and retains none, so the fourth budget imposes no
condition on it.

Feasibility requires $K_Q+h_0\ge m$, exclusion of the native design requires $h_0<m$,
and exclusion of every full reduction requires $K_Q<m$. Under all three conditions any
feasible design reduces at least one term and retains at least one term, so it is
strictly mixed. Conversely, if feasible designs exist and all are strictly mixed,
neither endpoint is feasible and the same three conditions hold.
\end{proof}

\subsection{Pair-Star Closed Form (Theorem 3)}
\label{sup:proofs:star}

\begin{proof}
Suppose $q$ unit-weight terms are assigned to one pair. Their threshold is $q$, the
penalty is $q+\varepsilon$, and the auxiliary degree is $q+2$, so the pair satisfies the
penalty and degree budgets exactly when $q\le\lfloor M_0-\varepsilon\rfloor$ and
$q\le\lfloor\Delta_0\rfloor-2$. A used pair therefore holds at most $C$ terms. If $C=0$
no pair holds a term, and if $r_0=0$ no pair may be used, so $K_Q=0$ in either case.
Assume $C\ge 1$ and $r_0\ge 1$.

Two leaves $\{a_i,b_i,w_t\}$ and $\{a_i,b_i,w_{t'}\}$ of star $i$ share only the core
$\{a_i,b_i\}$, every other pair belongs to one leaf, and stars share no pairs because
their vertex sets are disjoint. The core of star $i$ can therefore reduce at most
$\min\{\ell_i,C\}$ terms and a private pair exactly one.

Consider a valid assignment with $r\le r_0$ active pairs and let $I$ index the stars
whose cores are active. Counting one term per active pair, an active core of star $i$
adds at most $g_i=\min\{\ell_i,C\}-1$ further terms, so $|D|\le r+\sum_{i\in I}g_i$. At
most $\rho_0=\min\{r_0,s\}$ cores are active, all gains are nonnegative, and replacing
the sum by the $\rho_0$ largest gains and $r$ by $r_0$ can only increase the bound. No
more than $m$ terms can be reduced, which gives the upper bound
$|D|\le\min\{m,\,r_0+\sum_{j\le\rho_0}g_{(j)}\}$.

To attain it, activate the $\rho_0$ cores with the largest gains and assign
$\min\{\ell_i,C\}$ leaves to each, which uses $\rho_0$ auxiliaries and reduces
$\rho_0+\sum_{j\le\rho_0}g_{(j)}$ terms within both budgets. If auxiliaries remain,
reduce uncovered leaves through distinct private pairs, one term per auxiliary, until
the budget is used or no leaf remains. The total equals the bound, and the construction
is valid when every gain is zero.
\end{proof}

\subsection{Infinite Mixed-Only Families (Corollaries 1 and 2)}
\label{sup:proofs:families}

\begin{proof}[Proof of Corollary 1]
The objective has $L$ star terms and $H$ isolated terms on disjoint vertex sets.
Assigning all $L$ star terms to the core and retaining the $H$ isolated terms uses one
auxiliary with threshold $L$, penalty $L+\varepsilon$, and degree $L+2$, and retains $H$
terms. Its resource vector equals the budget $(1,L+\varepsilon,L+2,H)$, so the design is
feasible and exact. The native design retains $L+H>H$ terms and fails the retained-term
budget. Any full reduction needs one pair for the star and one pair inside each isolated
term, and disjoint vertex sets prevent sharing, so it needs at least $1+H\ge 2$
auxiliaries and fails the auxiliary budget. The argument holds for all $L,H\ge 1$.
\end{proof}

\begin{proof}[Proof of Corollary 2]
All counts refer to the reference synthesis of Section~IV-A of the main text, whose
gadget order and layer rule determine the depth.

Under $x_i=(I-Z_i)/2$, each term $x_ax_bx_{w_t}$ produces the weight-two strings
$Z_aZ_b$, $Z_aZ_{w_t}$, and $Z_bZ_{w_t}$ and the weight-three string $Z_aZ_bZ_{w_t}$.
Summing over the $L$ leaves collects $Z_aZ_b$ once with coefficient $L/8$, while the
other strings are distinct with coefficients $\pm 1/8$. The native star therefore has
$2L+1$ weight-two and $L$ weight-three strings. With $y$ for $x_ax_b$ and
$M=L+\varepsilon$, the reduced star is $y\sum_tx_{w_t}+M\,\PR(x_a,x_b,y)$, whose strings
are $Z_aZ_b$, $Z_aZ_y$, $Z_bZ_y$, and $Z_{w_t}Z_y$ with coefficients $M/4$, $-M/2$,
$-M/2$, and $1/4$, all nonzero. It has $L+3$ weight-two strings and none of weight three.
A weight-$k$ gadget costs $2(k-1)$ CNOTs \cite{CampbellDahl2022}, so the native star
costs $8L+2$ and the reduced star $2L+6$, and each isolated triple costs $10$. Hence
$G^{\mathrm{native}}=8L+2+10H$ and $G^{\mathrm{mixed}}=2L+6+10H$.

For depth, the pair gadgets of the native star are processed in the order $(a,b)$,
$(a,w_1),\dots,(a,w_L)$, $(b,w_1),\dots,(b,w_L)$, each occupying two layers on its pair.
After all pair gadgets, wire $a$ ends at layer $2L+2$, wire $b$ at $2L+4$, and wire $w_t$
at $2t+4$. The first triple gadget $(a,b,w_1)$ emits the CNOT gates $a\to w_1$, $b\to w_1$, $b\to w_1$, $a\to w_1$.
For $L\ge 2$, these occupy layers $2L+3$, $2L+5$, $2L+6$, $2L+7$, since wire $b$ is occupied through layer $2L+4$. Each further triple adds four consecutive
layers on the core wires, so the native star depth is $(2L+7)+4(L-1)=6L+3$. The reduced
star processes $(a,b)$, $(a,y)$, $(b,y)$, then $(w_t,y)$ for each leaf, finishing at
layers $2$, $4$, $6$, and $6+2t$, so its depth is $2L+6$. An isolated native triple has
depth $10$, and disjoint components combine by maximum. Since $2L+6\ge 10$ for $L\ge 2$,
the native depth is $6L+3$ and the mixed depth is $2L+6$.

The mixed design meets the gate and depth budgets with equality, uses one auxiliary so
that its width equals $n+1$, and has a single penalty $L+\varepsilon$, with a unique
consistent lift by Proposition~2. The native design exceeds the mixed gate count by
$6L-4$ and the mixed depth by $4L-3$, so it violates both budgets. Every full
direct-pair reduction needs at least $H+1$ auxiliaries by the argument of Corollary~1,
so its logical width is at least $n+H+1>n+1$ and it violates the width budget.
\end{proof}

\subsection{Reference Gate Count of Reduced Spin Objectives}
\label{sup:proofs:spincount}

The counts above concern Boolean monomials. A cubic spin interaction $K s_a s_b s_w$ with $s_i = 1 - 2x_i$ is a single weight-three Pauli string, because the quadratic and linear terms of its Boolean expansion cancel in Pauli form, so an isolated interaction costs four CNOT gates natively. For an isolated pair-star with $\ell$ leaves and couplings $K_t \in \{-1,+1\}$, substituting $y = x_a x_b$ in the cubic terms leaves the Boolean quadratic terms in place. Writing $S = \sum_t K_t$ and $M > T$, the nonzero weight-two strings are $\{a,w_t\}$ and $\{b,w_t\}$ with coefficient $K_t$, $\{y,w_t\}$ with coefficient $-2K_t$, $\{a,b\}$ with coefficient $S + M/4$, and $\{a,y\}$ and $\{b,y\}$ with coefficient $-M/2$. These are $3\ell + 3$ strings, so the reduced star costs $6\ell + 6$ CNOT gates against $4\ell$ natively, and reduction of an isolated star never lowers the reference gate count.

The statement does not extend to every pure cubic spin objective, because signed terms that share a core variable can cancel the original-to-leaf strings. Let $E = (s_0 - s_1)(s_2 - s_3)(s_4 - s_5 + s_6 - s_7)$ on eight variables. The native cost layer has 16 weight-three strings and 64 CNOT gates. Reducing every cubic term through the core pairs $\{0,2\}$, $\{0,3\}$, $\{1,2\}$, and $\{1,3\}$ is exact by Proposition~2, with $T_p = 16$ and $M_p = 17$ in unscaled units, or $T_p = 2$ and $M_p = 2.2$ under the scale and penalty rule of Section~S2-C. The 16 original-to-leaf supports each receive two contributions of opposite sign and cancel, the four core strings carry coefficient $M_p/4$, and the remaining 16 auxiliary-to-leaf and 8 core-to-auxiliary strings are nonzero, so the reduced layer has 28 weight-two strings and 56 CNOT gates on 12 qubits at either scale. The example shows that sign structure alone can reverse the isolated-star ordering. It does not establish routed feasibility, and random $\pm 1$ couplings on six variables with five supports rarely produce overlapping cores with the alternating signs that the cancellation requires.

\subsection{Remarks on Higher-Degree Objectives}
\label{sup:proofs:higherdegree}

The composition theorem applies to objectives of any degree, and an additional
correctness audit confirmed its implementation on 24 polynomials of degree four and
five, each with four overlapping high-order monomials and affine terms. Across 91
sequential shared substitutions, exhaustive checks over 42,752 assignments of the
current variables confirmed that $M=\norminf{A}$ preserves values, that
$M=\norminf{A}+1/4$ gives a unique consistent lift, and that $M=\norminf{A}-1/4$
produces a counterexample. The 45,056 assignments of the final composed representations
passed all pointwise and unique-lift checks, and an independent rational-polynomial
compiler agreed with the implementation in all 744 compilations of 248 pair-star
instances. These are truth-table checks on small instances rather than an efficient
implementation for arbitrary degree.

Computing the sharp threshold can itself require exponential Boolean optimisation. For
a residual $A(v)=\sum_Ba_Bv_B$ the bound $M\ge\sum_B|a_B|$ is sufficient and easy to
compute but need not be sharp. The residual $A=uv-uw$ has $\norminf{A}=1$ although its
coefficient sum is $2$, and $A=uvw-uvz$ shows the same gap in a degree-five
substitution. Sharpness is a property of the specified Rosenberg construction and not
of every quadratization method.

The branch-and-bound comparator of Section~IV-D of the main text was also tested with
early stopping on synthetic eight-variable instances with three to eleven cubic terms,
using the reference compiler and node, evaluation, and wall-time limits. Complete
certification was obtained for three and five terms, and evaluation limits were reached
from seven terms onward, with remaining score gaps between $0.24$ and $0.43$. Exhaustive
enumeration confirmed that every reported interval contained the optimum, and
interruptions at zero nodes retained a valid interval with no feasible design. These runs
confirm correct bound accounting and show the rapid growth of the search, and they make
no claim of scalability. Certificates remain relative to the declared design space,
score, compiler, and numerical feasibility rule.

\section{Resource Feasibility under Compilation}
\label{sup:compilation}

This section specifies the compilation protocols of Section~V-B and V-C of the main
text and reports the extended resource results. All quantities are defined in
Section~IV-A of the main text. The instance is the statistical unit throughout, and
budget cells, routing seeds, and target devices that share an instance are not treated
as independent samples.

\subsection{Reference Synthesis and Device Protocol}
\label{sup:compilation:reference}

The reference synthesis places original qubits in index order followed by active
auxiliaries in lexicographic pair order, combines equal Pauli strings, removes only
exact zero coefficients, and keeps the angle $\gamma$ as a free parameter. Strings are
processed in increasing weight and then lexicographically. For a support
$T=\{q_1<\dots<q_k\}$ the gadget is
\begin{equation}
\mathrm{CX}_{q_1,q_k}\cdots\mathrm{CX}_{q_{k-1},q_k}\;R_z^{(q_k)}(2\gamma h_T)\;
\mathrm{CX}_{q_{k-1},q_k}\cdots\mathrm{CX}_{q_1,q_k},
\label{eq:S:gadget}
\end{equation}
which implements $e^{-i\gamma h_TZ_T}$ with $2(k-1)$ CNOT gates
\cite{CampbellDahl2022,Verchere2023}. With $p_{k,\sigma}$ the number of nonzero strings
of weight $k$, the cubic family has reference cost $G^{\mathrm{ref}}_{2q,\sigma}=2p_{2,\sigma}+4p_{3,\sigma}$
and no SWAP gates. Two-qubit depth is computed by scanning the emitted CNOT sequence
$(a_j,b_j)$ in order with all wire layers initialised to zero, placing gate $j$ at layer
$\lambda_j=1+\max\{d_{a_j},d_{b_j}\}$, and updating both wires to $\lambda_j$. The depth
is the largest final wire layer. This schedule preserves gate order on each wire and is
the earliest-layer schedule for the fixed sequence. It counts no single-qubit
durations and performs no cross-gadget cancellation, so the reference cost is a
property of this synthesis and not a lower bound.

A device-level protocol $\Pi$ fixes the circuit construction rules, the circuit stage
(one cost layer or the complete circuit), the coupling map, basis gates, compiler
version, layout and routing passes, seed bundle, and aggregation rule
\cite{Li2019,Zulehner2019}. The physical width $Q^\Pi_\sigma$ counts occupied input
sites and every site touched by compiled operations, so $Q^\Pi_\sigma\ge N_\sigma$ with
equality for the reference synthesis. A logical width failure therefore excludes
physical feasibility, whereas passing the logical check does not guarantee that the
routed circuit fits.

\subsection{Candidate-Set Construction and Per-Study Filters}
\label{sup:compilation:candidates}

A candidate is an action vector that keeps each cubic term native or assigns it to one
of its three pairs, and terms assigned to the same pair share an auxiliary. The
experiment's penalty rule converts each action vector into its exact representation. The
candidate sets and their proposal seed are fixed before any QAOA outcome is known.
Algorithm~\ref{alg:S:select} states the common evaluation and selection procedure. A compiled record is reused only within the same protocol context, and reuse does not add candidate budget.
Three outcomes carry the label mixed-only and are kept distinct, as in Section~IV-D of the main text. A candidate-set mixed-only cell contains a feasible strictly mixed candidate and no feasible endpoint within the evaluated set. A certified mixed-only cell additionally excludes every fully reduced assignment of the direct-pair class under the fixed penalty rule by a complete check of logical width and penalty. A joint mixed-only success requires the selected mixed circuit to meet both the resource budgets and the quality target while neither compared endpoint meets both.

\begin{algorithm}[t]
\caption{Evaluation and selection within a fixed candidate set}
\label{alg:S:select}
\begin{algorithmic}[1]
\Require Candidate set $\mathcal{B}$, protocol $\Pi$, score weights and scales, optional budget tuple $b$, tie key $k$.
\Ensure Best eligible candidate, or a report that none was found in $\mathcal{B}$.
\State $E\gets\emptyset$
\ForAll{distinct action vectors $\alpha\in\mathcal{B}$}
  \State Build the exact representation $F_\alpha$ with the fixed penalty rule
  \State Record $a_\alpha$, $N_\alpha=n+a_\alpha$, and $M_{\max,\alpha}$
  \State Obtain or reuse the compiled record for $(\alpha,\Pi)$ and record $Q^\Pi_\alpha$, $G_{2q,\alpha}$, $D_{2q,\alpha}$
  \State Compute $J_\alpha$ from \eqref{eq:S:score}
  \If{no budget filter is requested, or the record satisfies every bound in $b$}
    \State $E\gets E\cup\{\alpha\}$
  \EndIf
\EndFor
\If{$E=\emptyset$} \State \Return none found in $\mathcal{B}$
\Else \State \Return $\argmin_{\alpha\in E}(J_\alpha,k(\alpha))$ in lexicographic order
\EndIf
\end{algorithmic}
\end{algorithm}

All current experiments rank candidates by
\begin{equation}
J=\frac{1}{4}\Bigl(\frac{a}{4}+\frac{G}{160}+\frac{D}{120}+\frac{M}{8}\Bigr),
\label{eq:S:score}
\end{equation}
where $a$ is the auxiliary count, $G$ the CX count, $D$ the CX-only depth, and $M$ the
largest penalty. The physical width is tested against the width budget and does not
replace $a$ in the score. Because the circuit stage changes $G$, $D$, and $Q$, scores from different experiments are not pooled.

\emph{Cost-layer compilation experiment.} The full domain assigns every cubic term to a
pair, and the unrestricted domain also permits native actions, so it contains both
endpoints. Designs whose logical width exceeds the 12-site device are removed before
proposal. Each candidate set has $\min\{32,\text{number of capacity-feasible full
designs}\}$ members, namely 32 for the random families and 27 for the constructed
family. Both sets contain the minimum-auxiliary, maximum-reuse, and distance-aware full
assignments, the unrestricted set also contains the native design, shared-pair groups,
and prefixes of a reuse ordering, and seeded permutations fill the remaining places.
The minimum-auxiliary rule tests the effect of minimising auxiliaries alone, and the maximum-reuse rule tests a simple pair-sharing heuristic. The resource-optimised full rule tests the value of allowing terms to remain native under an equal candidate budget, and the distance-aware rule selects the shortest original-bit pair on the target and resolves ties by reuse. The distance rule is a simple hardware-inspired comparator
and not a reproduction of a named algorithm. Every set is evaluated on each target
and synthesis, the score winner uses lexicographic action-vector ties, and the budget
maps test every compiled member separately at each $(Q,G,D,M)$ tuple. All 35 full
assignments of a random instance are enumerated for logical width and penalty, and all
27 capacity-feasible full assignments of a constructed instance are compiled.

\emph{Joint resource and quality experiment.} The full and strictly mixed domains use equal
budgets on the complete depth-one circuit, and selection uses budget filtering followed
by the score. The optional diagnostic comparison of Section~\ref{sup:diagnostic} uses
the same strictly mixed set.

\emph{Routing-seed follow-up.} The saved full and unrestricted sets are reused without
new proposals. Mixed-only classification tests every saved candidate at each budget
tuple and seed, and a separate winner-stability statistic uses the lowest score within
device capacity without the hard budgets.

Exact penalties certify the pointwise objective equality of each candidate. They do
not certify that a finite set contains the best compiled design, and a missing
feasible candidate means only that none was found in that set. Since
$Q^\Pi_\sigma\ge N_\sigma$, a design with $N_\sigma>Q_{\max}$ cannot fit, so when every
full assignment fails logical width or the penalty bound, all full assignments are
excluded within the direct-pair class. A full assignment that passes the logical
checks may still fail after routing.

\subsection{Joint-Budget Grid and Certification Rule}
\label{sup:compilation:grid}

Code, generators, candidate counts, score weights, compiler settings, and the budget
grid were frozen locally before data generation. The 30 Max-3SAT and 30 cubic
spin-glass instances have six variables and five distinct cubic supports, with signs
sampled independently and spin objectives divided by eight after conversion to bits.
The 30 constructed instances have nine variables, a four-leaf pair-star, one isolated
cubic term, random vertex labels, and weights drawn independently from
$\{0.75,1,1.25\}$. Because this family has only 27 full assignments that fit the
device, a technical amendment recorded before any compiler result set both budgets to 27 for that family. All representations use $M_p=1.1\,T_p$. Qiskit 2.5.2 \cite{Qiskit2024}
compiles to $\{\mathrm{CX},R_Z,\sqrt{X},X\}$ with optimisation level 3, SABRE routing
\cite{Li2019}, identity initial placement, and seed 1729, on line-12, ring-12, and
$3\times 4$ grid-12 targets of equal capacity. The canonical synthesis orders gadgets
by size and then lexicographically, and the reuse synthesis groups commuting gadgets by
a frequent root and orders their parity supports greedily. No gadget barriers are
inserted, so the compiler may optimise across gadgets. Resources cover one cost layer
and exclude preparation and mixers. The experiment used $31{,}577$ compiler calls and
$32{,}760$ scheduled candidate evaluations, with independent checks of the logical
phases and routed-circuit semantics.

\begin{figure}[t]
\centering
\includegraphics[width=0.95\textwidth]{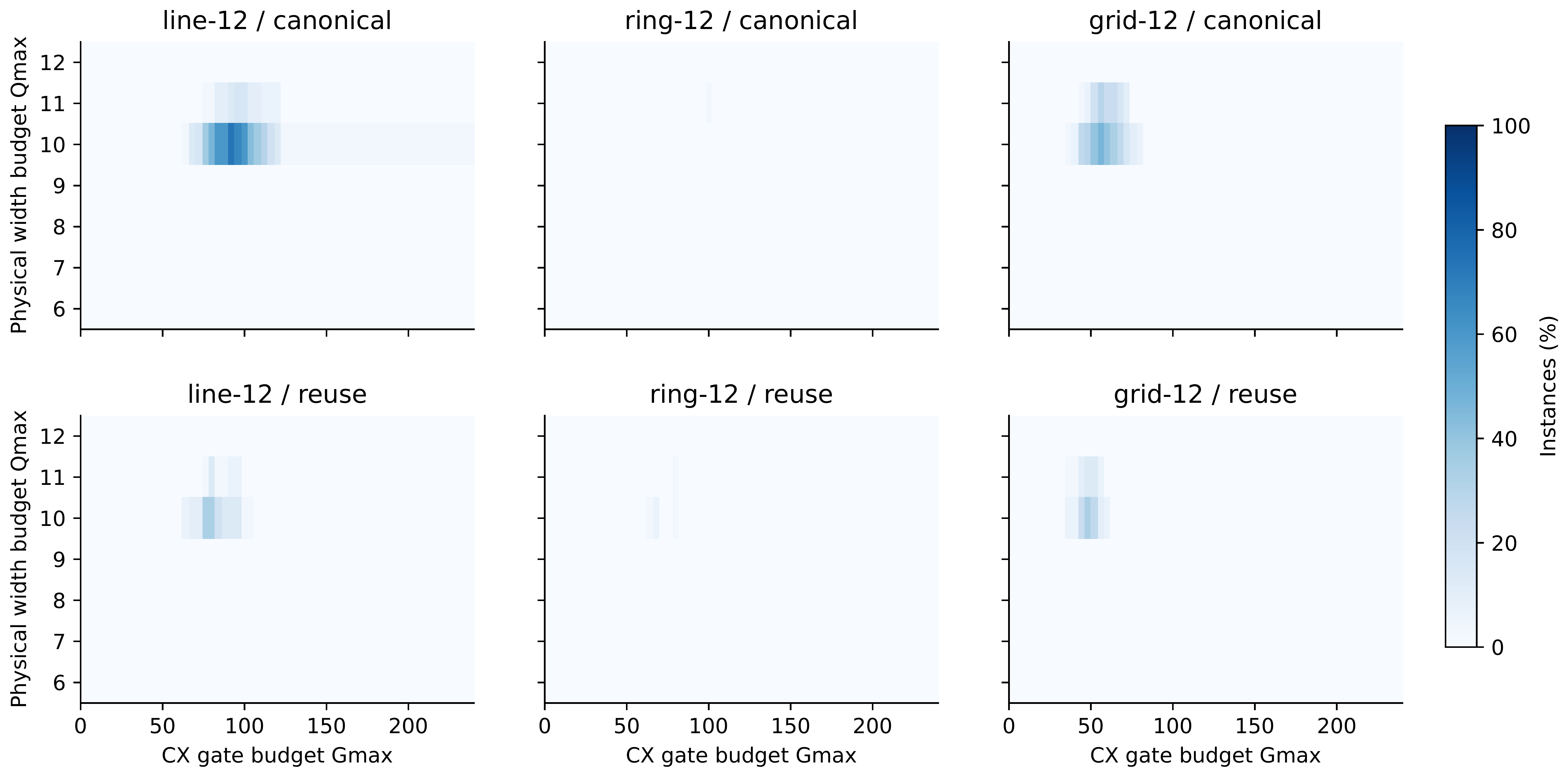}
\caption{Fraction of the 30 constructed instances with a certified mixed-only cell, on
the slice $B_D=96$ and $B_M=4.4$, for the three targets under canonical (top) and reuse
(bottom) synthesis.}
\label{fig:S:certified}
\end{figure}

The frozen grid is $Q_{\max}\in\{6,\dots,12\}$, $B_G\in\{0,4,\dots,240\}$,
$B_D\in\{24,48,72,96,120,160,240\}$, and $B_M\in\{0,1.1,2.2,3.3,4.4,6.6,8.8\}$, giving
$20{,}923$ cells. Each cell is labelled by the classes with a feasible candidate. A
cell is certified as mixed-only when a compiled strictly mixed candidate fits, the
native design fails the gate or depth budget while fitting the width, and a complete check of logical width and penalty excludes every full assignment.
Fig.~\ref{fig:S:certified} shows the incidence of certified cells across the 30 constructed instances on a fixed depth and penalty slice.

\subsection{Synthesis Sensitivity, Pareto Overlap, and Matched Controls}
\label{sup:compilation:sensitivity}

The random control matches the auxiliary count and reduced-term count of the best
strictly mixed candidate and, where possible, its largest penalty. A distinct
penalty-matched control exists in 495 of 540 instance, target, and synthesis
settings, and in 15 settings no distinct control matches even the two counts.
Table~\ref{tab:S:grid_scores} reports median scores of every retained comparator on
grid-12, namely the native design, the minimum-auxiliary (MinA), maximum-reuse
(Reuse), distance-aware (Dist.), and resource-optimised (ResFull) full reductions, the
best strictly mixed candidate (Mixed), and the matched random control (Random).

\begin{table}[t]
\caption{Median compiled resource score \eqref{eq:S:score} on grid-12 over 30 instances per row. Lower is better.}
\label{tab:S:grid_scores}
\centering
\begin{tabular}{@{}llccccccc@{}}
\toprule
Family & Synthesis & Native & MinA & Reuse & Dist. & ResFull & Mixed & Random \\
\midrule
Max-3SAT & canonical & 0.176 & 0.434 & 0.455 & 0.509 & 0.415 & 0.271 & 0.292 \\
Max-3SAT & reuse & 0.140 & 0.446 & 0.471 & 0.525 & 0.431 & 0.238 & 0.259 \\
Cubic spin & canonical & 0.107 & 0.448 & 0.485 & 0.520 & 0.445 & 0.237 & 0.248 \\
Cubic spin & reuse & 0.096 & 0.472 & 0.513 & 0.546 & 0.470 & 0.225 & 0.235 \\
Constructed & canonical & 0.206 & 0.395 & 0.395 & 0.396 & 0.387 & 0.296 & 0.305 \\
Constructed & reuse & 0.160 & 0.393 & 0.393 & 0.395 & 0.389 & 0.264 & 0.268 \\
\bottomrule
\end{tabular}
\end{table}

The mean paired score difference between the best strictly mixed and the resource-optimised full candidate on grid-12 uses percentile bootstrap intervals from $10{,}000$ instance resamples. It is $-0.139$ $[-0.152,-0.126]$ for Max-3SAT under
canonical synthesis and $-0.189$ $[-0.202,-0.176]$ under reuse, $-0.210$ $[-0.223,-0.196]$ and $-0.239$ $[-0.253,-0.226]$ for cubic spin, and $-0.095$ $[-0.103,-0.086]$ and $-0.127$ $[-0.136,-0.119]$ for the constructed family. These intervals are unadjusted, compare the best designs found within each evaluation budget, and imply nothing about solution quality.

Changing the synthesis strategy changes rankings. On the same candidate identities,
the median Spearman correlation between canonical and reuse scores on line, ring, and
grid is $0.913$, $0.915$, and $0.949$ for Max-3SAT, $0.906$, $0.905$, and $0.955$ for
cubic spin, and $0.771$, $0.729$, and $0.812$ for the constructed family. A candidate
is Pareto optimal within a set when no other member is at least as good in every
resource of $(Q,G,D,M)$ and better in one. The median Jaccard similarity of the two
Pareto sets is $0.333$, $0.250$, and $0.333$ for Max-3SAT, $1$ on all targets for cubic
spin, and $0.321$, $0.351$, and $0.355$ for the constructed family. The winning full
assignment changes on 6 to 14 of 30 Max-3SAT, 10 to 13 of 30 spin, and 0 to 4 of 30
constructed instances depending on the target. The Pareto overlaps depend on the
candidate sets, which can differ because the distance-aware rule proposes different
designs, so they measure stability within the sampled sets and do not isolate
connectivity.

\subsection{Routing-Seed Follow-Up}
\label{sup:compilation:seeds}

The follow-up reuses the 30 Max-3SAT and 30 constructed instances and their saved
candidate sets, and it adds no candidates after the seed results are examined. The
$20{,}292$ candidate, target, and synthesis rows, comprising $11{,}330$ full, $8{,}602$
strictly mixed, and 360 native rows, are compiled at seeds $\{1729, 2718, 31415, 57721,
65537\}$, giving $101{,}460$ tasks. The compiler settings are those of
Section~\ref{sup:compilation:grid} with an explicit SABRE pass using the decay
heuristic, eight trials, and no placement search, so the five seeds share the same
routing settings. The archived cost-layer run used implicit routing settings and is
not the control; the new seed-1729 run supplies it. The budget grid is unchanged. A
point is TRUE when at least one exact strictly mixed candidate fits and no native or
full candidate in the set fits, FALSE when complete records exclude that condition,
and UNKNOWN when evidence is missing. All tasks completed and no point is UNKNOWN. For
each instance, target, and synthesis, retention is the size of the intersection of the
five TRUE sets divided by the size of the seed-1729 set, and pooled retention divides
the summed counts.

\begin{table}[t]
\caption{Mixed-only budget points within the saved candidate sets, pooled over both synthesis strategies and 30 instances per family.}
\label{tab:S:retention}
\centering
\begin{tabular}{@{}llrrr@{}}
\toprule
Family & Target & Seed 1729 & All five seeds & Retained \\
\midrule
Max-3SAT & Line-12 & 23,734 & 13,294 & 56.0\% \\
Max-3SAT & Ring-12 & 24,083 & 14,677 & 60.9\% \\
Max-3SAT & Grid-12 & 30,024 & 13,228 & 44.1\% \\
Constructed & Line-12 & 22,110 & 16,412 & 74.2\% \\
Constructed & Ring-12 & 7,946 & 6,683 & 84.1\% \\
Constructed & Grid-12 & 41,542 & 14,775 & 35.6\% \\
\midrule
Pooled & & 149,439 & 79,069 & 52.9\% \\
\bottomrule
\end{tabular}
\end{table}

\begin{figure}[t]
\centering
\includegraphics[width=0.85\textwidth]{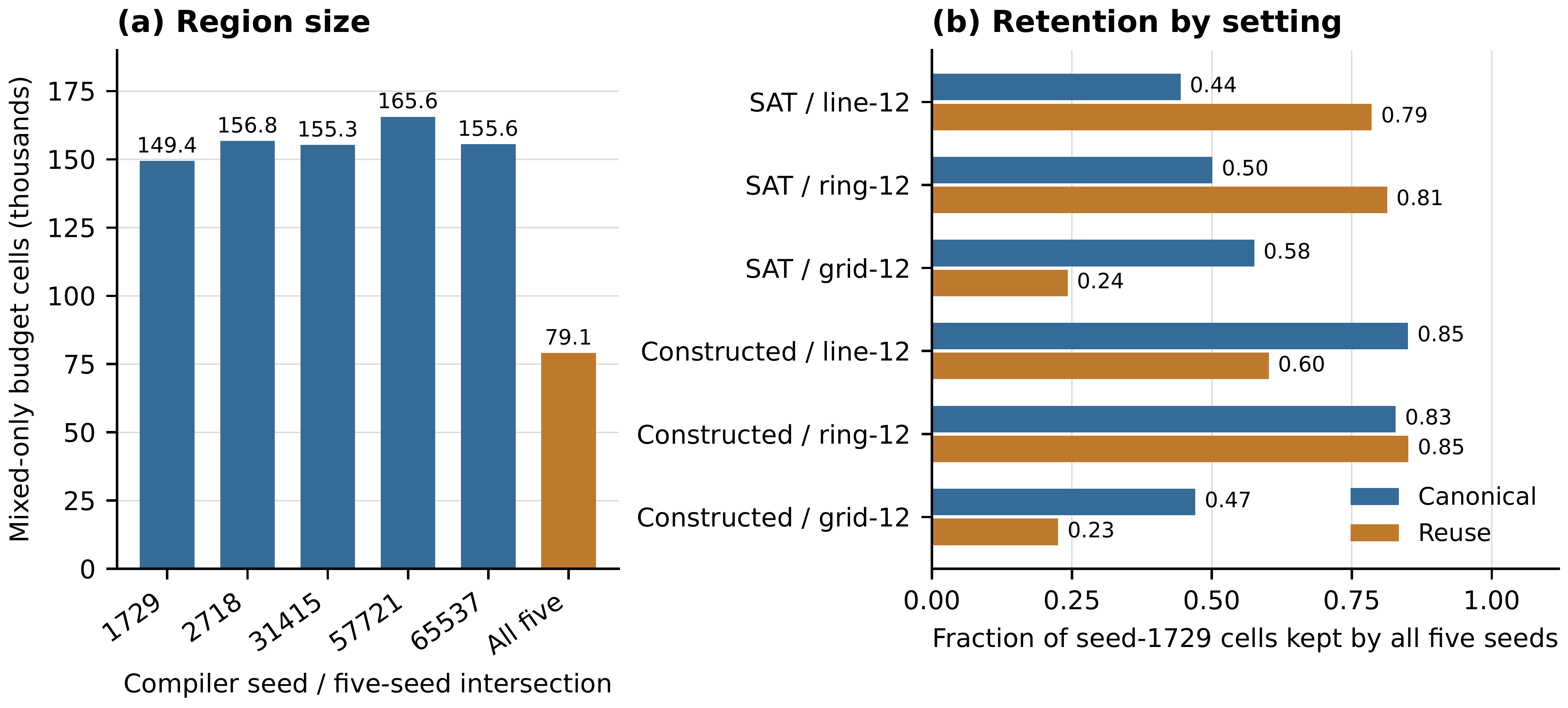}
\caption{Five-seed check within the saved candidate sets. (a) Mixed-only budget points
per seed and their intersection. (b) Fraction of seed-1729 points retained by all five
seeds by family, target, and synthesis. Each bar pools 30 instances.}
\label{fig:S:seeds}
\end{figure}

Table~\ref{tab:S:retention} and Fig.~\ref{fig:S:seeds} report retention. The total
number of TRUE points per seed is $149{,}439$, $156{,}824$, $155{,}289$, $165{,}626$, and
$155{,}638$, and the number of settings with at least one TRUE point is 255, 255, 262,
254, and 256 of 360. Similar totals do not show that the regions contain the same
points, because $70{,}370$ baseline points fail in at least one other seed while new
points appear elsewhere. Of the 360 settings, 15 have identical nonempty regions across
all seeds, 81 are identically empty, 35 have a nonempty baseline region with no common
point, and 229 show other changes including newly appearing regions. Across the
$1{,}440$ paired comparisons of each other seed with seed 1729, the median Spearman
correlation of candidate scores is $0.970$ with minimum $0.765$, and the median Jaccard
similarity of Pareto sets is $0.667$ with minimum $0.125$. The lowest-score candidate
within capacity changes in only 8 comparisons, but this selector chooses native in 353
of 360 settings, so the low rate describes native choices. 
Two further checks fix the design rather than the region. Keeping the seed-1729 score winner within capacity fixed and testing its other-seed circuits at its baseline-feasible points, the winner loses at least one feasible point in 320 of 1,440 comparisons, although this selector chooses native in 353 of 360 settings, so the statistic mainly describes native choices. At each baseline mixed-only cell, the feasible strictly mixed candidate with the lowest score, with ties broken by candidate identifier, is therefore selected and held fixed together with the budget. It remains feasible at all five seeds in 100,967 of the 149,439 cells (67.56\%), and if each seed may choose another mixed candidate, 120,838 cells (80.86\%) retain mixed feasibility. Mixed feasibility does not imply mixed-only status, since an endpoint may also become feasible, and the cells share instances, so these are descriptive fractions rather than independent trials.

\section{Finite-Depth QAOA Studies}
\label{sup:qaoa}

This section specifies the protocols of Sections~V-D and~V-E of the main text and reports the secondary optimisation results. The two QAOA experiments of this section use the unscaled spin objective $E(s) = \sum_{\{i,j,k\}} K_{ijk} s_i s_j s_k$ with $K_{ijk} \in \{-1,+1\}$, whose six-interaction instances take values in $[-6, 6]$, together with integer Max-3SAT clause weights, so all reported objectives and gains are in unscaled integer units. The cost-layer compilation experiment of Section~S2 instead divides the spin objective by eight, and penalty bounds at the two scales are not compared. The warm state, the matched mixer, the original-objective expectation $E_f$, and the auxiliary inconsistency $R_{\mathrm{inc}}$ are defined in Section~IV-E of the main text. Two circuit identities underlie the implementation.
Each mixer term factorises into single-qubit gates as $e^{+i\beta h_{\mathrm{M}}(\omega)}=R_y(\theta_\omega)R_z(2\beta)R_y(-\theta_\omega)$
with $\theta_\omega=2\arcsin\sqrt{\omega}$ \cite{Egger2021,TateCustom2023}, and each Rosenberg penalty contributes commuting one-qubit and two-qubit phases,
\begin{equation}
e^{-i\gamma M_p\PR(N_i,N_j,N_{y_p})}=
e^{-i\gamma M_pN_iN_j}\,e^{2i\gamma M_pN_iN_{y_p}}\,e^{2i\gamma M_pN_jN_{y_p}}\,e^{-3i\gamma M_pN_{y_p}},
\label{eq:S:penaltyphase}
\end{equation}
so that reducing a cubic term replaces a three-local phase by an auxiliary qubit with
lower-local penalty phases \cite{Schmidbauer2024}. The product warm state is the unique
ground state of the matched mixer with energy $-N_\sigma$ and spectral gap 2. When the
physical evolution is written as $\exp(-i\beta_{\mathrm{phys}}H_{\mathrm{M}})$, the
implementation sets $\beta_{\mathrm{phys}}=-\beta$ at every objective call so that the
warm and cold starts share optimiser coordinates. The null control requires uniform
probability on every qubit including the auxiliaries. The diagonal cost layer
preserves the consistency probability $1-R_{\mathrm{inc}}$, whereas the local mixer
can change it. Two alternative auxiliary initialisations are available for ablation but are not used in the reported experiments. The first is deterministic rounding $\hat{x}_i=\mathbf{1}\{\mu_i\ge 1/2\}$ with $\hat{y}_p=\hat{x}_i\hat{x}_j$, and the second computes each product coherently into its auxiliary by a circuit
$U_\rho|x\rangle|0\rangle^{\otimes|P_\rho|}=|x\rangle|\phi_\rho(x)\rangle$, which requires additional entangling gates and does not prevent a later local mixer from breaking the relations.

\subsection{Preparation and Evolution Protocol}
\label{sup:qaoa:protocol}

The code, seeds, hypotheses, and evaluation budgets were frozen locally before data
generation. Each family has 30 new independent instances with six variables and six
clauses or spin interactions, sampled uniformly subject to variable coverage and at least two nonzero cubic monomials, with Max-3SAT clause weights drawn from $\{1,\dots,5\}$. The experiment is noiseless and restricted to depth one.
Native, strictly mixed with $0<|D|<|\mathcal{C}_3(f)|$, and full representations are
compared. Mixed and full selection use the reference-compiler score
$J=(a/4+G/160+D/120+M/8)/4$ over up to 128 distinct seeded candidates, or the complete
domain when it is smaller, with penalties equal to the threshold plus one. The
consistency diagnostic is not used. When the native design is accepted it has the
lowest score on all 60 instances, so the mixed comparison is conditional on requiring a
strictly mixed design. Selection uses the all-to-all reference compiler and is separate from the topology-specific compilation experiment. The two experiments therefore do not establish a joint hardware and quality threshold on the same instances.

Every equal-effort arm uses the same three parameter starts per instance, drawn
uniformly on $[0,2\pi)\times[0,\pi)$, and exactly 120 objective evaluations per start.
COBYLA \cite{Powell1994} runs first, and if it stops early the remaining budget is filled
with uniform parameter samples from a fixed seed. The reported value for an instance,
representation, and warming rule is the lowest expectation over all 360 evaluations,
not the mean of the three restart optima. Gains are computed per instance and then
averaged, bootstrap resampling is over instances, and the initial state is not added as
an optimiser candidate, so a positive evolution gain is not guaranteed by construction.
All representations use the same pair and triple McCormick and
reformulation-linearisation relaxation \cite{SheraliAdams1990,McCormick1976}.
Original-bit probabilities are bounded to $[0.05, 0.95]$. Pair warming bounds the pair moments to the same interval, and independence warming multiplies the bounded original probabilities and bounds the product, which is the rule stated in Section~IV-E of the main text. Every representation has the same original-bit
distribution at warm depth zero, so the original expectations coincide there. Of
$172{,}800$ equal-effort evaluations, $17{,}959$ are budget-completion samples, and all
traces, early-stop messages, and evaluation counts are retained.

\subsection{Optimiser Comparisons and Convergence Behaviour}
\label{sup:qaoa:optimisers}

Two additional optimisation protocols test the dependence on effort and method. The
scan protocol evaluates a $65\times 65$ grid over the full $(\gamma,\beta)$ domain and
starts four Nelder--Mead refinements from the four lowest grid points, each with a
240-evaluation budget completed by the same uniform rule, for $5{,}185$ evaluations per
arm. The reported value is the lowest expectation over the four refinement traces, each
of which includes its grid start. Because the clause weights and couplings are integers and $M_p = T_p + 1$, every encoded energy is an integer, so the $2\pi$ period of the cost angle is exact at the QAOA scale, and the matched mixer has period $\pi$ up to a global phase. The second equal-effort protocol replaces COBYLA by
Nelder--Mead with the same three starts and $3\times 120$ evaluations and uses no
completion samples.

\begin{figure}[t]
\centering
\includegraphics[width=0.85\textwidth]{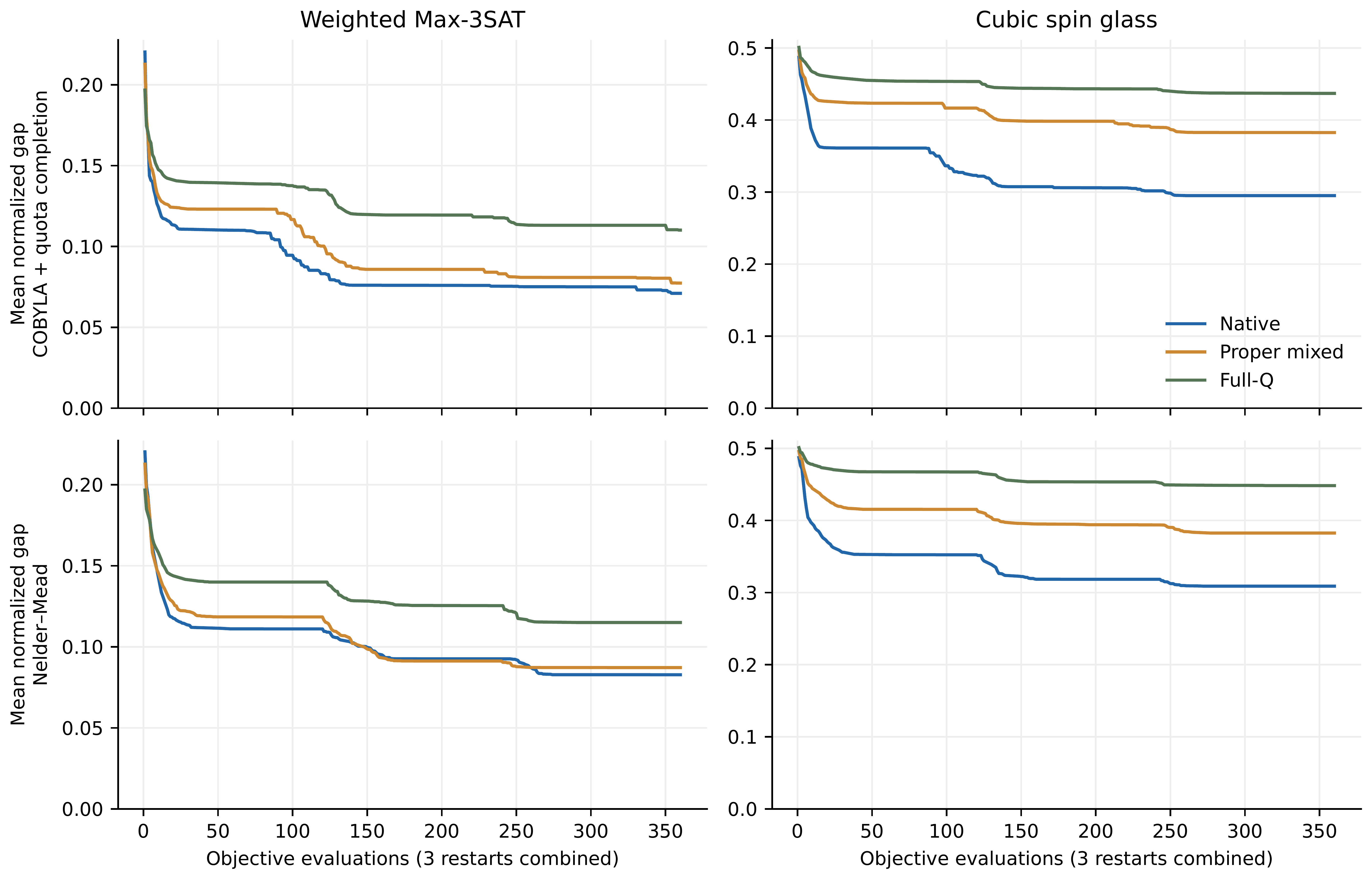}
\caption{Mean incumbent gap, normalised by the exact objective range, against
cumulative evaluations over three restarts for COBYLA with budget completion (top) and
Nelder--Mead (bottom), averaged over 30 instances per family with pair warming. Full-Q
denotes the fully reduced representation.}
\label{fig:S:convergence}
\end{figure}

For Max-3SAT, moving from COBYLA to the scan changes the best representation on 7 of
30 instances with a mean within-instance rank correlation of $0.717$, and replacing
COBYLA by Nelder--Mead changes it on 9 of 30. For cubic spin the corresponding counts
are 1 of 30 with rank correlation $0.917$ and 2 of 30. One optimiser at one budget
therefore does not fix the ranking for every instance, although the mean ordering is
stable. Fig.~\ref{fig:S:convergence} shows the incumbent normalised gap against
cumulative evaluations. Results for standard QAOA started from a classical bit string
\cite{Cain2023} do not apply directly to the matched product-state mixer used here.

A post-hoc paired comparison of the equal-effort outcomes defines $\Delta = E_{\mathrm{other}} - E_{\mathrm{native}}$, so that positive values favour native. For Max-3SAT the strictly mixed representation has mean difference 0.0659 with interval $[-0.0083, 0.1379]$ and the full representation 0.3741 with $[0.2388, 0.5074]$. For cubic spin the differences are 0.9115 with $[0.7414, 1.0866]$ and 1.4851 with $[1.3106, 1.6453]$. Intervals use 10,000 paired instance-bootstrap resamples with seed 3619, and the mixed representation beats native on 7 Max-3SAT and 1 spin instance.

\subsection{Classical Controls for the Preparation Gain}
\label{sup:qaoa:controls}

Three classical quantities accompany the preparation gain, namely deterministic rounding of the raw relaxation marginals with ties rounded to zero, the mean of $2{,}048$
independent samples from the bounded marginals, and the warm depth-zero expectation, which equals the sampling mean. For Max-3SAT the mean original objectives of rounding,
sampling, and the prepared state are $2.267$, $1.786$, and $1.783$. For cubic spin they
are $-1.467$, $0.010$, and $0.000$, because the relaxation returns uniform marginals
for this family. These controls measure the preparation and are not a comparison with
an optimised classical solver.

Pair warming is compared with independence warming at the fixed equal-effort budget.
For Max-3SAT, pair warming lowers the mean objective by $0.046$ $[-0.005,0.112]$ for
strictly mixed and $0.119$ $[0.040,0.202]$ for full representations. For cubic spin the
changes are $-0.051$ $[-0.175,0.061]$ and $0.112$ $[-0.011,0.238]$. Three of the four
intervals include zero, so these data do not establish a general benefit of pair
moments.

\subsection{Joint Resource and Quality Protocol}
\label{sup:qaoa:joint}

The joint experiment saves its protocol, source hashes, and dependency versions before
generating instances. The 30 Max-3SAT and 30 spin objectives have six variables and are
generated as in Section~\ref{sup:qaoa:protocol}. The 12 mechanism objectives have nine
variables, a four-leaf pair-star, and one disjoint triple, with uniformly permuted
labels, cubic weights from $\{1,2,3\}$, and affine coefficients from
$\{-2,-1,0,1,2\}$. Integer energies justify the $2\pi$ cost-angle period. Penalties are
the threshold plus one and bounding uses $\delta=0.05$.

The complete depth-one circuit, consisting of preparation, cost evolution, and matched
mixer, is compiled on ring-12 with reuse synthesis, Qiskit 2.5.2 at optimisation level
3, SABRE routing, identity initial layout, seed 1729, and basis
$\{R_Z,\sqrt{X},X,\mathrm{CX}\}$. Width counts occupied input sites and all sites touched
by routing, which differs from the cost-layer scope of Section~\ref{sup:compilation}.
The full and strictly mixed candidate sets use the same budget, namely the smaller of 32 and the capacity-feasible design space, with 27 for the mechanism cohort. Full
candidates include assignments chosen for minimum auxiliary count, maximum reuse, and
short ring distances, and mixed candidates include shared-pair groups and prefixes of a
reuse ordering. For each budget the two searches independently select the lowest score
with lexicographic ties, and QAOA outcomes are never used to propose or select designs.

\begin{figure}[t]
\centering
\includegraphics[width=0.85\textwidth]{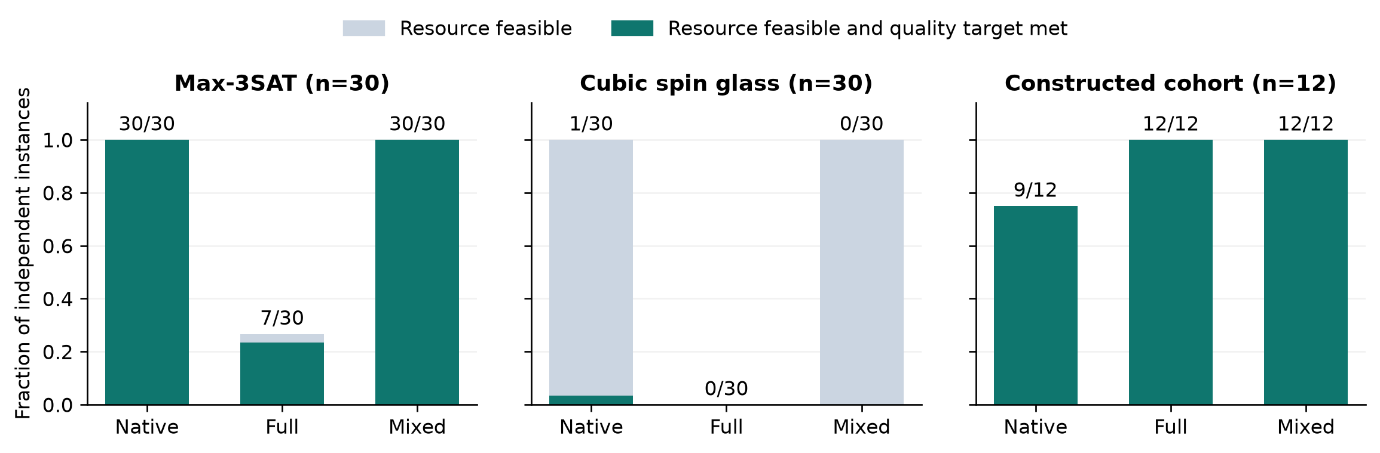}
\caption{Primary joint test at budget $(12,96,64,16)$ and $g\le 0.2$. Light bars are
resource-feasible instances and dark bars are instances that also meet the quality
target, for the three cohorts.}
\label{fig:S:joint}
\end{figure}

Every evaluated design receives three common starts with 120 Nelder--Mead evaluations
each and budget completion. Before any quality result is observed, a prespecified set of
designs is chosen for the larger scan, comprising the native, full, and three
diagnostic-constrained choices under capacity limits alone together with the full and
mixed choices at the primary budget. Each distinct design receives the $65^2$ grid and
four refinements of 240 evaluations, and designs shared between choices are evaluated
once. The primary test at budget $(12,96,64,16)$ uses scan-refined results. All
secondary grid counts use equal-effort results, including for designs that also have
scan-refined values. The secondary grid has $Q_{\max}\in\{8,10,12\}$,
$B_G\in\{64,96,128,160,192,256\}$, $B_D\in\{32,64,96,128\}$, and
$B_M\in\{2,4,8,16,32,64\}$, giving 432 cells, each evaluated at $g\le 0.1$, $0.2$, and
$0.3$. Cells are correlated sensitivity checks. A failed selected endpoint is distinguished from class-wide exclusion of all full assignments, which uses a complete check of logical width and penalty under the same direct-pair class and penalty rule.
Classical controls are the warm depth-zero expectation, deterministic rounding of the
raw marginals, and the mean and best of $2{,}048$ independent samples.
Fig.~\ref{fig:S:joint} shows the primary outcome by cohort. No primary instance has both endpoints excluded by resources, because native fits every random instance and full fits every mechanism instance, so the zero exclusive count combines endpoint success with mixed quality failure on the spin cohort.

\subsection{Uncertainty of the Zero Primary Count}
\label{sup:qaoa:uncertainty}

Zero observed successes do not establish a zero population rate. For a binomial
proportion with $k=0$ successes in $N$ trials, the one-sided 95\% upper confidence
limit is $1-0.05^{1/N}$, which equals $9.50\%$ for $N=30$ and $22.09\%$ for $N=12$.
Bootstrap resampling of an all-zero count returns a degenerate interval and does not
resolve this uncertainty. The primary result is therefore reported as an observed zero
together with these limits rather than as evidence of impossibility.

\section{Consistency Diagnostic}
\label{sup:diagnostic}

The consistency excess $\Efib(\sigma;\eta,\xi)$ of Eq.~(18) in the main text is the
expected encoded-energy excess of a representation under an independent product law
with original-bit probabilities $\eta$ and auxiliary probabilities $\xi$. It is
nonnegative by pointwise exactness. Raw relaxation moments and bounded preparation
probabilities define different product laws, and the definition makes no claim that
raw moments arise from a joint distribution. The diagnostic is optional. It does not
enter the primary selection of any experiment in the main text, and this section reports
its normalisation, its predictive value, and its effect on selection.

\subsection{Normalisation and Prepared-State Form}
\label{sup:diagnostic:form}

For a pair $p=\{i,j\}$, denote $t_p=\eta_i\eta_j$. The mismatch probability and the
expected Rosenberg penalty are
\begin{equation}
\nu_p=\xi_p(1-t_p)+(1-\xi_p)t_p,\qquad
\ell_p=t_p+\xi_p(3-2\eta_i-2\eta_j),
\label{eq:S:mismatch}
\end{equation}
where the expression separates the events $Y_p=1$ with $X_iX_j=0$ and $Y_p=0$ with $X_iX_j=1$,
and the second follows by taking expectations in the penalty of Eq.~(2) of the main
text. When $\xi_p=t_p$, the interaction correction in the closed form is zero. For the
fixed-additive-margin protocols a design-independent scale over nonempty cubic support
is
\begin{equation}
s_{\mathrm{fib}}=4\sum_{p\in\partial_2\mathcal{C}_3(f)}\Bigl(\sum_{A\supset p,\,A\in\mathcal{C}_3(f)}|c_A|+\varepsilon_p\Bigr),
\label{eq:S:scale}
\end{equation}
with $s_{\mathrm{fib}}=1$ when there are no cubic terms, in which case the unique
design is native and the excess is zero. The normalised diagnostic is
$\hat{E}_{\mathrm{fib}}=\Efib/s_{\mathrm{fib}}$. Two auxiliary bounds follow. The union
bound $\Lambda_\sigma=\min\{1,\sum_{p\in P_\rho}\nu_p\}$ bounds the probability that
any auxiliary is inconsistent, and it holds when mismatch events share original
variables. The ratio $R_{\mathrm{fib}}(\sigma)=\Efib(\sigma)/(4\sum_{p\in P_\rho}M_p)$
lies in $[0,1]$, because each pair's excess is nonnegative and at most
$\norminf{A_p}+3M_p\le 4M_p$. Both bounds refer to the initial product distribution and
assume nothing about independence after QAOA evolution.

\subsection{Cross-Validated Prediction of Auxiliary Inconsistency}
\label{sup:diagnostic:prediction}

The joint experiment of Section~\ref{sup:qaoa:joint} tests whether the diagnostic carries
predictive information beyond auxiliary count, compiled score, reduced-term count,
warm depth-zero gap, and family. Before any outcome is observed, a fixed seed selects
eight strictly mixed candidates from each of the 60 random instances, giving 480
probes. For each instance a linear model is fitted on the probes of all other instances
and used to predict the held-out probes, with the raw and prepared diagnostics added
separately and together. Intervals resample the held-out errors by instance without
refitting and are descriptive and unadjusted. The 12 mechanism instances are analysed
separately.

\begin{figure}[t]
\centering
\includegraphics[width=0.85\textwidth]{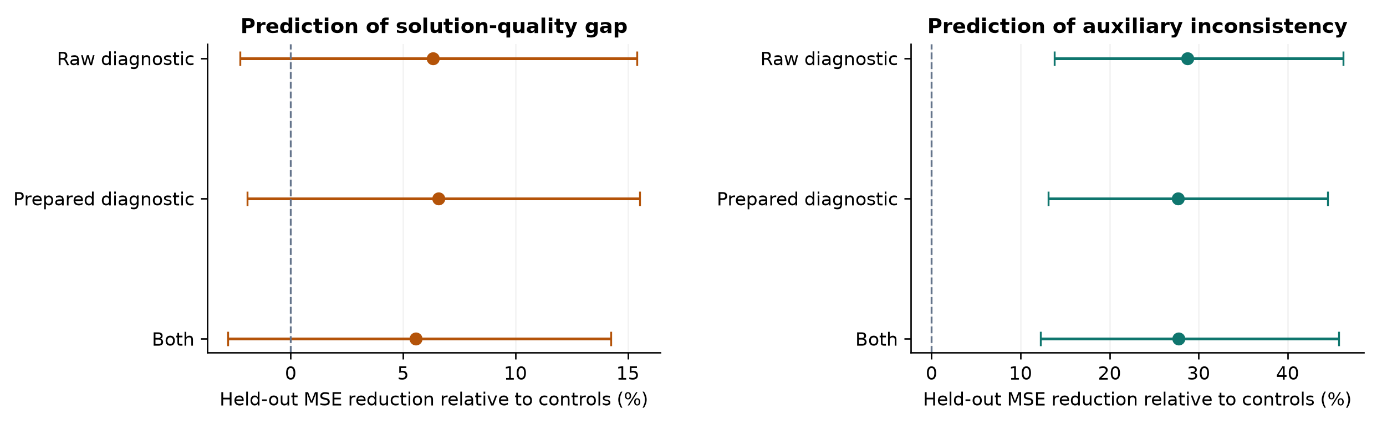}
\caption{Relative held-out MSE reduction from adding the raw, prepared, or both
diagnostics to the control model, for the quality gap (left) and auxiliary
inconsistency (right). Bars are instance-bootstrap 95\% intervals.}
\label{fig:S:diagnostic}
\end{figure}

Table~\ref{tab:S:prediction} and Fig.~\ref{fig:S:diagnostic} give the result. Adding the prepared-state diagnostic lowers the held-out mean squared error of the inconsistency prediction from 0.02863 to 0.02071, an improvement of 0.00792 with interval [0.00376, 0.01274], or 27.7\%. For the quality gap the error changes from 0.001488 to 0.001390 and the improvement interval [$-$0.00003, 0.00023] includes zero. The raw diagnostic gives a similar pattern. The diagnostic therefore helps to predict consistency for this model and instance distribution, whereas its value for predicting solution quality is not established. Prediction does not test whether filtering by the diagnostic improves selection.

\begin{table}[t]
\caption{Held-out prediction on 480 preselected candidates from 60 random instances.
Positive improvement is favourable.}
\label{tab:S:prediction}
\centering
\begin{tabular}{@{}lccc@{}}
\toprule
Outcome & Controls MSE & With prepared diagnostic & Improvement [95\% interval] \\
\midrule
Quality gap & 0.001488 & 0.001390 & 0.000098 $[-0.000029,\ 0.000231]$ \\
Inconsistency & 0.02863 & 0.02071 & 0.00792 $[0.00376,\ 0.01274]$ \\
\bottomrule
\end{tabular}
\end{table}

\subsection{Diagnostic-Constrained Selection}
\label{sup:diagnostic:selection}

Three selection rules are compared on the same strictly mixed candidate set with $Q\le 12$. They are no diagnostic, filtering by the raw-moment diagnostic, and filtering by the prepared-state diagnostic, each with threshold 0.02 and the shared scale $4\sum_p(\sum_{e\supset p}|c_e|+1)$ over all cubic pairs. An empty eligible set is
reported as unavailable, and no fallback or relaxed threshold is used. The two primary
comparisons for the random families use $10{,}000$ paired instance resamples and
$97.5$\% intervals.

All three rules select the same design on all 72 instances. The rules are not
identical, since of $1{,}920$ random-cohort candidates the raw rule rejects $1{,}243$,
the prepared rule rejects $1{,}329$, and they disagree on 86, while all 324 mechanism
candidates pass both. In every instance the lowest-score candidate passes both rules,
so filtering changes no selection. This is an observed absence of benefit for this
intervention and not a proof that the diagnostic can never affect selection.
Resampling identical zero differences cannot establish a population-zero effect. The
optional threshold rule therefore has no demonstrated algorithmic benefit in these
data, which is why the main text treats the diagnostic as descriptive.

\section{Baseline Studies on the Reference Compiler}
\label{sup:earlier}

The studies in this section preceded the four primary experiments of the main text. They
used the Pareto search and certification of Section~IV-D of the main text as the
selector, the all-to-all reference synthesis and a 12-qubit grid as the main targets,
and instance pools that were available during method development. They are therefore
exploratory, they are not pooled with the primary studies, and Section~V-F of the main
text summarises their conclusions. Each subsection states its protocol beside its
result. Unless stated otherwise, comparisons are paired within an instance, resource
summaries are medians of paired instance differences, QAOA summaries are means with
two-sided 95\% Student-$t$ intervals over instances, and repeated seeds and
restarts are not treated as independent samples.

\subsection{Instance Cohorts and Common Protocol}
\label{sup:earlier:cohorts}

Weighted Max-3SAT instances have three distinct variables per clause, independent
literal signs, no duplicate signed clauses, and integer weights drawn uniformly from
$\{1,\dots,5\}$, with clause loads $m_C/n$ near 2, 4, and 6. Cubic spin-glass instances
have hyperedges sampled without replacement with coefficients $K_{ijk}\in\{-1,+1\}$ and
loads $m_e/n$ near 1, 2, and 4. Both families use uniform proposals and anchor-pair
proposals, in which a pair is sampled first and combined with distinct third
variables, so that sparsity and pair reuse vary separately. Structural coordinates are
the cubic density $m_3/\binom{n}{3}$ and the pair-reuse score
$r_{\mathrm{pair}}=1-|\partial_2\mathcal{C}_3|/(3m_3)$, which is zero when every cubic
term has three private pairs. Three tiers were generated, namely an oracle tier of 30 instances per family with $n \in \{5,\dots,8\}$, a QAOA tier of 36 per family with $n \in \{6,8,10\}$ under a 12-qubit limit, and a compilation tier of 90 per family with $n \in \{8,12,16,20\}$. Each tier carries a 60/20/20 training, validation, and test split; selector settings were chosen on training and validation rows, but the pool was
available during protocol refinement. Four representations are compared, namely the
native design, a full quadratization using a minimum-cardinality pair cover with
deterministic ties, the Pareto-search selective design, and matched random controls
that match only the number of active auxiliaries, with five seeded controls per
instance. Compilation used Qiskit 2.4.2 with basis $\{R_z,\sqrt{X},X,\mathrm{CX}\}$,
optimisation level 1, SABRE layout and routing, no scheduling, and five transpiler
seeds. The targets were all-to-all coupling and a bidirectional $3\times 4$ grid, with a 12-qubit
line, a 12-qubit ring, and a 19-qubit heavy-hex graph added for the connectivity
extension. Penalties were $M_p=T_p+1$.

Table~\ref{tab:S:exploratory-conventions} records, for each exploratory study, the objective encoded in the Hamiltonian, the units of the optimised and reported objective, the penalty rule, and the cost-angle convention, checked against the archived implementation.

\begin{table}[t]
\caption{Conventions of the exploratory studies. Scale is the objective encoded in the Hamiltonian, and units are those of the optimised and reported objective.}
\label{tab:S:exploratory-conventions}
\centering
\small
\setlength{\tabcolsep}{3pt}
\begin{tabular}{@{}lllll@{}}
\toprule
Study &
\makecell[l]{Hamiltonian\\scale} &
\makecell[l]{Reporting\\units} &
\makecell[l]{Penalty\\rule} &
\makecell[l]{Angle initialisation\\and optimisation} \\
\midrule
\makecell[l]{Selector validation\\(S5-C)} &
\makecell[l]{Unnormalised native\\or penalised objective} &
\makecell[l]{not\\applicable} &
$M_p = T_p + 1$ &
\makecell[l]{not\\applicable} \\

\makecell[l]{Resource comparison\\(S5-D)} &
\makecell[l]{Unnormalised native\\or penalised objective} &
\makecell[l]{not\\applicable} &
$M_p = T_p + 1$ &
\makecell[l]{not\\applicable} \\

\makecell[l]{Noiseless QAOA\\(S5-E)} &
\makecell[l]{Unnormalised native\\or penalised objective} &
\makecell[l]{Original objective,\\unnormalised units} &
$M_p = T_p + 1$ &
\makecell[l]{Initial $\gamma\in[0,2\pi)$, $\beta\in[0,\pi)$\\
unconstrained COBYLA}
\\

\makecell[l]{Warm start\\(S5-F)} &
\makecell[l]{Unnormalised native\\or penalised objective} &
\makecell[l]{Original objective,\\unnormalised units} &
$M_p = T_p + 1$ &
\makecell[l]{Initial $\gamma\in[0,2\pi)$, $\beta\in[0,\pi)$\\
unconstrained COBYLA} \\

\makecell[l]{Simulated noise\\(S5-G)} &
\makecell[l]{Unnormalised native \\or penalised objective} &
\makecell[l]{Original objective,\\unnormalised units} &
$M_p = T_p + 1$ &
\makecell[l]{Noiseless angles reused\\ no re-optimisation} \\

\bottomrule

\end{tabular}

\end{table}

\subsection{Pareto Search and Branch-and-Bound Certification}
\label{sup:earlier:search}

Cubic supports are ordered as $A_1,\dots,A_m$ and each takes one of four actions,
native or one of its three pairs, giving a tree with $4^m$ leaves. A partial state fixes
the first $k$ actions and is evaluated through its native completion. The selector
first evaluates the native design and a fixed full reference. It then runs a width-$B$
search that keeps at most $B$ states per level, half by the scalar score
$\Jres=w_a a/s_a+w_G G/s_G+w_D D/s_D+w_M M_{\max}/s_M$ and half by a non-dominated ranking
of the resource vector with NSGA-II crowding \cite{DebEtAl2002}. Evaluations are cached
under a key that includes the complete compilation context. The main settings were
weights $(0.05,0.45,0.45,0.05)$, scales $(6,128,96,64)$, width $B=64$, and penalty margin
1. The budgets were 12 qubits, 256 gates, depth 192, and penalty 128 for the QAOA scope, together with a threshold $0.02$ on the normalised consistency excess from raw relaxation moments, which entered feasibility in these studies only.

Optional branch-and-bound certification restarts from the root, reuses the cache and
the best feasible design with score $U_J$, and maintains an unresolved frontier
\cite{LandDoig1960,LawlerWood1966}. A state is removed only when all completions are
proved infeasible by width or penalty, or when its admissible bound
$L_J(s)=w_a|P_s|/s_a+w_M M_{\max,s}/s_M$ exceeds $U_J$. Gate and depth contributions to
the bound are zero. With $L^{\mathrm{glob}}_J$ the minimum of $U_J$ and the frontier
bounds, $L^{\mathrm{glob}}_J\le J^\star\le U_J$, and equality certifies the optimal value
for the declared design space, score, compiler, and numerical feasibility rule. The
frontier covers every unevaluated design not excluded by a valid proof, so the
certificate holds at any stopping point, and exhausting the frontier without a feasible
design proves infeasibility. Floating-point bounds are rounded outward.

\subsection{Selector Validation and Sensitivity}
\label{sup:earlier:sensitivity}

\begin{figure}[t]
\centering
\includegraphics[width=0.85\textwidth]{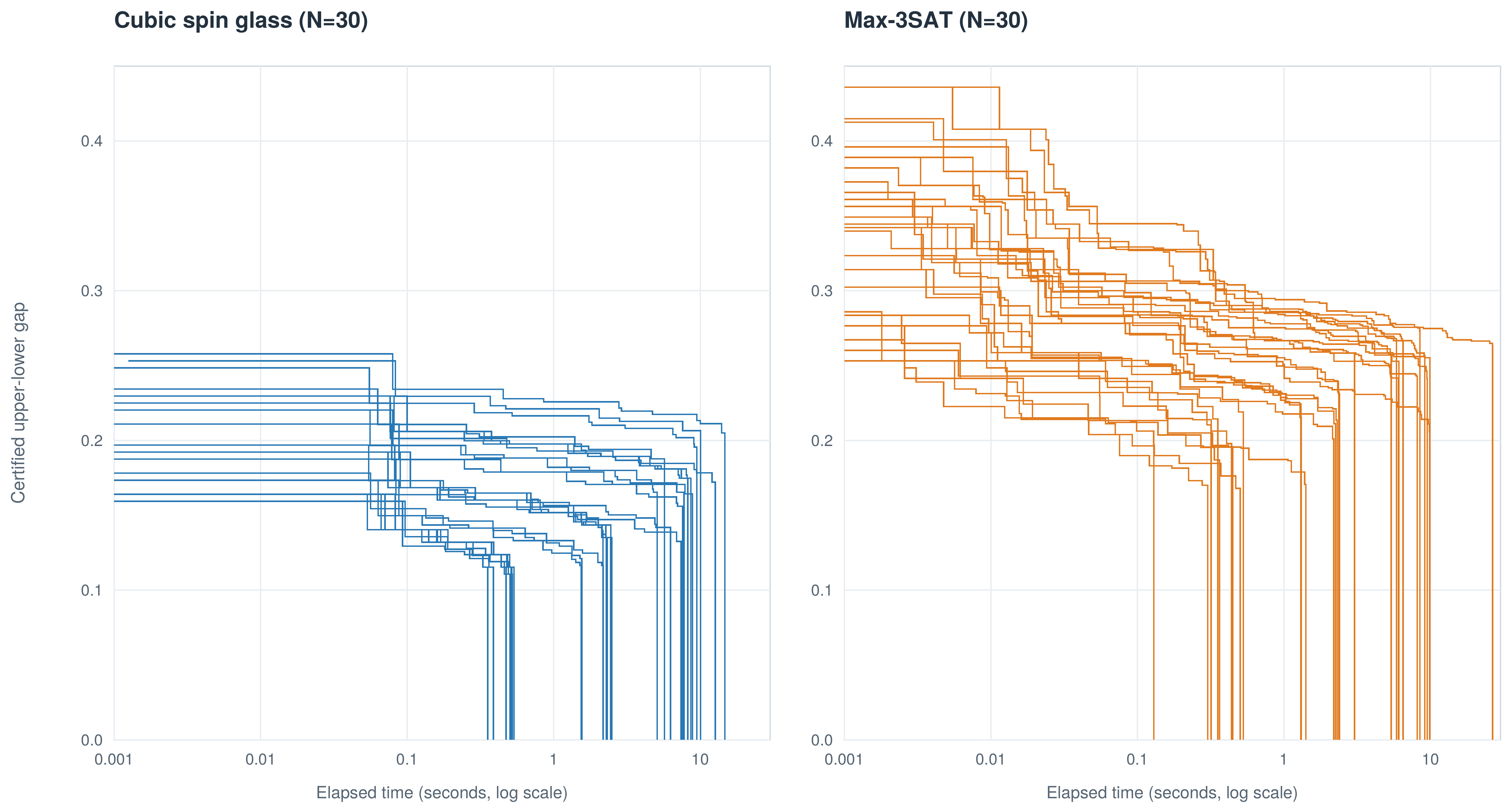}
\caption{Certified upper-minus-lower gap against elapsed time for the 60 oracle
instances. Every trace reaches zero.}
\label{fig:S:certification}
\end{figure}

On the 60 oracle instances, complete enumeration over all original and auxiliary assignments found no pointwise mismatch above the penalty threshold, ties at the threshold, and explicit failures below it. The checks covered $1{,}384$ representation, penalty,
and pair trials and $585{,}408$ joint assignments. Table~\ref{tab:S:oracle} compares
the selector with complete enumeration. The Pareto search matched the enumerated optimum on 58 of 60 instances with maximum score regret $0.0102$. Branch-and-bound
certification reached zero gap on all 60, with median runtimes of 2.5 s and 2.3 s for
the two families and a maximum of 26.5 s (Fig.~\ref{fig:S:certification}). Median node
counts of $5{,}461$ exceeded the median $4{,}096$ candidate evaluations of enumeration,
which is the observation behind the statement in Section~IV-D of the main text that
certification is an implementation check rather than a faster solver. A greedy
native-completion baseline matched only the 30 spin-glass optima, all of which were
native.

\begin{table}[t]
\caption{Oracle-tier validation on 60 instances. Regret is the selector score minus
the enumerated optimum.}
\label{tab:S:oracle}
\centering
\begin{tabular}{@{}lccc@{}}
\toprule
Method & Zero-regret cases & Mean regret & Maximum regret \\
\midrule
Complete enumeration & 60/60 & 0 & 0 \\
Pareto search & 58/60 & $2.30\times10^{-4}$ & $0.0102$ \\
Greedy native completion & 30/60 & $0.0194$ & $0.107$ \\
Branch-and-bound certification & 60/60 & 0 & 0 \\
\bottomrule
\end{tabular}
\end{table}

\begin{figure}[t]
\centering
\includegraphics[width=0.85\textwidth]{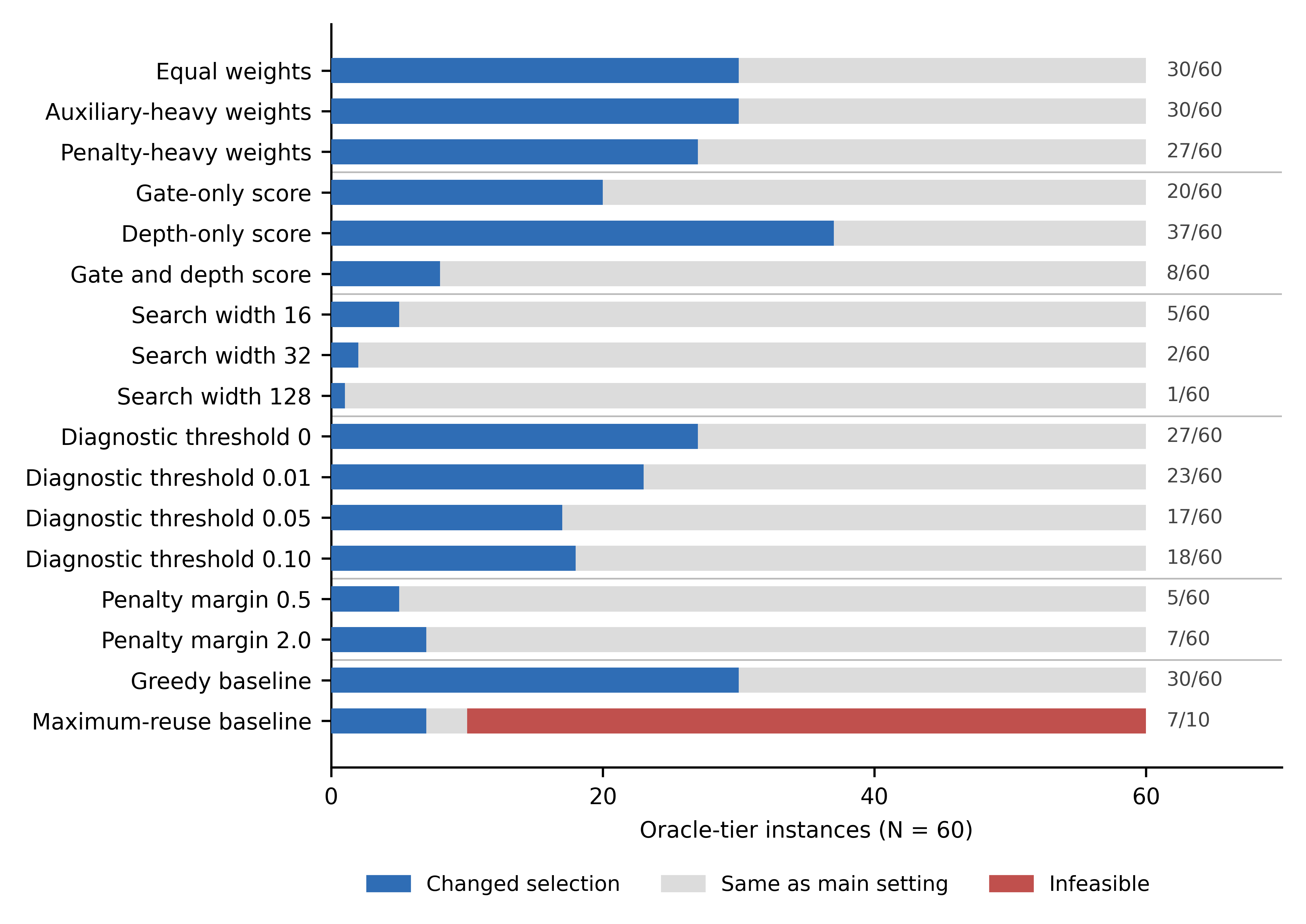}
\caption{Changed, unchanged, and infeasible selections under each alternative setting,
relative to the main setting, on 60 oracle instances.}
\label{fig:S:sensitivity}
\end{figure}

A one-factor-at-a-time experiment varied the selector settings on the same 60 instances
(Table~\ref{tab:S:sensitivity}, Fig.~\ref{fig:S:sensitivity}). Changing the resource
weights altered the selected design on 27 to 30 instances and a depth-only score on 37,
whereas search widths of 16, 32, and 128 altered it on 5, 2, and 1 and penalty margins of
0.5 and 2 on 5 and 7. A maximum-reuse rule was feasible on only 10 instances. The
scalar score is not comparable across weight settings, so the sweep measures the
stability of the choice rather than the merit of any weight vector. This sensitivity
to weights motivates the reporting of non-dominated feasible sets in the main text.

\begin{table}[t]
\caption{One-factor selector sensitivity on 60 oracle instances. Changed counts
representations that differ from the main setting.}
\label{tab:S:sensitivity}
\centering
\begin{tabular}{@{}llcc@{}}
\toprule
Setting & Alternative values & Changed & Feasible runs \\
\midrule
Resource weights & equal, auxiliary-heavy, penalty-heavy & 30, 30, 27 & 60/60 each \\
Resource mode & gate-only, depth-only, gate and depth & 20, 37, 8 & 60/60 each \\
Search width & 16, 32, 128 & 5, 2, 1 & 60/60 each \\
Diagnostic threshold & 0, 0.01, 0.05, 0.10 & 27, 23, 17, 18 & 60/60 each \\
Penalty margin & 0.5, 2.0 & 5, 7 & 60/60 each \\
Greedy baseline & deterministic & 30 & 60/60 \\
Maximum-reuse baseline & deterministic & 7 of 10 & 10/60 \\
\bottomrule
\end{tabular}
\end{table}

\subsection{Resource Comparison on 180 Instances}
\label{sup:earlier:resources}

\begin{figure}[t]
\centering
\includegraphics[width=0.85\textwidth]{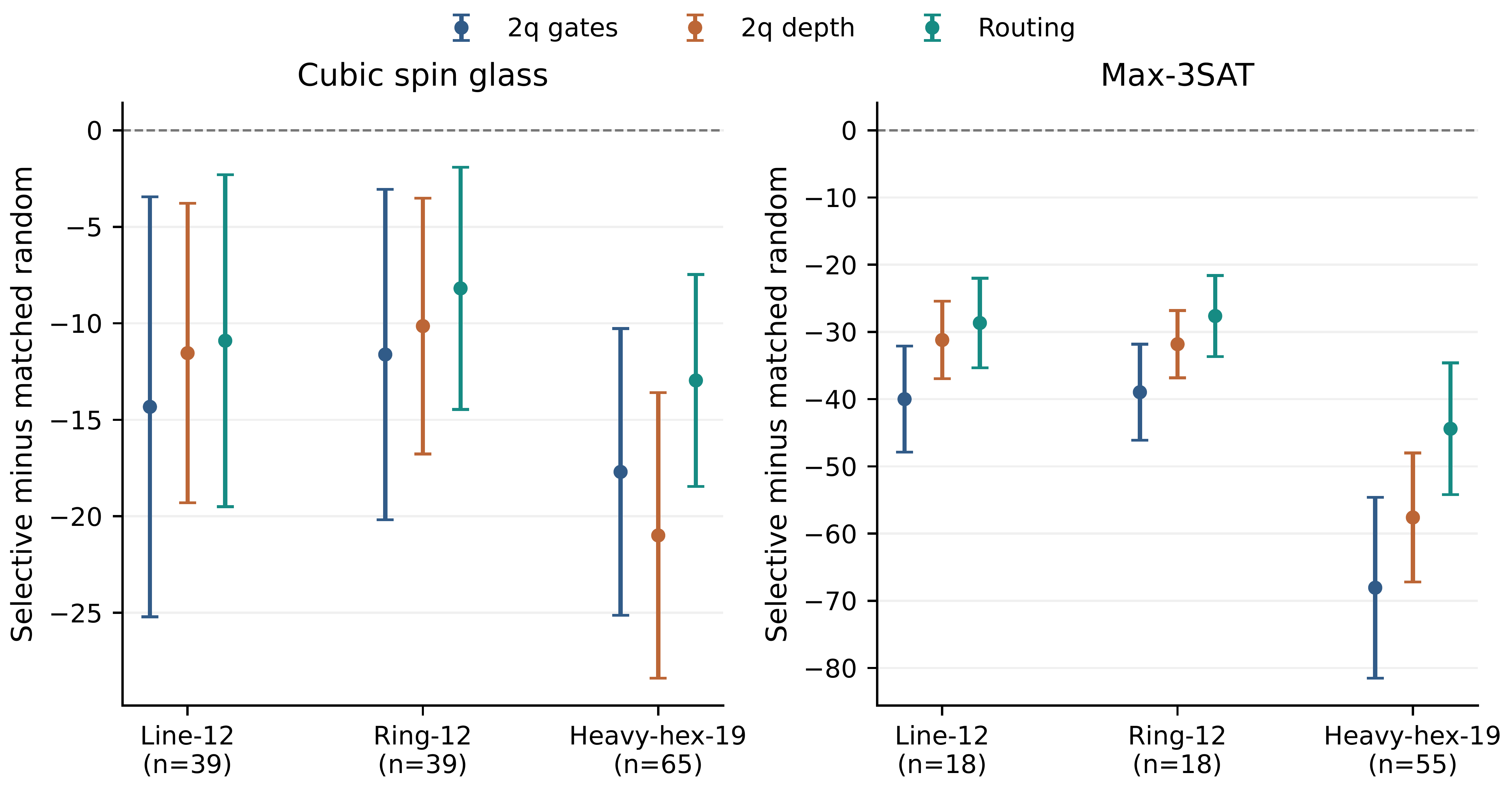}
\caption{Mean paired selective-minus-random gate, depth, and routing differences on
the line, ring, and heavy-hex targets, with 95\% Student-$t$ intervals over
common-fit instances.}
\label{fig:S:topologies}
\end{figure}

Table~\ref{tab:S:resources} reports median within-instance differences from the native
design on the all-to-all reference over all 180 compilation instances, after taking the
median over five transpiler seeds per design and the mean over the five random
controls. Selective reduction reduced Max-3SAT gates and depth substantially and
outperformed the auxiliary-matched random control, whereas for spin glasses it
reduced depth but added gates, and full reduction added many gates. On the 12-qubit
grid, 108 native, 21 full, and 57 selective designs fit, and only 21 instances fit all
four representations. On that common subset, selective Max-3SAT designs saved a median
of 10 gates and 21 depth units relative to native, whereas eight of nine spin-glass
selective designs were native. Line, ring, and heavy-hex targets reproduced the Max-3SAT advantage over the random control on every common-fit instance for depth and on all but one for gates. The paired medians were $-38.4$, $-36.3$, and $-55.8$ gates and $-33.3$, $-31.8$, and $-52.4$ depth units on 18, 18, and 55 instances respectively (Fig.~\ref{fig:S:topologies}). Spin-glass medians were zero on all three targets. The heavy-hex device has larger capacity, so these comparisons do not isolate connectivity
from capacity. A descriptive regime analysis over sparsity and pair-reuse bins found
selective designs helpful for 20 of 20 estimable Max-3SAT depth endpoints and 19 of 20
gate endpoints, with bins too small for a general rule.
Tables 11 to 13 of the main text stratify these records by problem size, report the common-fit counts on the grid, and give the recorded selector cost. Table 7 of the main text gives the signed resource margins of the two certified witnesses.

\begin{table}[t]
\caption{Median within-instance difference from the native design on the all-to-all
reference, 90 instances per family. Negative values are smaller circuits.}
\label{tab:S:resources}
\centering
\begin{tabular}{@{}llcc@{}}
\toprule
Family & Representation & $\Delta$ two-qubit gates & $\Delta$ two-qubit depth \\
\midrule
Max-3SAT & Fully reduced & $-37$ & $-72$ \\
 & Selective & $-25$ & $-60.5$ \\
 & Matched random & $+4$ & $-15.5$ \\
\addlinespace[2pt]
Cubic spin glass & Fully reduced & $+42$ & $-29$ \\
 & Selective & $+6$ & $-18$ \\
 & Matched random & $+8.2$ & $0$ \\
\bottomrule
\end{tabular}
\end{table}

\subsection{Noiseless QAOA Comparison}
\label{sup:earlier:qaoa}

Fourteen test-labelled instances, seven per family, were optimised with COBYLA using three restarts of at most 60 evaluations and tolerance $10^{-3}$. Depths $L=1$ and
$L=2$ were compared with grid gate budgets of 128 and 256, converted to depth by $L_\sigma(B)=\lfloor B/g_\sigma\rfloor$ with $g_\sigma$ the median one-layer gate count.
The loss was the original-objective expectation, and Table~\ref{tab:S:qaoa14} gives mean
paired differences from native. The native design had the lowest mean objective in
every setting. Four of eight selective-minus-native intervals included zero, the
Max-3SAT difference at $L=2$ was $1.935$ $[0.597,3.273]$ in favour of native, and the
spin-glass difference at $B=256$ was $0.821$ $[0.007,1.635]$. At $B=128$, 35 of 112
design groups had $L=0$ because no complete cost layer fitted the budget; these are
initial-state baselines. Restricting to the five spin-glass and three Max-3SAT
instances on which every design fitted a layer gave differences of $-0.072$
$[-0.544,0.400]$ and $0.982$ $[-0.890,2.854]$, both inconclusive. Auxiliary
inconsistency was about $0.8$ for full and selective Max-3SAT designs at all budgets.

\begin{table}[t]
\caption{Mean paired objective difference from the native design on 14 instances.
Positive values favour native.}
\label{tab:S:qaoa14}
\centering
\begin{tabular}{@{}llcccc@{}}
\toprule
Family & Representation & $L=1$ & $L=2$ & $B=128$ & $B=256$ \\
\midrule
Cubic spin glass & Fully reduced & 1.837 & 2.041 & 2.044 & 2.458 \\
 & Selective & 0.379 & 0.275 & 0.431 & 0.821 \\
 & Matched random & 0.394 & 0.271 & 0.448 & 0.543 \\
\addlinespace[2pt]
Max-3SAT & Fully reduced & 1.774 & 2.112 & 0.245 & 1.855 \\
 & Selective & 1.899 & 1.935 & 0.040 & 1.511 \\
 & Matched random & 1.390 & 1.192 & 0.385 & 1.516 \\
\bottomrule
\end{tabular}
\end{table}

\subsection{Warm-Start Comparison}
\label{sup:earlier:warmstart}

Four initialisation policies were compared with clipping $\delta=0.05$, namely cold start,
original-only warming with auxiliaries at probability $1/2$, original-plus-auxiliary
warming with bounded pair moments, and the independence rule $q_p=\mu_i\mu_j$. Uniform
probabilities on the whole register reproduce the cold start exactly at every
parameter vector, which served as a regression control. A coordinate correction
aligned the warm and cold mixer conventions and required the positive-depth cold
controls to be rerun in a different numerical environment, so historical and corrected
results are not pooled.

\begin{figure}[t]
\centering
\includegraphics[width=0.85\textwidth]{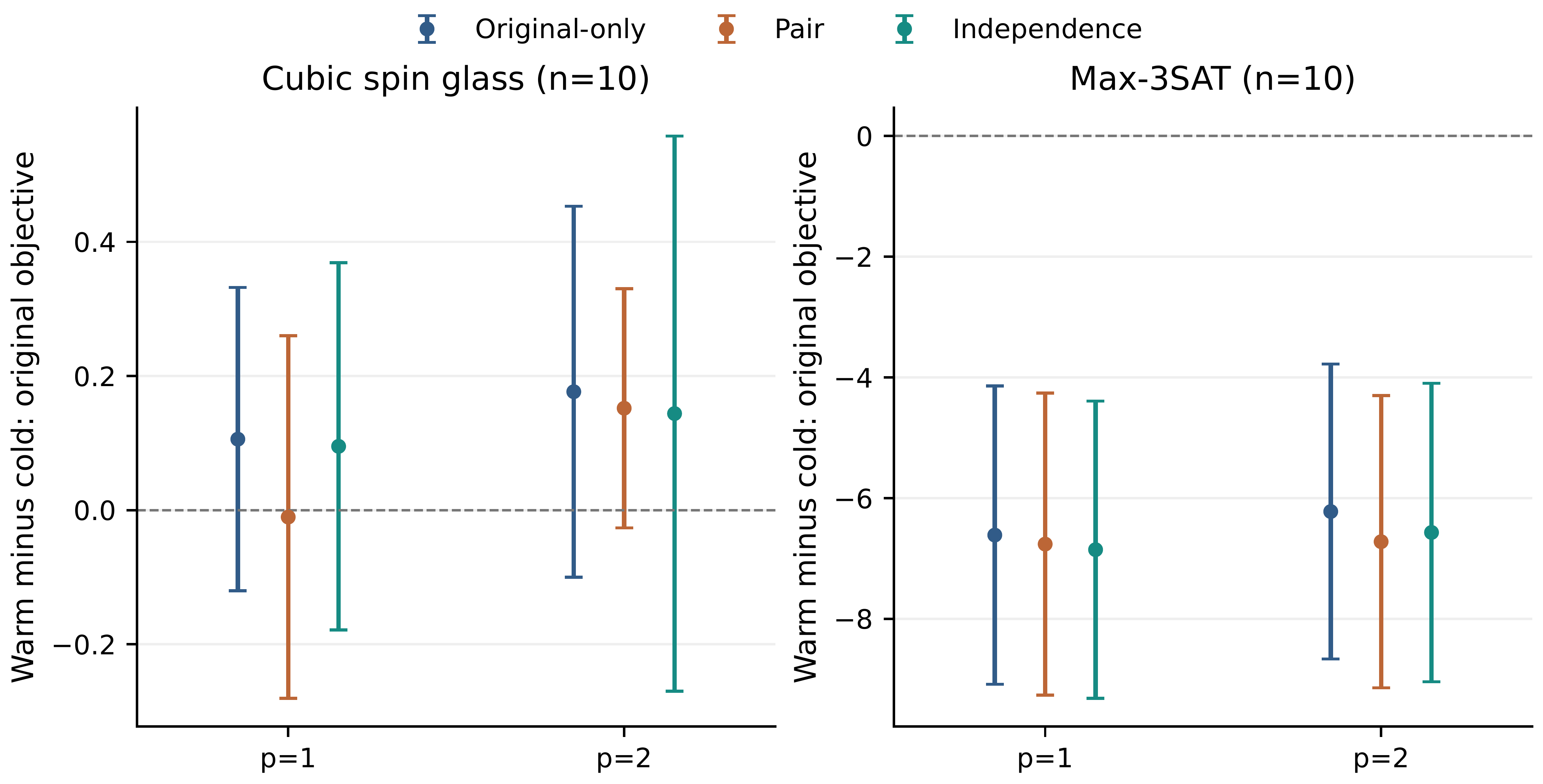}
\caption{Warm-minus-cold original objective for the three warm policies on the
screened set, with 95\% Student-$t$ intervals over ten instances per family.}
\label{fig:S:warm20}
\end{figure}

On the original 14-instance sample, 11 instances had weak relaxation signal with median
$|\mu_i-1/2|\le 0.05$, and all Max-3SAT pair-minus-cold intervals on the seven-instance
common subset included zero. A separate 20-instance set, ten per family, was screened
by relaxation signal alone before any QAOA outcome was inspected.
Table~\ref{tab:S:warm20} and Fig.~\ref{fig:S:warm20} give the corrected results. For Max-3SAT, all three warm policies improved on cold start by about $6.7$ objective units at both depths. Pair warming improved on original-only warming by $-0.149$ $[-0.219,-0.080]$ at $L=1$ and $-0.500$ $[-0.875,-0.124]$ at $L=2$, whereas
pair-minus-independence intervals included zero. Spin-glass differences remained
uncertain. This screened sample was selected for strong signal and does not represent
random instances.

\begin{table}[t]
\caption{Coordinate-corrected warm-start differences on 20 screened instances, ten per
family. Negative values favour the first policy.}
\label{tab:S:warm20}
\centering
\setlength{\tabcolsep}{3pt}
\begin{tabular}{@{}llccc@{}}
\toprule
Family & Depth & Pair minus cold & Pair minus original-only & Pair minus independence \\
\midrule
Spin glass & $L=1$ & $-0.010$ $[-0.281,0.260]$ & $-0.116$ $[-0.344,0.112]$ & $-0.105$ $[-0.320,0.109]$ \\
Spin glass & $L=2$ & $0.152$ $[-0.026,0.330]$ & $-0.025$ $[-0.228,0.179]$ & $0.008$ $[-0.289,0.304]$ \\
Max-3SAT & $L=1$ & $-6.761$ $[-9.263,-4.260]$ & $-0.149$ $[-0.219,-0.080]$ & $0.092$ $[-0.017,0.202]$ \\
Max-3SAT & $L=2$ & $-6.722$ $[-9.143,-4.300]$ & $-0.500$ $[-0.875,-0.124]$ & $-0.153$ $[-0.656,0.350]$ \\
\bottomrule
\end{tabular}
\end{table}

\subsection{Simulated-Noise Comparison}
\label{sup:earlier:noise}

Eighteen instances, nine per family, were selected from the QAOA pool by sparsity and
pair-reuse coverage before any noise outcome was examined. Circuits on the grid used
local depolarising noise after every $\sqrt{X}$, $X$, and CX gate with symmetric
readout error, at low rates $(0.001,0.01,0.01)$ and high rates $(0.003,0.03,0.03)$ for
one-qubit, two-qubit, and readout errors, and $4{,}096$ shots per evaluation.
Parameters optimised in the noiseless circuit were reused without re-optimisation.
Table~\ref{tab:S:noise} reports paired selective-minus-native differences, and noise worsened every representation. Full reduction changed least from its noiseless value
but started worse in both families and retained auxiliary inconsistency above $0.8$.
All spin-glass intervals included zero, and Max-3SAT favoured native without noise
and was inconclusive under noise. Smaller degradation therefore did not correspond to
better absolute performance, and the simulator parameters are representative rather
than calibrated.

\begin{table}[t]
\caption{Selective-minus-native objective differences in the noise experiment, nine
instances per family. Negative values favour selective reduction.}
\label{tab:S:noise}
\centering
\begin{tabular}{@{}llccc@{}}
\toprule
Family & Noise & Mean difference [95\% interval] & $d_z$ & Win/tie/loss \\
\midrule
Cubic spin glass & None & $0.129$ $[-0.868,1.126]$ & $0.099$ & 5/0/4 \\
 & Low & $-0.030$ $[-0.312,0.251]$ & $-0.083$ & 6/0/3 \\
 & High & $-0.036$ $[-0.089,0.016]$ & $-0.532$ & 4/3/2 \\
\addlinespace[2pt]
Max-3SAT & None & $0.749$ $[0.140,1.358]$ & $0.945$ & 3/0/6 \\
 & Low & $0.149$ $[-0.046,0.344]$ & $0.589$ & 1/0/8 \\
 & High & $-0.009$ $[-0.133,0.116]$ & $-0.053$ & 5/0/4 \\
\bottomrule
\end{tabular}
\end{table}

\section{Reproducibility Records}
\label{sup:reproducibility}

This section records the software environments, seeds, result locations, and revision
history of the experiments reported in the main text and in
Sections~\ref{sup:compilation} to \ref{sup:earlier}. All paths refer to the frozen 5 October 2026 reproduction archive of the project repository, whose manifest lists the configuration hashes, commit identifiers, and archive checksums for every record. A persistent identifier for the complete archive will be added before publication.

\subsection{Software Versions, Compiler Settings, and Seeds}
\label{sup:reproducibility:software}

Table~\ref{tab:S:software} lists the environment of each study. The primary experiments used Python 3.12 and device compilation used Qiskit 2.5.2 \cite{Qiskit2024}; the baseline
experiments of Section~\ref{sup:earlier} used Qiskit 2.4.2 and Qiskit Aer 0.17.2 on Windows,
with the warm-start correction executed on Linux. Compilation in the primary experiments
used the basis $\{R_z,\sqrt{X},X,\mathrm{CX}\}$, optimisation level 3, SABRE layout and
routing \cite{Li2019}, identity initial layout, and routing seed 1729, with the
additional seeds $\{2718, 31415, 57721, 65537\}$ in the follow-up. The baseline experiments
used optimisation level 1, SABRE, no scheduling, transpiler seeds 2026082721 to
2026082725, matched-random seeds 2026082711 to 2026082715, optimiser seeds 2026082731
to 2026082733, noise seeds 2026082741 to 2026082743, and measurement seeds 2026082751
to 2026082753. 
Relaxations were solved through the SciPy interface \cite{Virtanen2020} to the HiGHS solver \cite{HuangfuHall2018}, under the fixed variable order of singles, lexicographic
pairs, and sorted supported triples.
All scheduled seeds, restarts, and random controls are retained in every record. The primary QAOA experiments report the lowest expectation across the prescribed evaluations of an instance, whereas the baseline studies of Section~S5 average restart optima, and no seed or restart is discarded after the results are seen.

\begin{table}[t]
\caption{Execution environments of the reported studies.}
\label{tab:S:software}
\centering
\begin{tabular}{@{}llll@{}}
\toprule
Study & Platform & Python / NumPy / SciPy & Qiskit \\
\midrule
Cost-layer compilation & Windows & 3.12 / 2.5.3 / 1.18.1 & 2.5.2 \\
Routing-seed follow-up & Windows 11 & 3.12.10 / 2.5.3 / 1.18.1 & 2.5.2 \\
Preparation and evolution & Windows & 3.12 / 2.5.3 / 1.18.1 & 2.5.2 \\
Joint resource and quality & Linux & 3.12 / 2.5.3 / 1.18.1 & 2.5.2 \\
Joint-experiment verification audit & Windows 11 & 3.12.10 / 2.5.3 / 1.18.1 & 2.5.2 \\
Baseline studies (S5) & Windows & 3.12.10 / 2.5.2 / 1.18.1 & 2.4.2, Aer 0.17.2 \\
Warm-start correction (S5) & Linux & 3.12.14 / 2.3.5 / 1.17.0 & 2.4.2 \\
\bottomrule
\end{tabular}
\end{table}

\subsection{Result-to-Record Map}
\label{sup:reproducibility:map}

Table~\ref{tab:S:records} maps each reported result to its root in the
repository. Every result row carries its configuration hash, the manifest hash of its instance source, and the code commit of its run. The summaries in the main text are derived from the instance-level records by the aggregation rules stated in each protocol. Corrected derived outputs never replace unchanged raw records. The five-seed rows reuse the compilation objectives and do not constitute new instances.

\begin{table}[t]

\caption{Study roots for the reported results. Paths are relative to the frozen 5 October 2026 reproduction archive.}
\label{tab:S:records}
\centering
\small
\begin{tabular}{@{}p{0.19\textwidth}p{0.15\textwidth}p{0.60\textwidth}@{}}
\toprule
Result & Main-text location & Archive root and principal records \\
\midrule
90-instance compilation & Section V-B & \path{archives/revision_20260917/revision/experiments/compiled_results/}, with freeze, amendment, frozen instances, summary, and per-instance records \\
Routing-seed follow-up & Section V-C & \path{corrections/review_v2_20260920/final_analysis/}, with experiment configuration, instance manifest, compilation rows, and classification arrays \\
Preparation and evolution & Section V-D & \path{archives/revision_20260917/revision/qaoa_new/}, with protocol, freeze, instances, results, and paired contrasts \\
Joint resource and quality & Section V-E & \path{archives/feedback_iteration_v3/}, with protocol, freeze, execution results, analysis, and witness audit \\
Diagnostic prediction & Section~\ref{sup:diagnostic} & \path{archives/feedback_iteration_v3/}, 480 probes on the 60 random instances \\
Exactness and penalty checks & Section~\ref{sup:earlier:sensitivity} & \path{delivery/20260913/archives/URSS_E2_E5_LOCAL_RESULTS_20260913_020457_389.zip}, E1 exactness and penalty-trial records \\
Resource comparison & Section~\ref{sup:earlier:resources} & same archive, E2 merged rows and added-topology rows \\
Noiseless QAOA and warm start & Sections~\ref{sup:earlier:qaoa} and~\ref{sup:earlier:warmstart} & same archive, E3 and corrected E4 merged runs \\
Certification and sensitivity & Section~\ref{sup:earlier:sensitivity} & same archive, certification traces and selector ablation \\
Noise study & Section~\ref{sup:earlier:noise} & same archive, E6 merged runs \\
\bottomrule
\end{tabular}

\end{table}

\subsection{Fixed Protocol and Revision Log}
\label{sup:reproducibility:history}

Each primary experiment saved its protocol, generator, candidate counts, score weights,
compiler settings, and budget grids before any data were generated, and the freeze
files are part of the archive. The freeze was recorded locally and is not an external
preregistration. One technical amendment was recorded before any compiler result of
the compilation experiment, setting both candidate budgets to 27 for the constructed family
because only 27 full assignments fit the device. The joint experiment generated 72 unique objectives with no exact overlap against the 462 archived objectives of the earlier
cohorts, and it executed $4{,}560$ compiler calls and $1{,}476{,}585$ objective evaluations.

A verification audit of the joint experiment was performed on Windows 11 with pinned package
versions on 18 September 2026. It covered all 72 instances and $4{,}560$
pointwise-exact representations, $9{,}120$ product-law diagnostics, $1{,}086$ saved
parameter sets, $1{,}350$ grid samples, and $2{,}172$ trace samples, with a maximum
numerical difference of $1.6\times 10^{-13}$. Nine representative circuits were
recompiled and checked against an independently simulated state. All resource fields
used in budgets, scores, and selections matched exactly. Two of the nine circuits
differed by one gate in total gate count, which no decision uses, so the audit status
is a pass with compiler metadata drift, whereas the strict field-by-field replay
status is a fail. These checks support compatibility for the sampled circuits and do
not establish strict cross-platform compiler reproducibility. The one-gate secondary witness of Section~V-E of the main text was recompiled and checked by an independent statevector evaluation, which reproduced the saved expectation.

The baseline studies of Section~\ref{sup:earlier} passed through two recorded revisions. A warm-start coordinate correction aligned the warm and cold mixer conventions. It retained the original instances, designs, archived probabilities, initial parameters, seeds, budgets, and loss, validated 132 designs and $4{,}800$ saved-parameter replays, and reran $4{,}332$ positive-depth optimisations in a different numerical environment because optimiser equivalence across environments was not established. A subsequent selector repair changed 285 representation groups and
reoptimised 420 positive-depth jobs, and every new encoding passed the exactness
checks. Historical and corrected results are stored separately and are never pooled.
Original E4 warm-start runs used the first optimum returned by the relaxation solver,
whereas later selection used a deterministic optimum-face representative, and the two
are analysed separately. No experiment was rerun solely to produce a figure.


\begingroup

\endgroup


\begin{thebibliography}{10}
\providecommand{\url}[1]{#1}
\csname url@samestyle\endcsname
\providecommand{\newblock}{\relax}
\providecommand{\bibinfo}[2]{#2}
\providecommand{\BIBentrySTDinterwordspacing}{\spaceskip=0pt\relax}
\providecommand{\BIBentryALTinterwordstretchfactor}{4}
\providecommand{\BIBentryALTinterwordspacing}{\spaceskip=\fontdimen2\font plus
\BIBentryALTinterwordstretchfactor\fontdimen3\font minus \fontdimen4\font\relax}
\providecommand{\BIBforeignlanguage}[2]{{%
\expandafter\ifx\csname l@#1\endcsname\relax
\typeout{** WARNING: IEEEtran.bst: No hyphenation pattern has been}%
\typeout{** loaded for the language `#1'. Using the pattern for}%
\typeout{** the default language instead.}%
\else
\language=\csname l@#1\endcsname
\fi
#2}}
\providecommand{\BIBdecl}{\relax}
\BIBdecl

\bibitem{Weidenfeller2022}
J.~Weidenfeller, L.~C. Valor, J.~Gacon, C.~Tornow, L.~Bello, S.~Woerner, and D.~J. Egger, ``{Scaling of the quantum approximate optimization algorithm on superconducting qubit based hardware},'' \emph{Quantum}, vol.~6, 2022, {Art. no. 870}.

\bibitem{Harrigan2021}
M.~P. Harrigan \emph{et~al.}, ``{Quantum approximate optimization of non-planar graph problems on a planar superconducting processor},'' \emph{Nature Physics}, vol.~17, pp. 332--336, 2021.

\bibitem{Abbas2024}
A.~Abbas \emph{et~al.}, ``{Challenges and opportunities in quantum optimization},'' \emph{Nature Reviews Physics}, vol.~6, pp. 718--735, 2024.

\bibitem{Farhi2014}
E.~Farhi, J.~Goldstone, and S.~Gutmann, ``{A quantum approximate optimization algorithm},'' arXiv:1411.4028, 2014.

\bibitem{Zhou2020}
L.~Zhou, S.-T. Wang, S.~Choi, H.~Pichler, and M.~D. Lukin, ``{Quantum Approximate Optimization Algorithm: Performance, Mechanism, and Implementation on Near-Term Devices},'' \emph{Physical Review X}, vol.~10, 2020, {Art. no. 021067}.

\bibitem{BorosHammer2002}
E.~Boros and P.~L. Hammer, ``{Pseudo-boolean optimization},'' \emph{Discrete Applied Mathematics}, vol. 123, pp. 155--225, 2002.

\bibitem{Hadfield2021}
S.~Hadfield, ``{On the representation of Boolean and real functions as Hamiltonians for quantum computing},'' \emph{ACM Transactions on Quantum Computing}, vol.~2, no.~4, 2021, {Art. no. 18}.

\bibitem{Glos2022}
A.~Glos, A.~Krawiec, and Z.~Zimbor{\'a}s, ``{Space-efficient binary optimization for variational quantum computing},'' \emph{npj Quantum Information}, vol.~8, 2022, {Art. no. 39}.

\bibitem{CampbellDahl2022}
C.~Campbell and E.~Dahl, ``{QAOA of the highest order},'' in \emph{2022 IEEE 19th International Conference on Software Architecture Companion (ICSA-C)}, 2022, pp. 141--146.

\bibitem{Stein2023}
J.~Stein, F.~Chamanian, M.~Zorn, J.~N{\"u}{\ss}lein, S.~Zielinski, M.~K{\"o}lle, and C.~Linnhoff-Popien, ``{Evidence that PUBO outperforms QUBO when solving continuous optimization problems with the QAOA},'' in \emph{Proceedings of the Genetic and Evolutionary Computation Conference Companion}, 2023, pp. 2254--2262.

\bibitem{Bell2025}
K.~Bell, A.~Lowe, E.~Coles, N.~Ridgway, and G.~Marshall, ``{A comparison of quadratic and higher-order representations for QAOA},'' arXiv:2509.20127, 2025.

\bibitem{Koch2026}
F.~Koch, S.~Panahiyan, R.~Mukherjee, J.~Doetsch, and D.~Jaksch, ``{Resource-efficient quantum optimization via higher-order encoding},'' \emph{EPJ Quantum Technology}, vol.~13, 2026, {Art. no. 59}.

\bibitem{Verchere2023}
Z.~Verch{\`e}re, S.~Elloumi, and A.~Simonetto, ``{Optimizing variational circuits for higher-order binary optimization},'' in \emph{2023 IEEE International Conference on Quantum Computing and Engineering (QCE)}, vol.~1, 2023, pp. 19--25.

\bibitem{Rosenberg1975}
I.~G. Rosenberg, ``{Reduction of bivalent maximization to the quadratic case},'' \emph{Cahiers du Centre d'\'Etudes de Recherche Op\'erationnelle}, vol.~17, pp. 71--74, 1975.

\bibitem{BorosGruber2014}
E.~Boros and A.~Gruber, ``{On quadratization of pseudo-Boolean functions},'' arXiv:1404.6538, 2014.

\bibitem{Dattani2019}
N.~Dattani, ``{Quadratization in discrete optimization and quantum mechanics},'' arXiv:1901.04405, 2019.

\bibitem{Schmidbauer2024}
L.~Schmidbauer, K.~Wintersperger, E.~Lobe, and W.~Mauerer, ``{Polynomial reduction methods and their impact on QAOA circuits},'' in \emph{2024 IEEE International Conference on Quantum Software (QSW)}, 2024, pp. 35--45.

\bibitem{Schmidbauer2026}
L.~Schmidbauer, E.~Lobe, I.~Schaefer, and W.~Mauerer, ``{It's quick to be square: Fast quadratisation for quantum toolchains},'' \emph{ACM Transactions on Quantum Computing}, vol.~7, no.~3, pp. 1--46, 2026, {doi: 10.1145/3800943}.

\bibitem{Rovara2026}
D.~Rovara, L.~Burgholzer, and R.~Wille, ``{Quantum hardware-efficient selection of auxiliary variables for QUBO formulations},'' in \emph{Proc. Design, Automation and Test in Europe Conference (DATE)}, 2026, pp. 1--7, {doi: 10.23919/DATE69613.2026.11539407}.

\bibitem{Dragoi2026}
S.~Dr\u{a}goi, A.~Baiardi, and D.~J. Egger, ``{Approximate quadratization of high-order Hamiltonians for combinatorial quantum optimization},'' \emph{Physical Review Research}, vol.~8, 2026, {Art. no. 023159}.

\bibitem{Cerezo2021}
M.~Cerezo \emph{et~al.}, ``{Variational quantum algorithms},'' \emph{Nature Reviews Physics}, vol.~3, pp. 625--644, 2021.

\bibitem{Blekos2024}
K.~Blekos, D.~Brand, A.~Ceschini, C.-H. Chou, R.-H. Li, K.~Pandya, and A.~Summer, ``{A review on Quantum Approximate Optimization Algorithm and its variants},'' \emph{Physics Reports}, vol. 1068, pp. 1--66, 2024.

\bibitem{KolmogorovZabih2004}
V.~Kolmogorov and R.~Zabih, ``{What energy functions can be minimized via graph cuts?}'' \emph{IEEE Transactions on Pattern Analysis and Machine Intelligence}, vol.~26, no.~2, pp. 147--159, 2004.

\bibitem{FreedmanDrineas2005}
D.~Freedman and P.~Drineas, ``{Energy minimization via graph cuts: Settling what is possible},'' in \emph{Proc. IEEE Computer Society Conf. Computer Vision and Pattern Recognition (CVPR)}, vol.~2, 2005, pp. 939--946.

\bibitem{Ishikawa2011}
H.~Ishikawa, ``{Transformation of general binary MRF minimization to the first-order case},'' \emph{IEEE Transactions on Pattern Analysis and Machine Intelligence}, vol.~33, pp. 1234--1249, 2011.

\bibitem{AnthonyEtAl2017}
M.~Anthony, E.~Boros, Y.~Crama, and A.~Gruber, ``{Quadratic reformulations of nonlinear binary optimization problems},'' \emph{Mathematical Programming}, vol. 162, pp. 115--144, 2017.

\bibitem{Anthony2016}
------, ``{Quadratization of symmetric pseudo-Boolean functions},'' \emph{Discrete Applied Mathematics}, vol. 203, pp. 1--12, 2016.

\bibitem{BorosCramaRodriguezHeck2020}
E.~Boros, Y.~Crama, and E.~Rodr{\'i}guez-Heck, ``{Compact quadratizations for pseudo-Boolean functions},'' \emph{Journal of Combinatorial Optimization}, vol.~39, pp. 687--707, 2020.

\bibitem{VermaLewis2020}
A.~Verma and M.~Lewis, ``{Optimal quadratic reformulations of fourth degree pseudo-Boolean functions},'' \emph{Optimization Letters}, vol.~14, pp. 1557--1569, 2020.

\bibitem{VermaLewisKochenberger2021}
A.~Verma, M.~Lewis, and G.~Kochenberger, ``{Efficient QUBO transformation for higher degree pseudo Boolean functions},'' arXiv:2107.11695, 2021.

\bibitem{CramaEtAl2025}
Y.~Crama, S.~Elloumi, A.~Lambert, and E.~Rodr\'{\i}guez-Heck, ``{Quadratization and convexification in polynomial binary optimization},'' \emph{Journal of Combinatorial Optimization}, vol.~50, 2025, {Art. no. 28}.

\bibitem{CowtanPhase2020}
A.~Cowtan, S.~Dilkes, R.~Duncan, W.~Simmons, and S.~Sivarajah, ``{Phase Gadget Synthesis for Shallow Circuits},'' \emph{Electronic Proceedings in Theoretical Computer Science}, vol. 318, pp. 213--228, 2020.

\bibitem{CowtanUCC2020}
A.~Cowtan, W.~Simmons, and R.~Duncan, ``{A generic compilation strategy for the unitary coupled cluster ansatz},'' arXiv:2007.10515, 2020.

\bibitem{Amy2018}
M.~Amy, P.~Azimzadeh, and M.~Mosca, ``{On the CNOT-complexity of CNOT-PHASE circuits},'' \emph{Quantum Science and Technology}, vol.~4, 2018, {Art. no. 015002}.

\bibitem{Sriluckshmy2023}
P.~V. Sriluckshmy, V.~Pina-Canelles, M.~Ponce, M.~G. Algaba, F.~\v{S}imkovic, and M.~Leib, ``{Optimal, hardware native decomposition of parameterized multi-qubit Pauli gates},'' \emph{Quantum Science and Technology}, vol.~8, 2023, {Art. no. 045029}.

\bibitem{Basso2022}
J.~Basso, D.~Gamarnik, S.~Mei, and L.~Zhou, ``{Performance and limitations of the QAOA at constant levels on large sparse hypergraphs and spin glass models},'' in \emph{Proc. IEEE 63rd Annual Symposium on Foundations of Computer Science (FOCS)}, 2022, pp. 335--343.

\bibitem{BassoHighDepth2022}
J.~Basso, E.~Farhi, K.~Marwaha, B.~Villalonga, and L.~Zhou, ``{The quantum approximate optimization algorithm at high depth for MaxCut on large-girth regular graphs and the Sherrington--Kirkpatrick model},'' in \emph{Proc. 17th Conf. Theory of Quantum Computation, Communication and Cryptography (TQC)}, ser. LIPIcs, vol. 232, 2022, pp. 7:1--7:21.

\bibitem{PelofskeShortDepth2024}
E.~Pelofske, A.~B{\"a}rtschi, and S.~Eidenbenz, ``{Short-depth QAOA circuits and quantum annealing on higher-order Ising models},'' \emph{npj Quantum Information}, vol.~10, 2024, {Art. no. 30}.

\bibitem{Pelofske2024}
E.~Pelofske, A.~B{\"a}rtschi, L.~Cincio, J.~Golden, and S.~Eidenbenz, ``{Scaling whole-chip QAOA for higher-order Ising spin glass models on heavy-hex graphs},'' \emph{npj Quantum Information}, vol.~10, 2024, {Art. no. 109}.

\bibitem{Lotshaw2022}
P.~C. Lotshaw, T.~Nguyen, A.~Santana, A.~McCaskey, R.~Herrman, J.~Ostrowski, G.~Siopsis, and T.~S. Humble, ``{Scaling quantum approximate optimization on near-term hardware},'' \emph{Scientific Reports}, vol.~12, 2022, {Art. no. 12388}.

\bibitem{Li2019}
G.~Li, Y.~Ding, and Y.~Xie, ``{Tackling the qubit mapping problem for NISQ-era quantum devices},'' in \emph{Proc. 24th Int. Conf. Architectural Support for Programming Languages and Operating Systems (ASPLOS)}, 2019, pp. 1001--1014.

\bibitem{Zulehner2019}
A.~Zulehner, A.~Paler, and R.~Wille, ``{An efficient methodology for mapping quantum circuits to the IBM QX architectures},'' \emph{IEEE Transactions on Computer-Aided Design of Integrated Circuits and Systems}, vol.~38, no.~7, pp. 1226--1236, 2019.

\bibitem{Murali2019}
P.~Murali, J.~M. Baker, A.~Javadi-Abhari, F.~T. Chong, and M.~Martonosi, ``{Noise-adaptive compiler mappings for noisy intermediate-scale quantum computers},'' in \emph{Proc. 24th Int. Conf. Architectural Support for Programming Languages and Operating Systems (ASPLOS)}, 2019, pp. 1015--1029.

\bibitem{Alam2020}
M.~Alam, A.~Ash-Saki, and S.~Ghosh, ``{Circuit compilation methodologies for quantum approximate optimization algorithm},'' in \emph{2020 53rd Annual IEEE/ACM International Symposium on Microarchitecture (MICRO)}, 2020, pp. 215--228.

\bibitem{Holmes2020}
A.~Holmes, S.~Johri, G.~G. Guerreschi, J.~S. Clarke, and A.~Y. Matsuura, ``{Impact of qubit connectivity on quantum algorithm performance},'' \emph{Quantum Science and Technology}, vol.~5, 2020, {Art. no. 025009}.

\bibitem{Herrman2021depth}
R.~Herrman, J.~Ostrowski, T.~S. Humble, and G.~Siopsis, ``{Lower bounds on circuit depth of the quantum approximate optimization algorithm},'' \emph{Quantum Information Processing}, vol.~20, 2021, {Art. no. 59}.

\bibitem{Herrman2021}
R.~Herrman, L.~Treffert, J.~Ostrowski, P.~C. Lotshaw, T.~S. Humble, and G.~Siopsis, ``{Globally optimizing QAOA circuit depth for constrained optimization problems},'' \emph{Algorithms}, vol.~14, 2021, {Art. no. 294}.

\bibitem{Mirkarimi2024}
P.~Mirkarimi, I.~Shukla, D.~C. Hoyle, R.~Williams, and N.~Chancellor, ``{Quantum optimization with linear Ising penalty functions for customer data science},'' \emph{Physical Review Research}, vol.~6, 2024, {Art. no. 043241, doi: 10.1103/PhysRevResearch.6.043241}.

\bibitem{Ayodele2022}
M.~Ayodele, ``{Penalty weights in QUBO formulations: Permutation problems},'' in \emph{Evolutionary Computation in Combinatorial Optimization (EvoCOP 2022)}, ser. Lecture Notes in Computer Science, vol. 13222, 2022, pp. 159--174.

\bibitem{MontanezBarrera2024}
J.~A. Monta{\~n}ez-Barrera, D.~Willsch, A.~Maldonado-Romo, and K.~Michielsen, ``{Unbalanced penalization: a new approach to encode inequality constraints of combinatorial problems for quantum optimization algorithms},'' \emph{Quantum Science and Technology}, vol.~9, 2024, {Art. no. 025022, doi: 10.1088/2058-9565/ad35e4}.

\bibitem{Dominguez2023}
F.~Dominguez, J.~Unger, M.~Traube, B.~Mant, C.~Ertler, and W.~Lechner, ``{Encoding-independent optimization problem formulation for quantum computing},'' \emph{Frontiers in Quantum Science and Technology}, vol.~2, 2023, {Art. no. 1229471}.

\bibitem{Sawaya2023}
N.~P.~D. Sawaya, A.~T. Schmitz, and S.~Hadfield, ``{Encoding trade-offs and design toolkits in quantum algorithms for discrete optimization: coloring, routing, scheduling, and other problems},'' \emph{Quantum}, vol.~7, 2023, {Art. no. 1111}.

\bibitem{Chancellor2019}
N.~Chancellor, ``{Domain wall encoding of discrete variables for quantum annealing and QAOA},'' \emph{Quantum Science and Technology}, vol.~4, 2019, {Art. no. 045004}.

\bibitem{Tan2021}
B.~Tan, M.-A. Lemonde, S.~Thanasilp, J.~Tangpanitanon, and D.~G. Angelakis, ``{Qubit-efficient encoding schemes for binary optimisation problems},'' \emph{Quantum}, vol.~5, 2021, {Art. no. 454}.

\bibitem{Egger2021}
D.~J. Egger, J.~Mare\v{c}ek, and S.~Woerner, ``{Warm-starting quantum optimization},'' \emph{Quantum}, vol.~5, 2021, {Art. no. 479}.

\bibitem{TateSDP2023}
R.~Tate, M.~Farhadi, C.~Herold, G.~Mohler, and S.~Gupta, ``{Bridging classical and quantum with SDP initialized warm-starts for QAOA},'' \emph{ACM Transactions on Quantum Computing}, vol.~4, no.~2, 2023, {Art. no. 9}.

\bibitem{TateCustom2023}
R.~Tate, J.~Moondra, B.~Gard, G.~Mohler, and S.~Gupta, ``{Warm-started QAOA with custom mixers provably converges and computationally beats Goemans--Williamson's Max-Cut at low circuit depths},'' \emph{Quantum}, vol.~7, 2023, {Art. no. 1121}.

\bibitem{Sack2021}
S.~H. Sack and M.~Serbyn, ``{Quantum annealing initialization of the quantum approximate optimization algorithm},'' \emph{Quantum}, vol.~5, 2021, {Art. no. 491}.

\bibitem{Cain2023}
M.~Cain, E.~Farhi, S.~Gutmann, D.~Ranard, and E.~Tang, ``{The QAOA gets stuck starting from a good classical string},'' arXiv:2207.05089, 2023.

\bibitem{Hadfield2019}
S.~Hadfield, Z.~Wang, B.~O'Gorman, E.~G. Rieffel, D.~Venturelli, and R.~Biswas, ``{From the Quantum Approximate Optimization Algorithm to a Quantum Alternating Operator Ansatz},'' \emph{Algorithms}, vol.~12, 2019, {Art. no. 34}.

\bibitem{WangXY2020}
Z.~Wang, N.~C. Rubin, J.~M. Dominy, and E.~G. Rieffel, ``{XY mixers: Analytical and numerical results for the quantum alternating operator ansatz},'' \emph{Physical Review A}, vol. 101, 2020, {Art. no. 012320}.

\bibitem{Bartschi2020}
A.~B\"artschi and S.~Eidenbenz, ``{Grover mixers for QAOA: Shifting complexity from mixer design to state preparation},'' in \emph{2020 IEEE International Conference on Quantum Computing and Engineering (QCE)}, 2020, pp. 72--82.

\bibitem{HerrmanMultiAngle2022}
R.~Herrman, P.~C. Lotshaw, J.~Ostrowski, T.~S. Humble, and G.~Siopsis, ``{Multi-angle quantum approximate optimization algorithm},'' \emph{Scientific Reports}, vol.~12, 2022, {Art. no. 6781}.

\bibitem{DebEtAl2002}
K.~Deb, A.~Pratap, S.~Agarwal, and T.~Meyarivan, ``{A fast and elitist multiobjective genetic algorithm: NSGA-II},'' \emph{IEEE Transactions on Evolutionary Computation}, vol.~6, no.~2, pp. 182--197, 2002.

\bibitem{LandDoig1960}
A.~H. Land and A.~G. Doig, ``{An automatic method of solving discrete programming problems},'' \emph{Econometrica}, vol.~28, no.~3, pp. 497--520, 1960.

\bibitem{LawlerWood1966}
E.~L. Lawler and D.~E. Wood, ``{Branch-and-bound methods: A survey},'' \emph{Operations Research}, vol.~14, no.~4, pp. 699--719, 1966.

\bibitem{SheraliAdams1990}
H.~D. Sherali and W.~P. Adams, ``{A hierarchy of relaxations between the continuous and convex hull representations for zero-one programming problems},'' \emph{SIAM Journal on Discrete Mathematics}, vol.~3, no.~3, pp. 411--430, 1990.

\bibitem{McCormick1976}
G.~P. McCormick, ``{Computability of global solutions to factorable nonconvex programs: Part I. Convex underestimating problems},'' \emph{Mathematical Programming}, vol.~10, pp. 147--175, 1976.

\bibitem{Nusslein2023}
J.~N\"u{\ss}lein, S.~Zielinski, T.~Gabor, C.~Linnhoff-Popien, and S.~Feld, ``{Solving (Max) 3-SAT via quadratic unconstrained binary optimization},'' arXiv:2302.03536, 2023.

\bibitem{Nagies2025}
S.~Nagies, K.~T. Geier, J.~Akram, D.~Bantounas, M.~Johanning, and P.~Hauke, ``{Boosting quantum annealing performance through direct polynomial unconstrained binary optimization},'' \emph{Quantum Science and Technology}, vol.~10, 2025, {Art. no. 035008}.

\bibitem{Qiskit2024}
A.~Javadi-Abhari \emph{et~al.}, ``{Quantum computing with Qiskit},'' arXiv:2405.08810, 2024.

\bibitem{Powell1994}
M.~J.~D. Powell, ``{A direct search optimization method that models the objective and constraint functions by linear interpolation},'' in \emph{Advances in Optimization and Numerical Analysis}, S.~Gomez and J.-P. Hennart, Eds.\hskip 1em plus 0.5em minus 0.4em\relax Dordrecht, The Netherlands: Kluwer, 1994, pp. 51--67.

\bibitem{WangMaxCut2018}
Z.~Wang, S.~Hadfield, Z.~Jiang, and E.~G. Rieffel, ``{Quantum approximate optimization algorithm for MaxCut: A fermionic view},'' \emph{Physical Review A}, vol.~97, 2018, {Art. no. 022304}.

\bibitem{FarhiLocal2020}
E.~Farhi, D.~Gamarnik, and S.~Gutmann, ``{The quantum approximate optimization algorithm needs to see the whole graph: Worst case examples},'' arXiv:2005.08747, 2020.

\bibitem{Akshay2020}
V.~Akshay, H.~Philathong, M.~E.~S. Morales, and J.~D. Biamonte, ``{Reachability Deficits in Quantum Approximate Optimization},'' \emph{Physical Review Letters}, vol. 124, 2020, {Art. no. 090504}.

\end{thebibliography}

\begin{thebibliography}{10}
\providecommand{\url}[1]{#1}
\csname url@samestyle\endcsname
\providecommand{\newblock}{\relax}
\providecommand{\bibinfo}[2]{#2}
\providecommand{\BIBentrySTDinterwordspacing}{\spaceskip=0pt\relax}
\providecommand{\BIBentryALTinterwordstretchfactor}{4}
\providecommand{\BIBentryALTinterwordspacing}{\spaceskip=\fontdimen2\font plus
\BIBentryALTinterwordstretchfactor\fontdimen3\font minus \fontdimen4\font\relax}
\providecommand{\BIBforeignlanguage}[2]{{%
\expandafter\ifx\csname l@#1\endcsname\relax
\typeout{** WARNING: IEEEtran.bst: No hyphenation pattern has been}%
\typeout{** loaded for the language `#1'. Using the pattern for}%
\typeout{** the default language instead.}%
\else
\language=\csname l@#1\endcsname
\fi
#2}}
\providecommand{\BIBdecl}{\relax}
\BIBdecl

\bibitem{Rosenberg1975}
I.~G. Rosenberg, ``{Reduction of bivalent maximization to the quadratic case},'' \emph{Cahiers du Centre d'\'Etudes de Recherche Op\'erationnelle}, vol.~17, pp. 71--74, 1975.

\bibitem{BorosGruber2014}
E.~Boros and A.~Gruber, ``{On quadratization of pseudo-Boolean functions},'' arXiv:1404.6538, 2014.

\bibitem{VermaLewis2020}
A.~Verma and M.~Lewis, ``{Optimal quadratic reformulations of fourth degree pseudo-Boolean functions},'' \emph{Optimization Letters}, vol.~14, pp. 1557--1569, 2020.

\bibitem{VermaLewisKochenberger2021}
A.~Verma, M.~Lewis, and G.~Kochenberger, ``{Efficient QUBO transformation for higher degree pseudo Boolean functions},'' arXiv:2107.11695, 2021.

\bibitem{Schmidbauer2026}
L.~Schmidbauer, E.~Lobe, I.~Schaefer, and W.~Mauerer, ``{It's quick to be square: Fast quadratisation for quantum toolchains},'' \emph{ACM Transactions on Quantum Computing}, vol.~7, no.~3, pp. 1--46, 2026, {doi: 10.1145/3800943}.

\bibitem{Karp1972}
R.~M. Karp, ``{Reducibility among combinatorial problems},'' in \emph{Complexity of Computer Computations}, R.~E. Miller and J.~W. Thatcher, Eds.\hskip 1em plus 0.5em minus 0.4em\relax New York, NY, USA: Plenum, 1972, pp. 85--103.

\bibitem{CampbellDahl2022}
C.~Campbell and E.~Dahl, ``{QAOA of the highest order},'' in \emph{2022 IEEE 19th International Conference on Software Architecture Companion (ICSA-C)}, 2022, pp. 141--146.

\bibitem{Verchere2023}
Z.~Verch{\`e}re, S.~Elloumi, and A.~Simonetto, ``{Optimizing variational circuits for higher-order binary optimization},'' in \emph{2023 IEEE International Conference on Quantum Computing and Engineering (QCE)}, vol.~1, 2023, pp. 19--25.

\bibitem{Li2019}
G.~Li, Y.~Ding, and Y.~Xie, ``{Tackling the qubit mapping problem for NISQ-era quantum devices},'' in \emph{Proc. 24th Int. Conf. Architectural Support for Programming Languages and Operating Systems (ASPLOS)}, 2019, pp. 1001--1014.

\bibitem{Zulehner2019}
A.~Zulehner, A.~Paler, and R.~Wille, ``{An efficient methodology for mapping quantum circuits to the IBM QX architectures},'' \emph{IEEE Transactions on Computer-Aided Design of Integrated Circuits and Systems}, vol.~38, no.~7, pp. 1226--1236, 2019.

\bibitem{Qiskit2024}
A.~Javadi-Abhari \emph{et~al.}, ``{Quantum computing with Qiskit},'' arXiv:2405.08810, 2024.

\bibitem{Egger2021}
D.~J. Egger, J.~Mare\v{c}ek, and S.~Woerner, ``{Warm-starting quantum optimization},'' \emph{Quantum}, vol.~5, 2021, {Art. no. 479}.

\bibitem{TateCustom2023}
R.~Tate, J.~Moondra, B.~Gard, G.~Mohler, and S.~Gupta, ``{Warm-started QAOA with custom mixers provably converges and computationally beats Goemans--Williamson's Max-Cut at low circuit depths},'' \emph{Quantum}, vol.~7, 2023, {Art. no. 1121}.

\bibitem{Schmidbauer2024}
L.~Schmidbauer, K.~Wintersperger, E.~Lobe, and W.~Mauerer, ``{Polynomial reduction methods and their impact on QAOA circuits},'' in \emph{2024 IEEE International Conference on Quantum Software (QSW)}, 2024, pp. 35--45.

\bibitem{Powell1994}
M.~J.~D. Powell, ``{A direct search optimization method that models the objective and constraint functions by linear interpolation},'' in \emph{Advances in Optimization and Numerical Analysis}, S.~Gomez and J.-P. Hennart, Eds.\hskip 1em plus 0.5em minus 0.4em\relax Dordrecht, The Netherlands: Kluwer, 1994, pp. 51--67.

\bibitem{SheraliAdams1990}
H.~D. Sherali and W.~P. Adams, ``{A hierarchy of relaxations between the continuous and convex hull representations for zero-one programming problems},'' \emph{SIAM Journal on Discrete Mathematics}, vol.~3, no.~3, pp. 411--430, 1990.

\bibitem{McCormick1976}
G.~P. McCormick, ``{Computability of global solutions to factorable nonconvex programs: Part I. Convex underestimating problems},'' \emph{Mathematical Programming}, vol.~10, pp. 147--175, 1976.

\bibitem{Cain2023}
M.~Cain, E.~Farhi, S.~Gutmann, D.~Ranard, and E.~Tang, ``{The QAOA gets stuck starting from a good classical string},'' arXiv:2207.05089, 2023.

\bibitem{DebEtAl2002}
K.~Deb, A.~Pratap, S.~Agarwal, and T.~Meyarivan, ``{A fast and elitist multiobjective genetic algorithm: NSGA-II},'' \emph{IEEE Transactions on Evolutionary Computation}, vol.~6, no.~2, pp. 182--197, 2002.

\bibitem{LandDoig1960}
A.~H. Land and A.~G. Doig, ``{An automatic method of solving discrete programming problems},'' \emph{Econometrica}, vol.~28, no.~3, pp. 497--520, 1960.

\bibitem{LawlerWood1966}
E.~L. Lawler and D.~E. Wood, ``{Branch-and-bound methods: A survey},'' \emph{Operations Research}, vol.~14, no.~4, pp. 699--719, 1966.

\bibitem{Virtanen2020}
P.~Virtanen \emph{et~al.}, ``{SciPy 1.0: Fundamental algorithms for scientific computing in Python},'' \emph{Nature Methods}, vol.~17, pp. 261--272, 2020.

\bibitem{HuangfuHall2018}
Q.~Huangfu and J.~A.~J. Hall, ``{Parallelizing the dual revised simplex method},'' \emph{Mathematical Programming Computation}, vol.~10, pp. 119--142, 2018.

\end{thebibliography}
\end{document}